\documentclass[fleqn,usenatbib]{mnras}

\usepackage{cancel}
\usepackage[dvipsnames]{xcolor}
\usepackage{ae,aecompl}
\usepackage{natbib}
\usepackage{graphicx}
\usepackage{array}
\usepackage{amsmath}	
\usepackage{amssymb}
\usepackage{subcaption}
\usepackage{newtxtext,newtxmath}
\usepackage[T1]{fontenc}
\usepackage{multirow}

\newcommand*{\mr}{\mathrm}
\newcommand*{\msun}{\mathrm{M}_\odot}

\newcommand{\pren}[1]{\left(#1\right)}

\newcommand{\kpc}{\mathrm{kpc}}

\title{Probing Dark Matter with Strong Gravitational Lensing through an Effective Density Slope}

\author[A. Ç. Şengül and C. Dvorkin]{
Atınç Çağan Şengül$^{1}$\thanks{sengul@g.harvard.edu}
and Cora Dvorkin$^{1}$\thanks{cdvorkin@g.harvard.edu}\\
$^{1}$Harvard University, Department of Physics, Cambridge, Massachusetts, 02138, U.S.A.
}

\pubyear{2022}
\hypersetup{final}
\begin{document}
\label{firstpage}
\pagerange{\pageref{firstpage}--\pageref{lastpage}}
\maketitle

\begin{abstract}
Many dark matter (DM) models that are consistent with current cosmological data show differences in the predicted (sub)halo mass function, especially at sub-galactic scales, where observations are challenging due to the inefficiency of star formation. Strong gravitational lensing has been shown to be a useful tool for detecting dark low-mass (sub)halos through perturbations in lensing arcs, therefore allowing the testing of different DM scenarios. However, measuring the total mass of a perturber from strong lensing data is challenging. Over or underestimating perturber masses can lead to incorrect inferences about the nature of DM. In this paper, we argue that inferring an effective slope of the dark matter density profile, which is the power-law slope of perturbers at intermediate radii, where we expect the perturber to have the largest observable effect, is a promising way to circumvent these challenges. Using N-body simulations, we show that (sub)halo populations under different DM scenarios differ in their effective density slope distributions. Using realistic mocks of Hubble Space Telescope observations of strong lensing images, we show that the effective density slope of perturbers can be robustly measured with high enough accuracy to discern between different models. We also present our measurement of the effective density slope $\gamma=1.96\substack{+0.12 \\ -0.12}$ for the perturber in JVAS B1938+666, which is a $2\sigma$ outlier of the cold dark matter scenario. More measurements of this kind are needed to draw robust conclusions about the nature of dark matter.
\end{abstract}


\section{Introduction \label{sec:intro}}

Cosmological observations and galaxy dynamics have showed us that about 84\% of all matter in the universe is composed of dark matter (DM), whose presence can be
inferred via its gravitational influence. However, the Standard Model of particle physics lacks a description of it. The nature and interactions of DM remain one of the greatest puzzles of fundamental physics in modern times. 

The standard Cold Dark Matter (CDM) paradigm predicts that dark matter
is clustered into gravitationally bound systems called halos. These halos are expected to roughly follow a power-law spectrum from galaxy clusters to ultra-faint dwarf galaxies \citep{1985ApJ...292..371D,NFW,NFW2,1999ApJ...524L..19M,Green:2003un,Wang:2019ftp}. While the CDM model provides a good description of the universe at large cosmological scales \citep{planck2018,boss_matter_power,des_matter_power,cosmic_shear,lyman_alpha,large_scales}, smaller scales remain largely untested. Moreover, many theoretical models describing the microphysics of dark matter disagree with CDM at small scales \citep{2001ApJ...556...93B,PhysRevD.72.063510,milky_way_satellites,wdm_at_small,Bullock:2017xww,2018PhR...761....1B,Tulin:2017ara}.

Strong gravitational lensing is a promising observable that can be used as a probe of dark matter fluctuations at all scales, in particular those below that of the typical dwarf galaxies. Furthermore, in the coming years, we expect to have an increase in observed lensed systems, with tens of thousands expected to be discovered with upcoming optical imaging surveys \citep{Collett:2015roa,Bechtol:2019acd,2019ApJS..243...17J,2021ApJ...909...27H}.

The (sub)halo mass function, which is the number of dark matter (sub)halos as a function of mass, is commonly used to probe dark matter scenarios, in particular its low-mass end, where many theoretical DM models predict a suppression in the number of expected subhalos or halos along the line of sight \citep{wdm_mass_function,small_scale_hmf}. So far, there have been a handful of claimed detections of low-mass substructure in strong lens systems \citep{sdss_substructure,2012Natur.481..341V,hezaveh_alma_substructure} through perturbations in lensed arcs, as well as line-of-sight halos \citep{sengul2021}. The lack of detection of (sub)halos together with the few detections have been used to constrain the properties of DM \citep{unrealistic_tidal_assumption,ritondale_lack_of_detection}. At smaller physical scales where direct detection is not possible, the power spectrum of the collective perturbations of low-mass perturbers in the convergence field, either from subhalos \citep{Hezaveh:2014aoa,Cyr-Racine:2015jwa,DiazRivero:2017xkd,Cyr-Racine:2018htu,DiazRivero:2018oxk,Brennan:2018jhq} or line-of-sight halos \citep{CaganSengul:2020nat}, has been proposed as a powerful probe of CDM. The magnification ratio of multiple images has also been proposed as a probe of small-scale structure in the lens \citep{10.1046/j.1365-8711.1998.01319.x,2001ApJ...563....9M} and has been used to detect subhalos \citep{2002ApJ...572...25D,10.1046/j.1365-8711.2003.06055.x,2006MNRAS.367.1367A,10.1111/j.1365-2966.2011.19729.x,10.1111/j.1365-2966.2012.20484.x,MacLeod_2013,nierenberg_subhalo,10.1093/mnras/stx1400} and line-of-sight halos \citep{Gilman:2019vca}. Additionally, the collective effects of the large population of low-mass (sub)halos have been harnessed by machine learning techniques that attempt to extract information that may be buried in the observations \citep{Brewer:2015yya,Daylan:2017kfh,Brehmer:2019jyt,Ostdiek:2020mvo,Ostdiek:2020cqz,Wagner-Carena:2022mrn} as well as other types of techniques \citep{Birrer:2017rpp}.

When working with gravitational lensing as a probe, the total extended mass of a perturber is not an optimal parameter, since it relies on assumptions about the extended mass distribution \citep{vegetti_vogelsberger,unrealistic_tidal_assumption,better_mass}. Unusually dense perturbers will have a high lensing effect (despite their possible low mass), and vice versa. Hence, predicting the total mass from strong lensing measurements can be misleading. The substructure detected in the SDSSJ0946+1006 lens system \citep{sdss_substructure} is a good example of these challenges. Its total mass ($3.5 \times 10^{9}\, \msun$) was inferred by assuming a tidal truncation due to the host halo. Considering only its mass, this detection is consistent with predictions of CDM. However, a recent analysis \citep{sdss_slope} of the subhalo has shown that this substructure is inconsistent with CDM with an unusually steep density slope. A robust way of measuring perturber masses has been proposed in Refs. \cite{better_mass,Minor:2020bmp} where they define a ``perturbation radius", which is the distance from the perturber to the critical curve in the direction in which the perturbation to the magnification is largest. The effective perturber lensing mass (this is, the projected mass within the perturbation radius, divided by the log-slope of the main lens' 2D density profile) has been shown to be a robust estimator of the mass of the perturber under changes in the density profile \citep{better_mass}. In addition to measuring the perturber mass, we can probe the density of the perturber as it falls with increasing radius. More specifically, the effect of perturbers in strong lensing systems is quite sensitive to the slope of the average convergence of the perturber at a characteristic scale, where we expect the effect of the perturber to have the largest observable effect. We call this the {\it effective} density slope. This effective slope is expected to vary under different dark matter scenarios, so a measurement of the effective slope would provide fertile grounds for discerning between different dark matter models.

Measuring the effective density slope may also shed light on a discrepancy claimed between the observations of dwarf galaxies and CDM simulations, known as the ``cusp-core'' problem \citep{cusp_vs_core1,cusp_vs_core2,deBlok:2001hbg,Gentile:2004tb,2010AdAst2010E...5D}. The cusp-core problem refers to the fact that the observed central regions of galaxies (inferred from rotation curves) tend to have a lower amplitude and shallower slope of the density profile than the ones predicted from CDM simulations.  When the effects of baryons are neglected, the CDM halos show a self-similarity over a large mass range in simulations. These halos can be approximated by the Navarro-Frenk-White (NFW) profile \citep{NFW,NFW2}, with a ``cusp'' at its center, where the density ($\rho$) diverges with the distance $r$ to the center as $\rho \propto r^{-1}$. In contrast, the rotational speed of the stars in dwarf galaxies is observed to be roughly linear with radius, which implies a shallow inner slope, a ``core''. The origin of this problem is yet unclear, and a number of solutions have been proposed to this problem, ranging from new physics in the form of warm dark matter (WDM) \citep{2001ApJ...556...93B,wdm_halos2}, self-interacting dark matter (SIDM) \citep{sidm_base,velocity_dependent_sidm,sidm_core,sidm_core2}, ultra-light dark matter \citep{PhysRevLett.85.1158} or fluid dark matter \citep{2000ApJ...534L.127P} to the effect of baryons \citep{Navarro:1996bv,Gnedin:2001ec,10.1111/j.1365-2966.2004.08424.x,2006Natur.442..539M,cuspy_no_more,2012MNRAS.424.1275B,2012MNRAS.421.3464P,2015MNRAS.454.2981C}.  
A measurement of the effective density slope through gravitational lensing in dwarf galaxies should provide a powerful probe of CDM, since these systems are largely devoid of stars \citep{Fitts:2016usl,2017MNRAS.467.2019R}. 

An observed diversity in rotation curves of dwarf galaxies has also been claimed to be poorly reproduced in simulations \citep{diversity}. This could be due to a limitation of numerical simulations in accounting for supernovae feedback or star formation mechanisms, or it could also be due to non-circular motions in the gas not accounted for \citep{10.1093/mnras/sty2687,10.1093/mnras/sty354}. 
Ref. \cite{diversity} shows that systems for which the rotation curves do not agree with simulations also present a deficit in the enclosed mass compared to CDM simulations. They also show that for many of these systems the inner slopes are similar while the rotation curves are very different.  Different solutions in the form of dark matter self interactions and baryonic flows have been suggested to solve this problem \citep{sidm_solves_div,sidm_cross_section,Kaplinghat:2019dhn,baryons_on_diversity}. However, these solutions each have shortcomings, hence it remains an open question.

In this work, we suggest the use of gravitational lensing to measure the power-law slope of the average convergence in an intermediate regime to discern between possible dark matter scenarios. This could help shed light on the problems mentioned above and give us new insights into the nature of dark matter. 
We will restrict our analysis to low-mass systems ($\lesssim$ $10^9 \msun$), where the effect of baryons is expected to be small \citep{Fry:2015rta,2022arXiv220605338P}. 

This paper is organized as follows. In \S\ref{sec:region_of_max} we discuss how we determine the characteristic scale at which the power-law slope of a perturber is measured with lensing. In \S\ref{sec:sims} we analyze the density profiles obtained from N-body simulations of different DM scenarios and extract their effective power-law slopes. We test the accuracy and robustness of the measurements of the effective power-law slopes of perturbers using realistic mocks of strong lensing images in \S\ref{sec:HSTsims}. In \S\ref{sec:JVAS} we present a measurement of the effective slope of the density profile of a perturber previously found in the strong lens system JVAS B1938+666. In \S\ref{sec:conclusions} we summarize the results and the implications of our analysis. We include more details of our analysis in Appendices \ref{sec:inner profile}, \ref{sec:model_params}, and \ref{sec:residuals_JVAS}. 

\section{Region of Maximum Observability}\label{sec:region_of_max}

In general, our ability to probe the density profile of a perturber at very small radii is limited by resolution and noise. On the other hand, at much larger radii, the effect of the perturber is weaker and can be absorbed into the smooth model of the lens. Between these inner and outer radii, there is a region where we are most sensitive to the changes in the average convergence, which we will call the {\it region of maximum observability} (RMO). The effect on a lensing image of two perturbers with different density profiles is indistinguishable if they have overlapping average convergences within the RMO. 

When modeling a gravitationally lensed image with a perturber, one has to simultaneously model the main lens, the external shear, and the source light. In general, it is not possible to measure the average convergence of a perturber without assuming it has some functional form. Only then, one can measure the parameters of that functional form. Determining the sensitivity of the images to the average convergence at various radii becomes tractable if we consider the much simpler problem of measuring a perturber from a noisy image when we know the main lens mass distribution, the perturber position, and the source-light distribution $I_s(\vec y)$, where $\vec y$ is a 2D vector representing the angular position on the lens plane. All the uncertainties we derive will therefore be lower bounds on the true uncertainties from modeling where all the lens and source parameters are varied.

Our goal is to write a function that maps the average convergence of a perturber at various radii to the pixel brightness changes that it causes. We start by calling the surface brightness distribution on the image plane $I(\vec x)$, where $\vec x$ is a 2D vector representing an angular position on the image plane (we choose the center of the perturber to be $\vec x = 0$ without loss of generality). The unlensed and lensed brightness distributions are related by $I(\vec x) = I_s(\vec y)$, where the source plane position is given by the lens equation $\vec y = \vec x - \vec \alpha(\vec x)$, with $\alpha(\vec x)$ being the angular deflection due to the lens. In gravitationally lensed systems, $I(\vec x)$ is not directly measured. Instead, pixel values $p_{i,j}$ (with some noise which we will not explicitly write for clarity) of an $N_x \times N_y$ grid with $i \in \{1,2,...,N_x\}$ and $j \in \{1,2,...,N_y\}$ are measured. Each pixel value is given by
\begin{equation}\label{eq:pixel_values}
    p_{i,j} = \int_{i,j} d^2\vec x\, \bar I(\vec x),
\end{equation}
where $\int_{i,j}$ represents the integral over the the area of the pixel $(i,j)$ and $\bar I$ is the surface brightness convolved with the point spread function (PSF) $\delta$:
\begin{align}
    \bar I(\vec x) &= \int_{\mathbb{R}^2} d^2\vec u \, I(\vec u)\,\delta(\vec x - \vec u)  \label{eq:psf_convo} \\
    &=\int_{\mathbb{R}^2} d^2\vec u \, I_s(\vec v)\,\delta(\vec x - \vec u), \label{eq:psf_convo_source}
\end{align}
We can substitute $I_s(\vec v)$ by $I(\vec u)$ in Eq. \eqref{eq:psf_convo} to obtain Eq. \eqref{eq:psf_convo_source}, where $\vec v = \vec u - \vec \alpha(\vec u)$. 

In general, the maximum angular deflection due to typical perturbers considered in galaxy-galaxy lensing studies is much smaller than the average pixel size. For example, the angular deflection of an NFW perturber with mass $M_{200} = 10^{9}\, \msun$ ($M_{200}$ is the total mass enclosed within a radius at which the average density of the halo is $200$ times the critical density of the universe), concentration $c_{200} = 12$, lens redshift $z_{l} = 0.5$ and source redshift $z_{s} = 1.5$ is $|\vec \alpha_p|  < 0.0030''$. With that in mind, we split the angular deflection into two terms: $\vec \alpha = \vec \alpha_m + \vec \alpha_p$, where $\vec \alpha_m$ is the angular deflection due to the main lens and $\vec \alpha_p$ is the angular deflection due to the perturber. Since $|\vec \alpha_p| \ll |\vec \alpha_m|$, we expand $I_s(\vec v)$ around $\vec v_m \equiv \vec u - \vec \alpha_m(\vec u)$, the position on the source plane that gets lensed to $\vec u$ by the main lens, and only keep the terms up to linear order in $\vec \alpha_p$:
\begin{align}
    \bar I(\vec x) &= \int_{\mathbb{R}^2} d^2\vec u\, \left[I_s(\vec v_m)-\vec \alpha_p(\vec u) \cdot \nabla I_s (\vec v_m)\right] \delta(\vec x-\vec u) \label{eq:brightness}.
\end{align}

Integrating the first term gives the unperturbed brightness, which we denote $\bar I_0(\vec x)$. We can write the effect of the perturber as differences of pixel values $\Delta p_{i,j} \equiv p_{i,j} - p^{0}_{i,j}$, where $p^{0}_{i,j}$ are the pixel values that we would observe if the perturber did not exist. These differences can be obtained by substituting the second term of Eq. \eqref{eq:brightness} into Eq. \eqref{eq:pixel_values},
\begin{equation}
    \Delta p_{i,j} = \int_{i,j} d^2\vec x \int_{\mathbb{R}^2} d^2\vec u\,  \left[-\vec \alpha_p(\vec u)\cdot \nabla I_s(\vec v_m)\right]\delta(\vec x - \vec u).
    \label{eq:deltap}
\end{equation}
We see that pixel value differences caused by a perturber are larger if the source has a large directional derivative along the direction of the angular deflection caused by the perturber. We can approximate the integral over the pixel area in Eq. \eqref{eq:deltap} as the integrand evaluated at the center of the pixel multiplied by the pixel area. We can also approximate the convolution integral as a discrete sum, which gives
\begin{equation}
    \Delta p_{i,j} \approx \Delta x^2\sum_{k,l} \left[-\vec \alpha_p(\vec x_{k,l})\cdot \nabla I_s(\vec y_{m,k,l})\right] \delta_{i-k,j-l},
\end{equation}
where $\vec x_{k,l}$ is the center of the pixel $(k,l)$, $\vec y_{m,k,l}  = \vec x_{k,l} - \vec \alpha_m(\vec x_{k,l})$ is the position on the source plane that maps onto $\vec x_{k,l}$ after being lensed by only the main lens, and $\delta_{i-k,j-l}$ is the discretized PSF. 

For an axially symmetric perturber, the angular deflection is given by $\vec \alpha_p(\vec x) = \tilde \kappa (|\vec x|)\, \vec x$, where $\tilde \kappa (|\vec x|)$ is the average convergence of the perturber within the angular radius $|\vec x|$. We finally obtain a linear mapping from the average convergence values of the perturber at various radii to the pixel differences it causes:
\begin{equation}\label{eq:linear_relation}
    \Delta p_{i,j} \approx \Delta x^2\sum_{k,l} \left[-\vec x_{k,l}\cdot \nabla I_s(\vec y_{m,k,l})\right] \delta_{i-k,j-l}\, \tilde \kappa (|\vec x_{k,l}|).
\end{equation}
Let $K$ be the vector of model parameters where each element $K_i$ is the average convergence at some radii $r_i$. Let $D$ be the data vector where each element is a pixel difference $\Delta p_{i,j}$ of some pixel $(i,j)$. We now have a linear map $L$ that takes a set of average convergences to their pixel differences, $D = LK$. The elements of the matrix $L$ can be obtained from the coefficients that multiply $\tilde \kappa(|\vec x_{i,j}|)$ in Eq. \eqref{eq:linear_relation}. Assuming Gaussian pixel errors given by a covariance matrix $C$, we obtain uncertainties for each average convergence element, $K_i$, as
\begin{equation}\label{eq:kappa_noise}
    \sigma_i = \sqrt{[A^{-1}]_{i,i}} \quad \text{where} \quad A \equiv L^T C^{-1} L.
\end{equation}
We will refer to these uncertainties as {\it linear errors}, and they can be used as lower bounds on the true uncertainties. We define the RMO as the region where the relative linear errors in average convergence are less than twice the lowest relative error. We will show in \S \ref{sec:HSTsims} that with this definition of the RMO we can consistently predict the effective power-law slope of a given perturber profile. We see from Eq. \ref{eq:kappa_noise} that the linear errors, and consequently the RMO, is a function of the main lens, the perturber position, and the source-light distribution. Sharper source-light distributions with steeper gradients and higher image resolution broaden the RMO to include the more inner regions of the perturber profile.

\begin{figure}
    \centering
    \includegraphics[width=\linewidth]{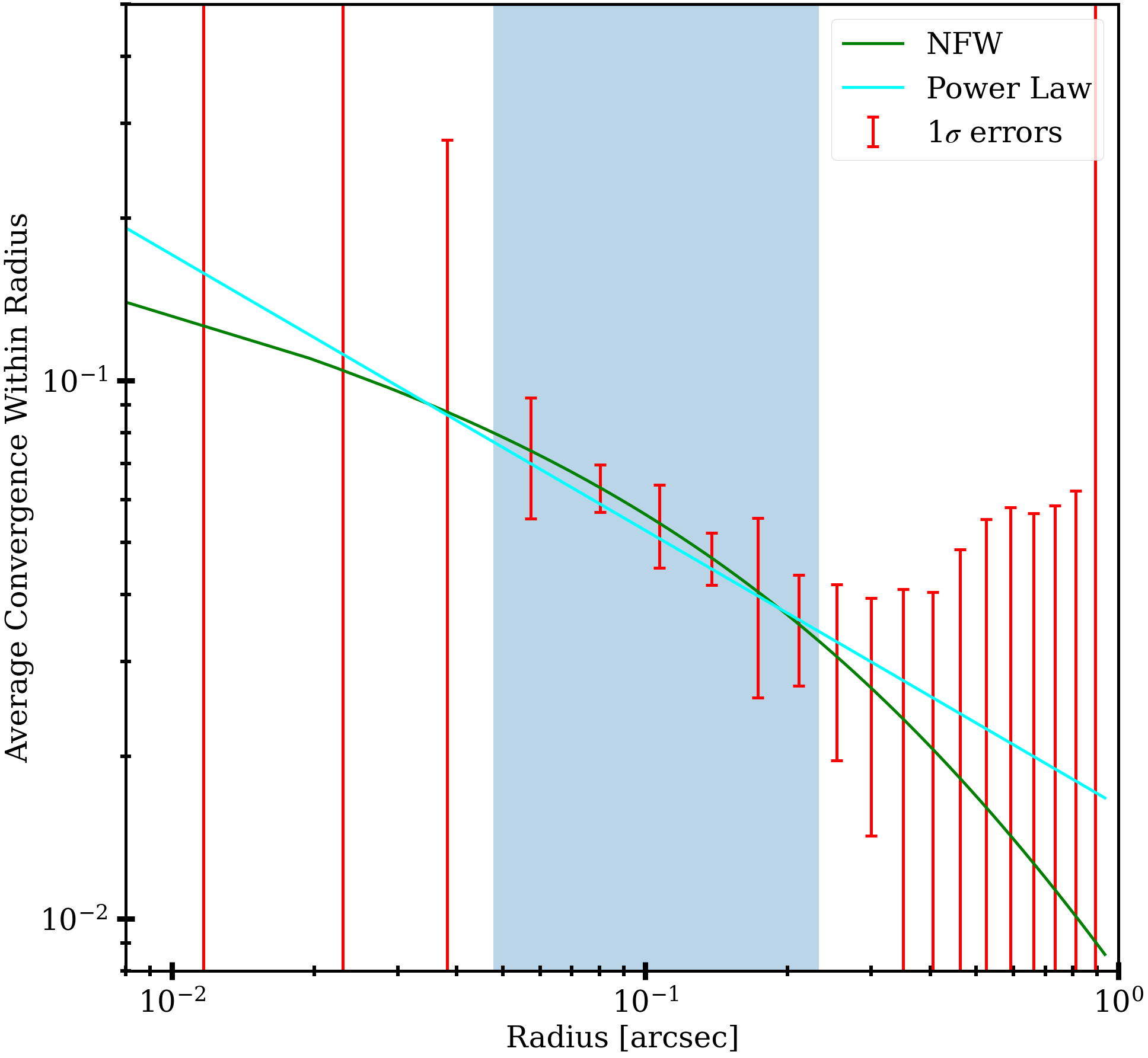}
    \caption{Sensitivity to the average convergence at various radii for an NFW perturber with $M_{200} = 10^{9}\, \msun$ and $c_{200} = 12$, where the main lens is at redshift $z=0.5$ and the source is at $z=1.5$. The green line shows the true NFW profile of the perturber. The cyan line shows the power-law best fit. The red error bars are the 1$\sigma$ linear errors obtained from Eq. \eqref{eq:kappa_noise}. The shaded-blue region shows the RMO, where the errors are less then twice the lowest one.}
    \label{fig:conv_vs_errors}
\end{figure}

Fig. \ref{fig:conv_vs_errors} we show an example of a result of this calculation, which assumes a PSF with a full width at half maximum (FWHM) of 0.049'', a pixel size of 0.025'', a source with a Sersic profile, a main lens with an Elliptical Power-Law mass profile, and a perturber located near one of the bright images of the source created by the main lens. We see that for scales that are smaller than the pixel size, we are not able to measure the average convergence. This means that the very inner region ($r$ < 0.04'') of any perturber placed on this location cannot be measured. For larger scales ($r$ > 0.3'') the average convergence is again not well constrained since the effect of the perturber gets weaker as you get further from it. Between these two limits lies the RMO, where our errors are small enough to determine the profile of the perturber. However, because this region only spans roughly half a decade, an NFW perturber, which has a varying power-law slope, behaves effectively like a single power-law whose slope is the average slope of the true profile within the RMO.

\section{Density Profile for Different Dark Matter Models}
\label{sec:sims}

In this section, we will analyze the density profile of high-resolution N-body simulations under different dark matter microphysics. The simulations used are the ETHOS (Effective Theory of Structure Formation) suite, originally presented in Refs. \cite{2016MNRAS.460.1399V,Cyr-Racine:2015ihg}. In particular, five different dark matter scenarios are analyzed: a cold dark matter scenario and four other non-standard scenarios that include dark matter-dark radiation interactions, that produce a cutoff in the primordial power spectrum, and dark matter self-interactions, that affect the non-linear structure formation (under the label of ETHOS1-4). The parameters used in each simulation are described in Table \ref{tab:CDM-ETHOS}. In particular, the parameters of the ETHOS4 simulation have been found to give a reasonable fit to observations of the Milky Way satellite population \citep{2016MNRAS.460.1399V}. 

\begin{table*}
\centering
\begin{tabular}{||c|c|c|c|c|c|c|c|c||}
\hline 
Name&$\alpha_\chi$&$\alpha_\nu$& $m_\phi$ [Gev/$c^2$] & $m_\chi$ [Gev/$c^2$]&$a_4$ [h/Mpc]&$\langle\sigma\rangle_{30}/m_\chi$[cm$^2$/g]&$\langle\sigma\rangle_{200}/m_\chi$ [cm$^2$/g]& $\langle\sigma\rangle_{1000}/m_\chi$ [cm$^2$/g] \\
\hline 
\hline
ETHOS1 & 0.071 & 0.123 & 0.723 & 2000 &14095.65 & 4.98 & 0.072& 0.0030\\
\hline
ETHOS2 & 0.016 & 0.03 & 0.83 & 500 &1784.05 & 9.0 & 0.197&0.00097\\
\hline 
ETHOS3 & 0.006 & 0.018 & 1.15 & 178 &305.94 & 16.9& 0.48&0.0028\\
\hline
ETHOS4& 0.5 & 1.5 & 5.0  & 3700 &286.09 & 0.16& 0.022&0.00075\\
\hline
\end{tabular}
\caption{Relevant parameters of the ETHOS (Effective Theory of Structure Formation) models considered in this work. Here $\alpha_\chi$ is the coupling between the mediator and dark matter, $\alpha_\nu$ is the coupling between the mediator and a massless neutrinos-like fermion, $m_\phi$ is the mass of the mediator and $m_\chi$ is the mass of the DM particle. The parameter $a_4$ is the coefficient of the power-law expansion in redshift of the
DM drag opacity caused by the DM-DR interaction, which determines the linear initial matter power spectrum (see Ref. \protect \cite{Cyr-Racine:2015ihg}). The parameters $\langle$$\sigma$$\rangle$$_{30}$, $\langle$$\sigma$$\rangle$$_{200}$ and $\langle$$\sigma$$\rangle$$_{1000}$ are the DM self-interaction cross-sections at different velocities: $30$, $200$ and $1000$ km/s and describe the velocity-dependent cross sections. We refer the reader to Ref. \protect \cite{2016MNRAS.460.1399V} for more details on these models.}
\label{tab:CDM-ETHOS}
\end{table*}

\begin{figure}
    \centering
    \includegraphics[width=0.99\linewidth]{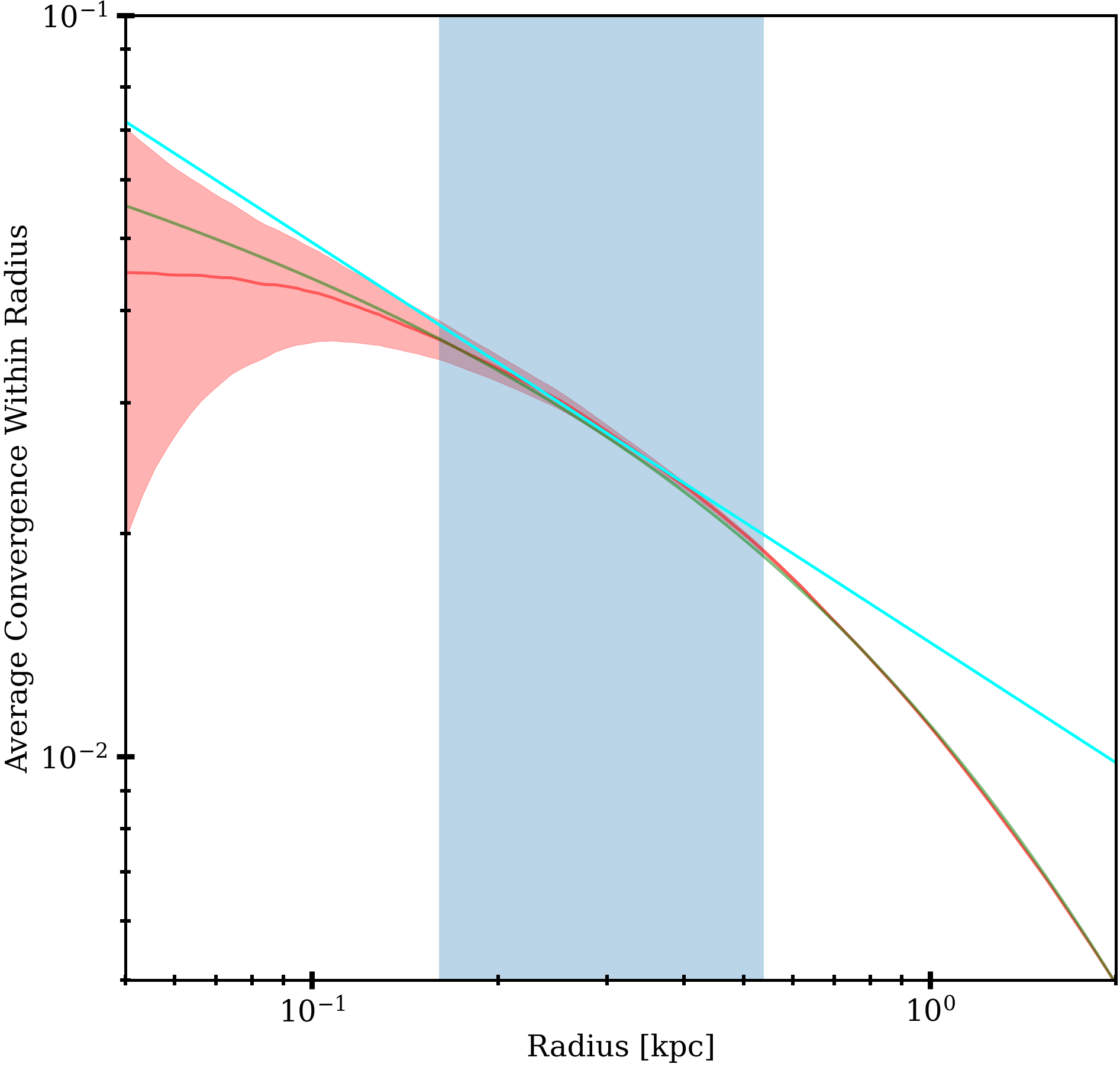}
    \caption{Illustrative example of the average convergence of a simulated CDM subhalo at $z=1.47$ (in red line). The $1\sigma$ uncertainty due to the Poisson noise is shown as the red-shaded region. The subhalo follows a truncated NFW profile with concentration $c_{200}=12.7$ and mass $M_{200}=4.3\times 10^8\msun$ (the fit for the average convergence is shown in green line). The shaded-blue region shows the RMO within which we will be measuring the effective density slope in this work. The RMO depends on the configuration of the strong lensing system that the subhalo is perturbing. For this case, we assumed that it is placed in the JVAS B1938+666 system (which we describe in \S \ref{sec:JVAS}). The power-law best fit, which is only applied within the RMO, is shown in the cyan line.}
    \label{fig:density_profile}
\end{figure}

For each dark matter scenario, we will extract the radial profiles of the perturbers in the simulation at different redshifts. In Fig. \ref{fig:density_profile}, we show an example of the normalized density profile of a CDM subhalo at redshift $z=1.47$. We then calculate the average convergence of each profile as a function of radii, which is what the angular deflections of a perturber depend on. The power-law slope of the average convergence profile in the RMO (shown as the blue-shaded region in Fig. \ref{fig:density_profile}) is what we will call the effective power-law slope for each perturber. We should note that the RMO will be different for each perturber detected via strong lensing, depending on the resolution, main lens, source light, and the perturber position. Therefore, the RMO calculation needs to be done on a case-by-case basis. For the analysis in this section, we will assume that the main lens is at $z=0.88$ and the source is at $z=2.079$, as in the JVAS B1938+666 system. The rest of the properties of the main lens and the source are given by the best-fit parameters of the JVAS B1938+666 lens system, as reported in Ref. \cite{sengul2021}.

\begin{figure}
    \centering
    \includegraphics[width=0.99\linewidth]{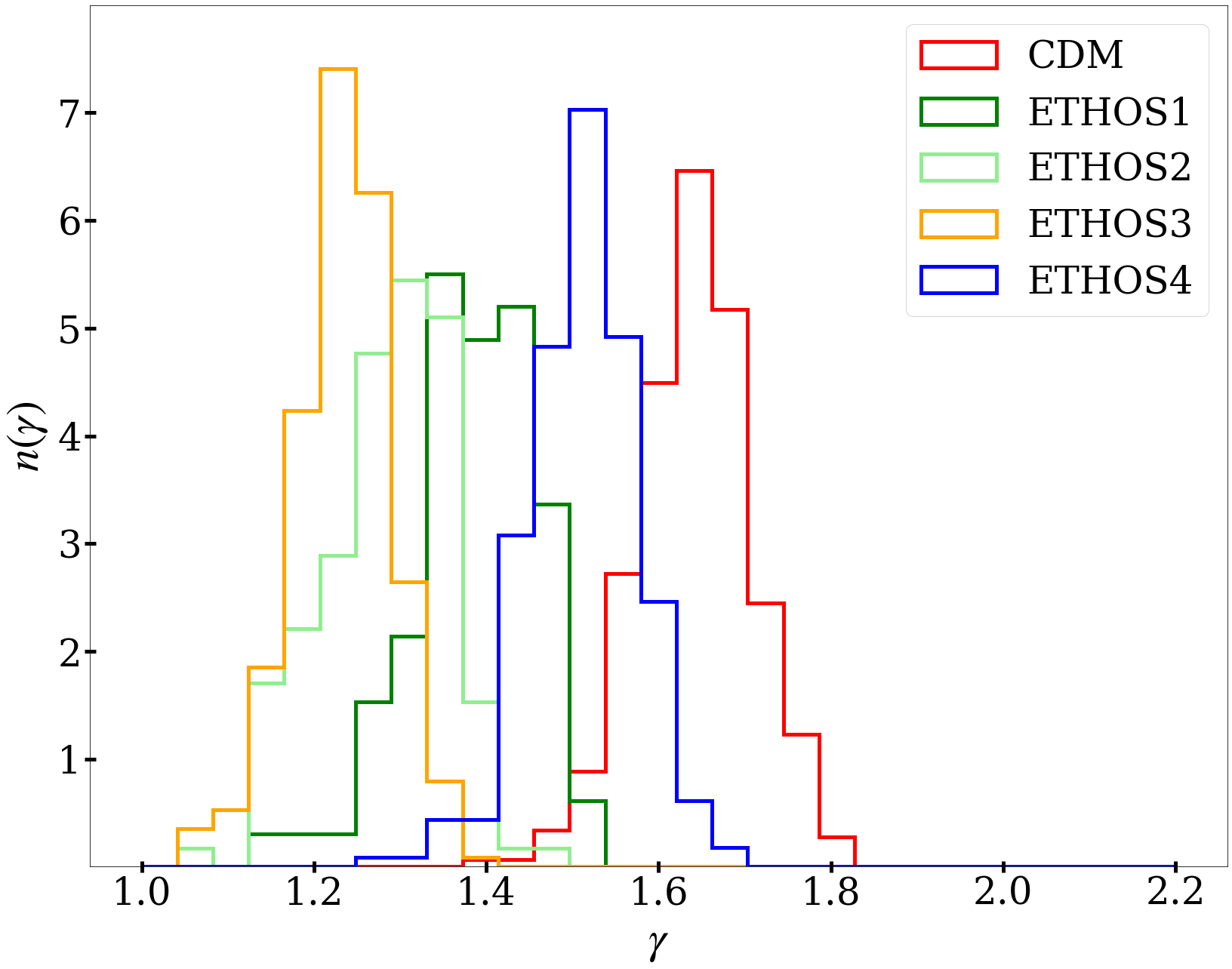}
    \caption{Normalized distributions of the effective density profile slopes ($\gamma$) of (sub)halos with masses $5\times 10^{8} \msun < M < 2\times 10^{9} \msun$ for all 5 DM scenarios (CDM, ETHOS1-4) analyzed in this work, at $z=1.47$. The y-axis shows the fractional abundance of (sub)halos. We see that the CDM population has the steepest mean effective slope.}
    \label{fig:gammas_of_models}
\end{figure}

We sort the particles of each perturber into radial bins and fit a power-law density profile ($\rho \propto r^{-\gamma}$) around the perturber radius to obtain the effective power-law slope, $\gamma$,  for each perturber. We assume that the number of particles in each radial bin is a Poissonian realization of an underlying power-law density profile. The details of this analysis can be found in Appendix \ref{sec:inner profile}. We only considered perturbers that are more massive than $5\times 10^{7}\, \msun$ to ensure that the radial bins are sufficiently populated to have an accurate effective power-law slope estimate. The particle mass of the simulations we use is $2.8\times 10^{4}\, \msun$, which ensures that the smallest perturbers consist of more than $10^3$ particles. 
We observe significant differences between the distribution of effective slopes in different DM scenarios. In Fig. \ref{fig:gammas_of_models} we show the effective density slope distributions of the different DM models at redshift $z=1.47$. The CDM population has a mean of $\gamma_{\rm CDM} = 1.64$ with a standard deviation of $\sigma(\gamma_{\rm CDM})=0.07$. In the ETHOS1-4 models, the dark matter interactions with dark radiation and with itself result in a shallower profile. The ETHOS4 distribution has the largest overlap with CDM with a mean of $\gamma_{\rm ETHOS4} = 1.51$ and a standard deviation of $\sigma(\gamma_{\rm ETHOS4}) = 0.06$. 

\begin{figure}
    \centering
    \includegraphics[width=0.99\linewidth]{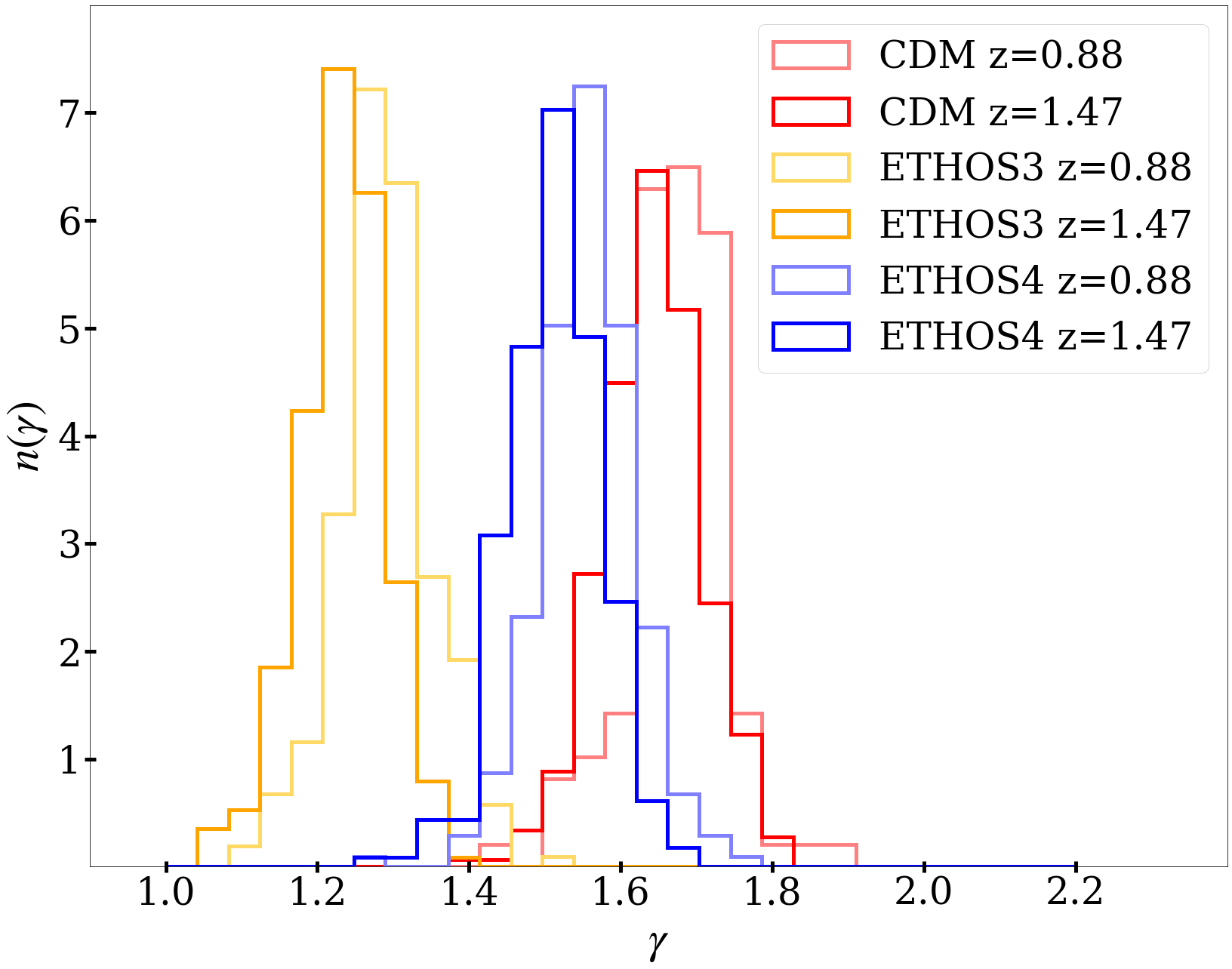}
    \caption{Redshift evolution of the normalized distributions of the effective density slopes ($\gamma$) of (sub)halos with masses $5\times 10^{8} \msun < M < 2\times 10^{9} \msun$ for CDM, ETHOS3, and ETHOS4. We see that for all scenarios the slope distributions shift towards lower values for higher redshifts.}
    \label{fig:z_dependence}
\end{figure}

We find a redshift dependence in the effective slope distributions of the different DM models, with a shift towards lower values of the slope as we increase the redshift. In Fig. \ref{fig:z_dependence}, we show the distribution of $\gamma$ for CDM, ETHOS3, and ETHOS4, at two redshifts. The mean $\gamma$ of the CDM population shifts by 0.4$\sigma$ (from $\gamma = 1.67$ to $\gamma=1.64$ when going from $z=0.88$ to $z=1.47$, while the mean of the ETHOS4 distribution shifts by $0.6\sigma$ (from $\gamma = 1.55$ to $\gamma=1.51$).
The redshift dependence for ETHOS3 is stronger, which is expected since it has a higher DM self-interaction cross section. The mean $\gamma$ in the ETHOS3 model, goes from $\gamma=1.29$ at $z=0.88$ to $\gamma=1.23$ at $z=1.47$, shifting by $1\sigma$. This is likely due to the effect of tidal stripping (see the discussion below). This effect needs to be accounted for when using the effective density slope to probe different dark matter scenarios.

\begin{figure}
    \centering
    \includegraphics[width=0.99\linewidth]{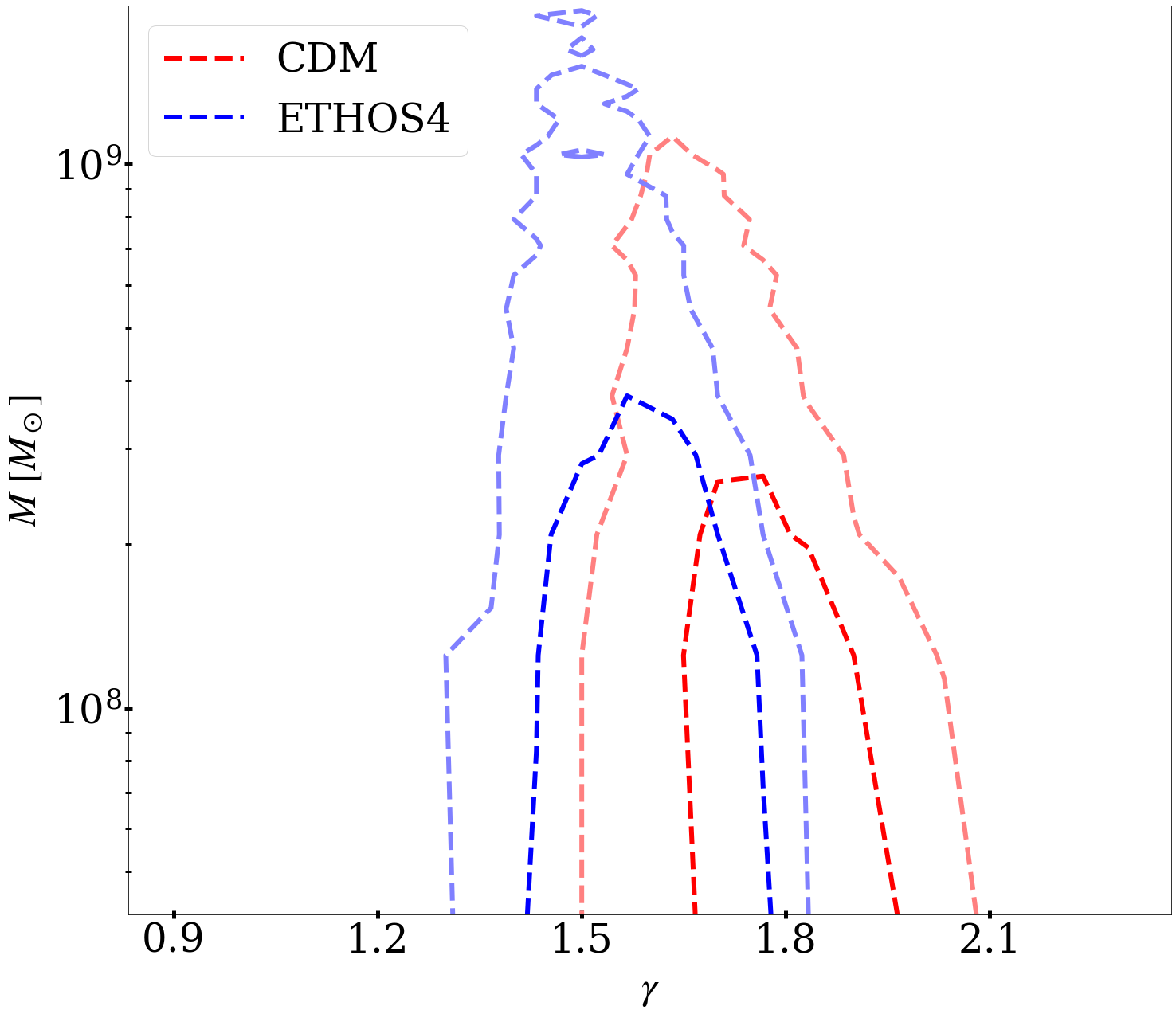}
 \caption{Distribution of mass vs. the effective density profile slope of CDM (in red) and ETHOS4 (in blue) perturbers at redshift $z=1.47$. Dashed contours enclose 68\% (darker color) and 95\% (lighter color) (which correspond to $1\sigma$ and $2\sigma$) of the perturbers around the mean, respectively.}
    \label{fig:mass_vs_gamma}
\end{figure}

In Fig. \ref{fig:mass_vs_gamma} we show the distribution of the masses and the effective power-law slopes of the perturbers at redshift $z=1.47$ for two dark matter scenarios: CDM and ETHOS4.
We note that lower masses have higher effective slopes on average. This mass dependence is stronger for CDM which makes the slope distributions at lower masses, with their means separated by $\approx 1.5\sigma$ for the case of CDM and ETHOS4. Due to the overlap in the distribution of slopes, we will not be able to distinguish between CDM and ETHOS4 with a single measurement. As we detect more and more perturbers, we will be able to determine from which underlying distribution the measured slopes are sampled.

\begin{figure}
    \centering
    \includegraphics[width=0.99\linewidth]{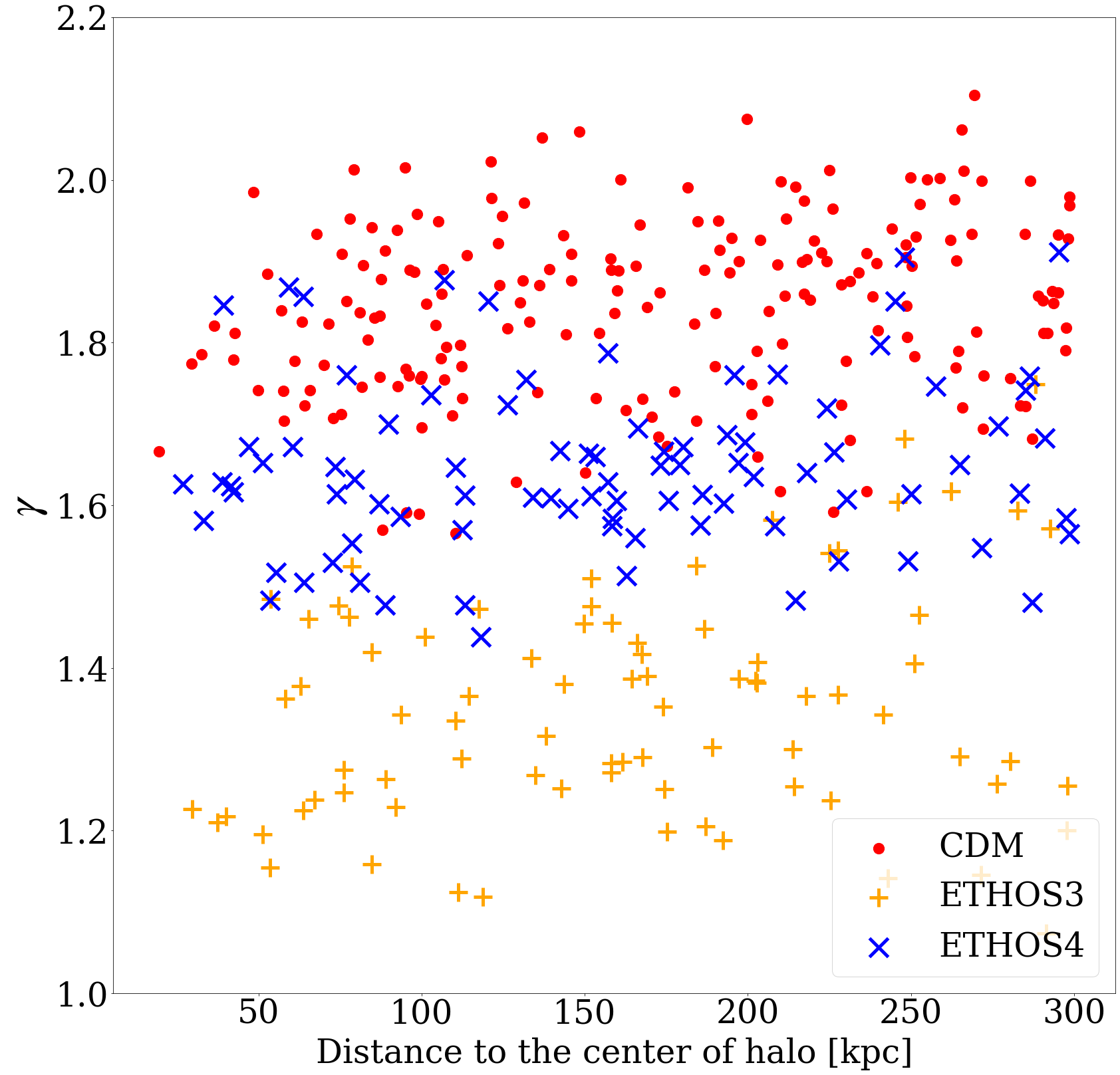}
 \caption{Effective density profile slope ($\gamma$) vs. the distance to the center of the host halo of the subhalos in the most massive halos at each of the CDM, ETHOS3, and ETHOS4 simulations at redshift $z=0.88$. The host halo has a total mass of $M = 1.3 \times 10^{12}\,\msun$ in each scenario.}
    \label{fig:gamma_vs_radius}
\end{figure}

We further explore the distribution of the slopes and distances to the host center for the subhalos of the most massive halo in each set of simulations (in Fig. \ref{fig:gamma_vs_radius}) and find a weak correlation between the two, most likely caused by the tidal forces of the host galaxy \citep{10.1046/j.1365-8711.1998.01285.x}. We find for all the models studied that the subhalos that are closer to the center of their host have lower slopes on average. This could be understood as follows: Tidal stripping can cause subhalos to lose over 90\% of their total mass. During this mass loss, the shape of the inner profile of the subhalo is preserved while the mass in the outer radii gets tidally stripped \citep{2010MNRAS.406.1290P,2019MNRAS.485..189O}. A subhalo that ends up with a certain mass $M$ after tidal stripping will have a larger scale radius compared to a subhalo with the same mass $M$ that did not undergo tidal stripping, since the former started as a much more massive subhalo. Measuring the density slope within the same radii interval for both of these subhalos will therefore give different values. For the tidally stripped subhalo, we will be probing the inner region of an originally more massive halo that lost its mass. These inner regions have a shallower density slope compared to the slope at larger radii. This correlation needs to be taken into account when using gravitational lensing measurements to discern between different dark matter scenarios.

\section{Measuring the Effective Density Slope from HST-like noisy Mocks}
\label{sec:HSTsims}

We made realistic mock images of gravitational lenses at the resolution and sensitivity of Hubble Space Telescope (HST) observations to demonstrate the feasibility of measuring the effective density slope of perturbers. The images are $60\times 60$ pixels with a pixel size of $0.025''$, which gives a field of view of $1.5'' \times 1.5''$. We assume a noise level of 10 HST orbits (33,000 seconds) of exposure and a Gaussian PSF with a full width at half maximum (FWHM) of $0.049''$. The sky brightness is 22.3 (AB magnitude) and the zero-point magnitude is 25.96.

In each image, we placed a perturber at a randomly chosen position near the critical curve, no dimmer than $50\%$ of the brightest pixel. We model the main lens with an Elliptical Power-Law (EPL) mass profile given by Ref. \cite{barkana_spemd}:
\begin{equation}\label{eq:epl}
    \kappa(x,y) = \frac{3-\gamma}{2}\pren{\frac{\theta_E}{\sqrt{qx^2 + y^2/q}}}^{\gamma-1},
\end{equation}
where $\gamma$ is the negative power-law slope of the 3D mass distribution (see Appendix \ref{sec:inner profile} for more details), $\theta_E$ is the Einstein radius, $x$ and $y$ are the angular coordinates aligned with the major and minor axes of the lens, and $q$ is the ratio of the major and minor axes. For the perturber, we assume an NFW profile, and we will fit it with an EPL mass profile with cylindrical symmetry, $q=1$. In addition, we include an external shear to capture the effect of weak lensing along the line of sight.

\begin{figure}
    \centering
    \includegraphics[width=0.99\linewidth]{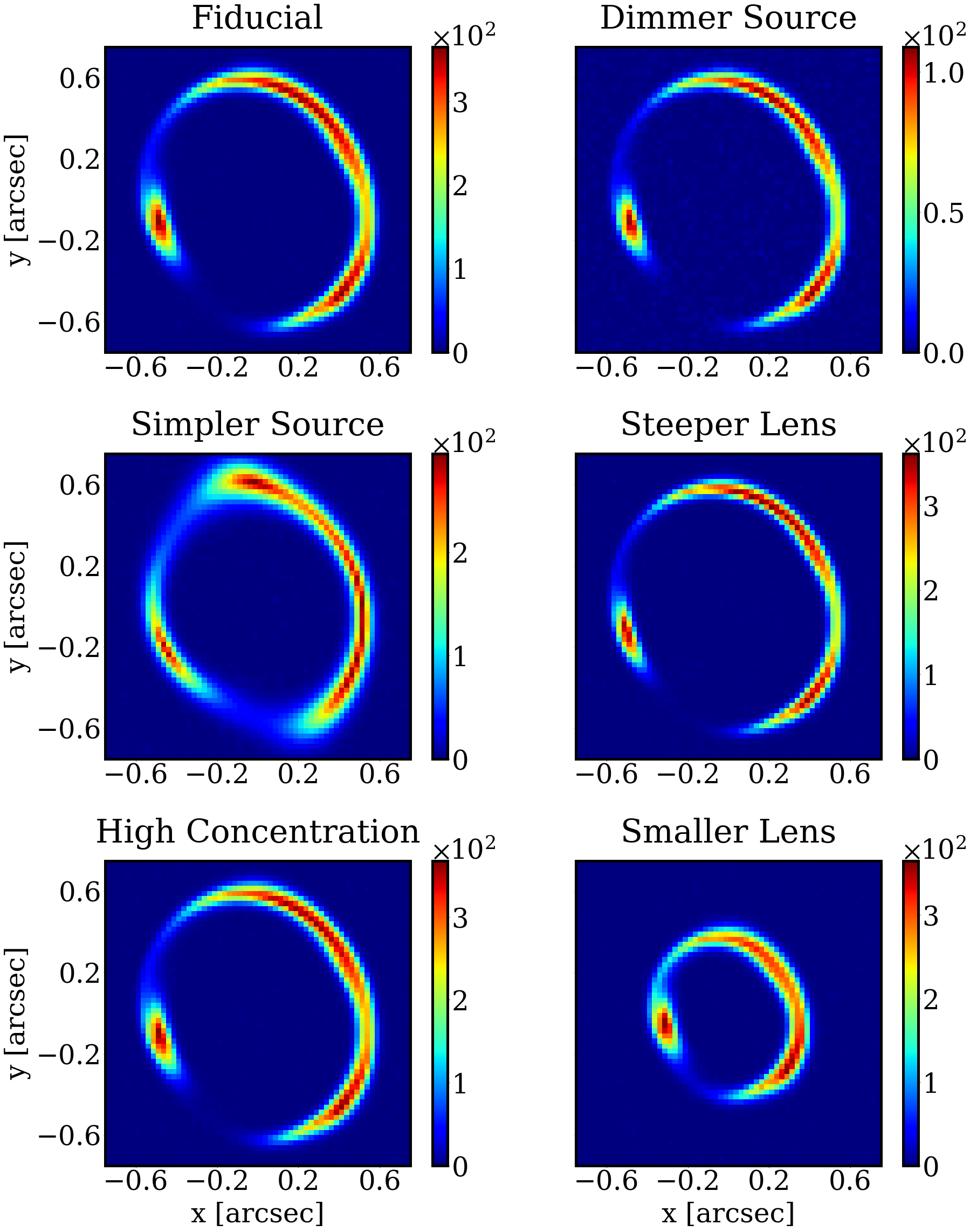}
    \caption{Mock images used in our analysis. The parameters of each image are described in Table \ref{tab:table_1}.}
    \label{fig:mock_images}
\end{figure}

We produced 6 mock images capturing possible variations in the models, which we list in Table \ref{tab:table_1} and show in Fig. \ref{fig:mock_images}. We set the lens redshift to $z=0.5$ and the source redshift to $z=1.5$ for all mock images. We analyze the mocks in this work using the publicly available code \texttt{lenstronomy} \citep{lenstronomy_shape,Birrer:2021wjl}.

\begin{table}
    \centering
    \begin{tabular}{||c|c|}
    \hline
       Model & Properties  \\
    \hline
    \hline
        {\it Fiducial} & $\theta_{E,\mr{main}} = 0.55''$ \\
        & $\gamma_{\mr{main}} = 2.0$\\
        & $M_\mr{200, perturber} = 10^{9}\, \msun$ \\
        & $c_{200}=12$\\
        & $m_\mr{AB} = 22$ \\
        & $N_\mr{sersic} = 5$\\
    \hline
        {\it Dimmer source}& $m_\mr{AB} = 24$ \\
    \hline
        {\it Simpler source} & $N_\mr{sersic} = 1$ \\
    \hline
        {\it Steeper lens} & $\gamma_{\mr{main}} = 2.3$ \\
        & $M_\mr{200, perturber} = 1.27\times 10^{9}\, \msun$\\
    \hline
        {\it High-concentration perturber} & $c_{200}=30$ \\
        & $M_\mr{200, perturber} = 1.93\times 10^{8}\, \msun$\\
    \hline
        {\it Smaller lens} & $\theta_{E,\mr{main}} = 0.35''$ \\
        & $M_\mr{200, perturber} = 2.78\times 10^{9}\, \msun$\\
    \hline
    \end{tabular}
    \caption{The properties of each of the models analyzed in this work (shown in Fig. \ref{fig:mock_images}). For the non-fiducial models we only list the parameters that differ from those of the fiducial. The remaining model parameters are chosen from uniform random distributions (see Appendix \ref{sec:model_params} for more details on the these models). The effective perturber lensing mass is kept fixed (at $1.5\times10^7 \msun$) in all the models (this is why the {\it Steeper lens}, {\it High-concentration perturber} and {\it Smaller lens} models have a different value of $M_{200}$).  The effective perturber mass is the projected mass within the perturbation radius, divided by the log-slope of the main lens' 2-dimensional density profile \citep{better_mass}.}
    \label{tab:table_1}
\end{table}

We will now explain each of the cases analyzed and refer the reader to Appendix \ref{sec:model_params} for more details on these models. We use \texttt{dynesty} \citep{2020MNRAS.493.3132S}, a publicly available nested sampling algorithm package, to sample the posterior probability distribution of the model parameters and to obtain the best fits from the sampled points. The main lens, external shear, and the perturber parameters are varied simultaneously during the posterior sampling. The means and standard deviations from the sampling as well as the true values for each case are shown in Table \ref{tab:table_2}.

\subsection*{Fiducial model:}
For our fiducial case, we create a main lens with an Einstein radius of $\theta_{E,\mr{main}} = 0.55''$ and a power-slope of $\gamma_\mr{main} = 2.0$. The x and y components of the center of the main lens, the ellipticity of the main lens, and the external shear are all chosen randomly from uniform distributions in the range $[-0.1'',0.1'']$, $[-0.1,0.1]$, and $[-0.1,0.1]$, respectively. We place an NFW perturber with mass $M_{200,\mr{perturber}} = 10^{9}\,\msun$ and concentration $c_{200}=12$ at $(x,y)\approx(0.212'',0.465'')$ near a lensed arc of the source. 

We assume a source composed of 5 Sersic profiles with randomly chosen ellipticities, radii, and Sersic indices. The Sersic light profile is given by
\begin{equation}\label{eq:sersic}
    I(\mathbf x) = I_0 \exp\left[-k\left\{\left(\frac{\sqrt{x_1^2 + x_2^2/q^2}}{R_\mathrm{s}}\right)^{1/n_\mathrm{s}} - 1\right\}\right],
\end{equation}
where $R_\mathrm{s}$ is the Sersic radius, $I_0$ is the surface brightness at $R_\mathrm{s}$, $x_1$ and $x_2$ are the coordinates of the centroid, $q$ is the ratio of the major and minor axes, and $n_\mathrm{s}$ is the Sersic index. The normalizing constant $k$ ensures that $R_\mathrm{s}$ is the half-light radius, which is where the surface brightness is half of the maximum value at the center. To test how much the source-light complexity affects our ability to measure the effective density slope, we also have created a mock lens that has a single Sersic profile as a source (this is the {\it Simpler source} model).

We model the source light with shapelets \citep{shapelets,lenstronomy_shape,Birrer:2018vtm,DES:2019fny,sengul2021}, which are an orthonormal set of weighted Hermite polynomials. The source complexity is controlled by the shapelet-order parameter, $n_\mr{max}$, that determines the number of degrees of freedom $N$ via the relation $N = (n_\mr{max}+1)(n_\mr{max}+2)/2$. The size of the shapelet reconstruction is determined by the scaling parameter $\delta$. Given a set of lens-model and shapelet parameters, the brightness of each pixel in the image is a linear function of the shapelet coefficients. If one knows the covariance matrix of the noise of the pixels, the best fit shapelet coefficients can be obtained by a simple matrix inversion. The centroid and the width of the shapelet basis are freely varied like any other model parameter. The optimal $n_\mr{max}$ changes depending on the complexity of the light distribution of the source galaxy and the resolution of the image. Therefore, we need to find the best fit $n_\mr{max}$ for each image on a case-by-case basis. We do this by starting from a low value of $n_\mr{max} = 5$ and iterating by $1$ until the Bayesian information criterion (BIC) of the best fit is larger than the previous iteration. The BIC is given by $\mathrm{BIC} = k\ln(n) +\chi^2$, where $k$ is the number of parameters in the model (this includes the number of shapelet coefficients) and $n$ is the number of unmasked pixels. 

\begin{figure*}
    \begin{subfigure}[t]{0.49\textwidth}
        \includegraphics[width=\textwidth]{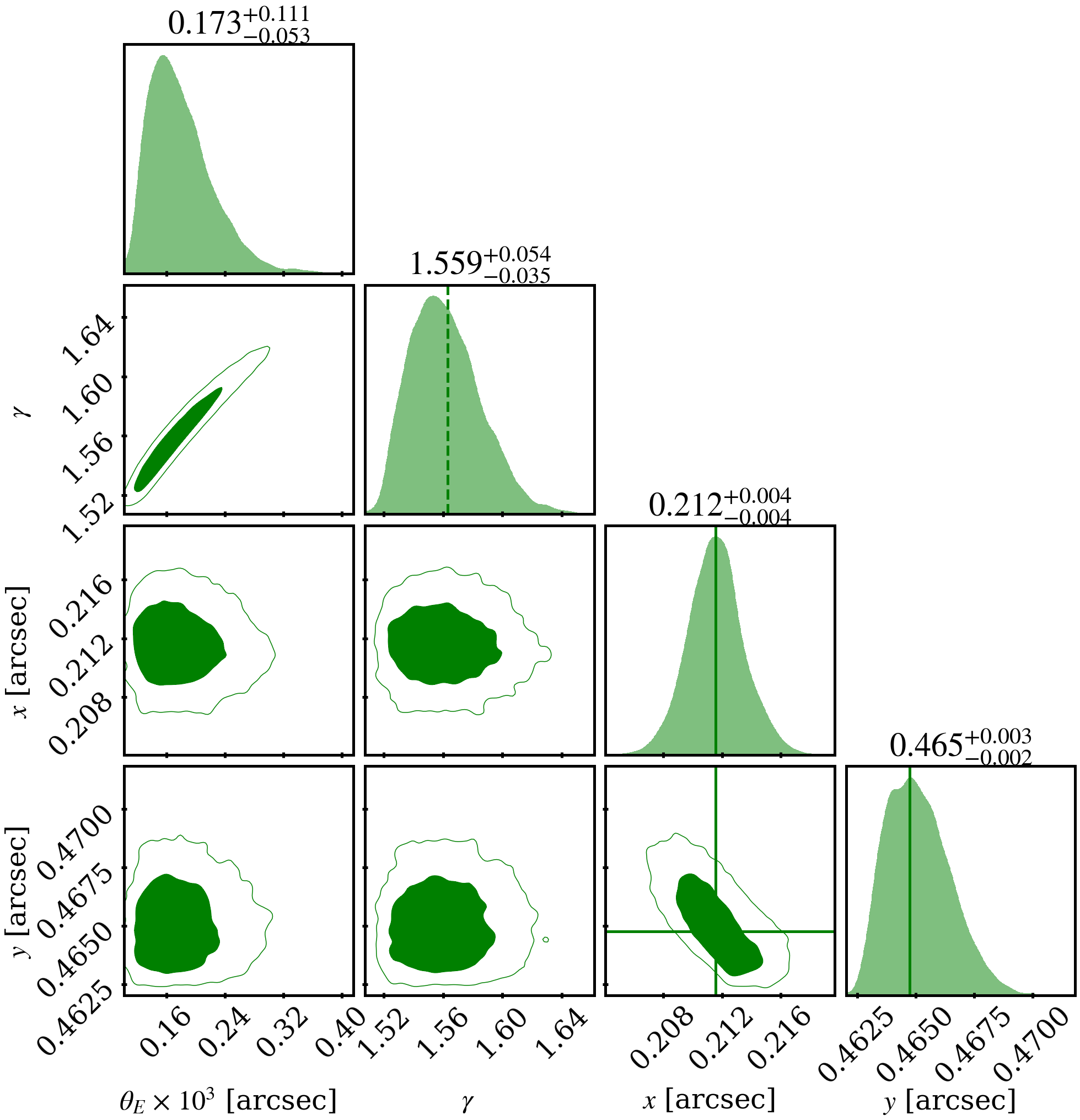}
        \caption{}
        \label{fig:fiducial_posterior}
    \end{subfigure}
    \hfill
    \begin{subfigure}[t]{0.49\textwidth}
            \centering
        \includegraphics[width=\textwidth]{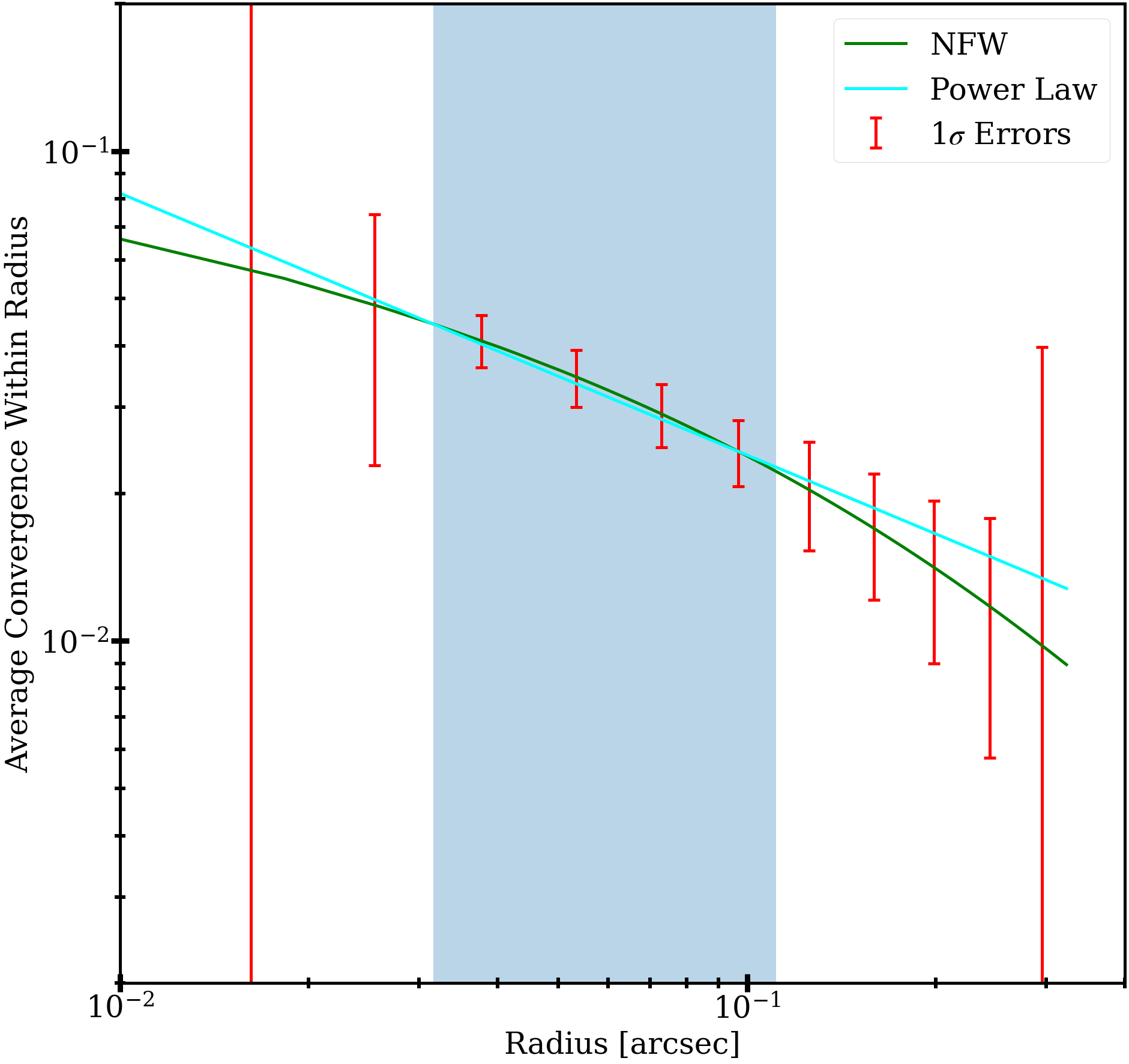}
        \caption{}
        \label{fig:fiducial_error}
    \end{subfigure}
         \label{fig: fiducial_figures}
    \caption{{\it  Fiducial} model: (a) Posterior probability distribution of the perturber parameters. The true position values are shown in solid green lines. The average power-law slope in the RMO is shown as the dashed green line. (b) Average convergence as a function of radius. The power-law best fit for the {\it Fiducial} model is shown in the cyan line. The true NFW profile is shown with solid green lines. The red error bars show the $1\sigma$ linear errors of the average convergence within each radii. The blue-shaded region corresponds to the RMO.}
\end{figure*}

We find that in the {\it Fiducial} model, we can measure a slope of $\gamma = 1.559 \substack{+0.054 \\ -0.035}$. The posteriors of the perturber parameters are shown in Fig. \ref{fig:fiducial_posterior}. The average convergence profile of the power-law best fit is compared with the true NFW profile in Fig. \ref{fig:fiducial_error} where we see that the best-fit slope agrees with the average slope within the RMO. The red error bars are the $1\sigma$ linear errors we get on the average convergences at each radius when all other model parameters are fixed. 
We note that this fiducial case is encouraging since we can measure $\gamma$ with an uncertainty of 3\%, which is smaller than the differences in the means of the different DM populations (which is at least 7\%, as shown in  Fig. \ref{fig:gammas_of_models}).

\subsection*{Dimmer source:}
\begin{figure*}
    \begin{subfigure}[t]{0.49\textwidth}
        \includegraphics[width=\textwidth]{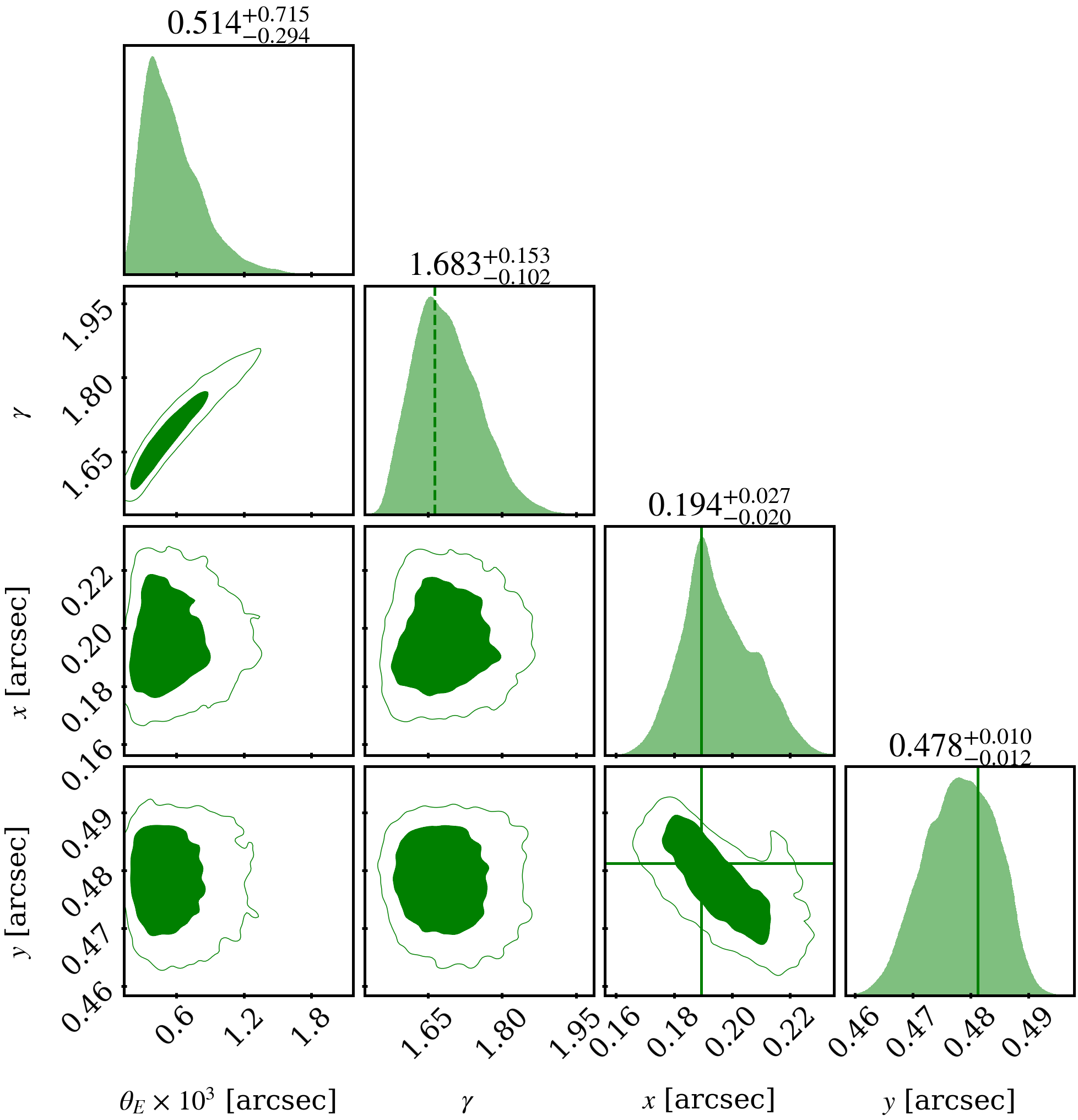}
        \caption{}
        \label{fig:dimmer_posterior}
    \end{subfigure}
    \hfill
    \begin{subfigure}[t]{0.49\textwidth}
            \centering
        \includegraphics[width=\textwidth]{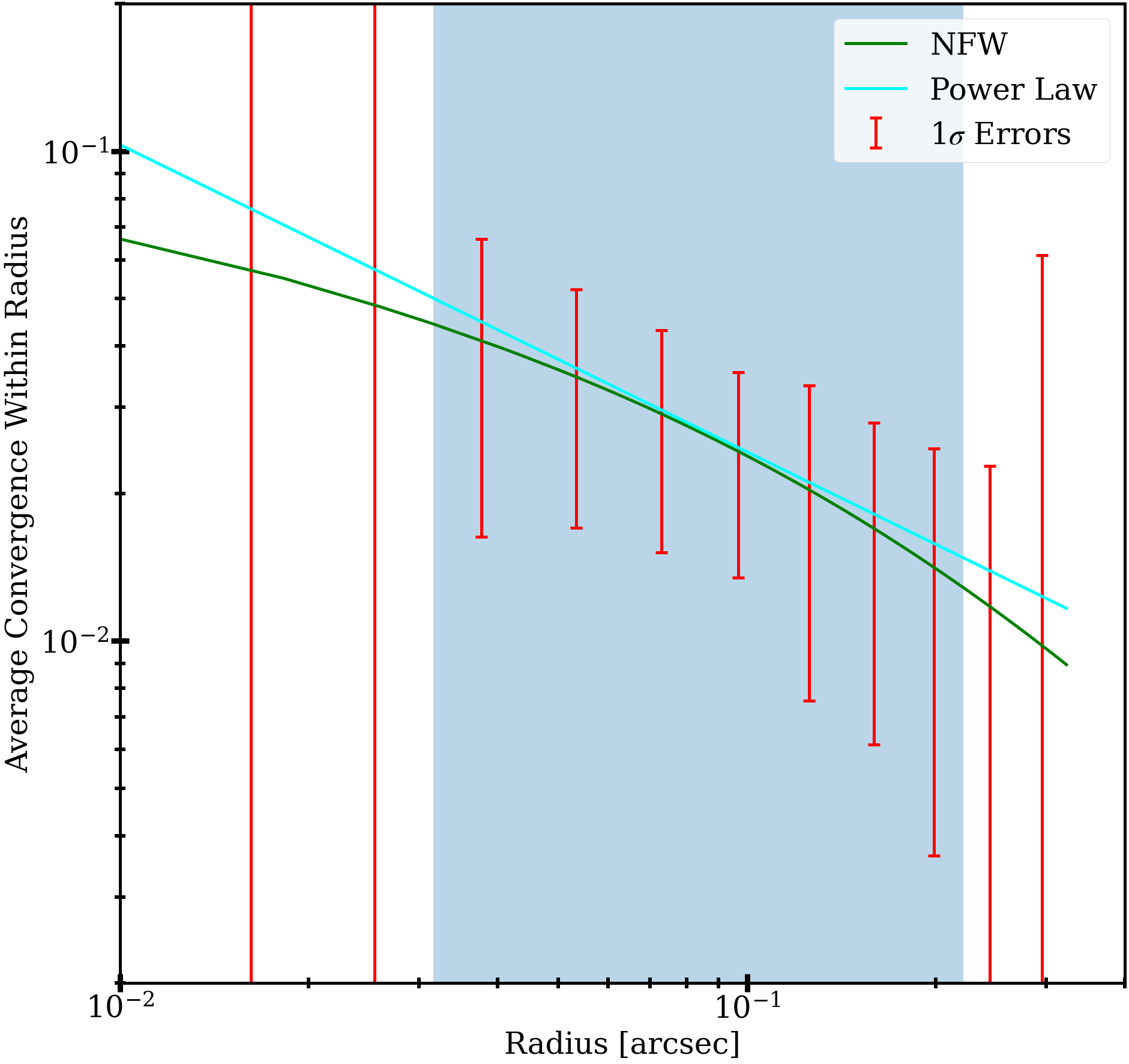}
        \caption{}
        \label{fig:dimmer_error}
    \end{subfigure}
    \label{fig:dimmer_figures}
    \caption{{\it  Dimmer source} model: (a) Posterior probability distribution of the perturber parameters. The true position values are shown in solid green lines. The average power-law slope in the RMO is shown as the dashed green line. (b) Average convergence as a function of radius. The power-law best fit for the {\it Dimmer source} model is shown in the cyan line. The true NFW profile is shown with solid green lines. The red error bars show the $1\sigma$ linear errors of the average convergence within each radii. The blue-shaded region corresponds to the RMO.}
\end{figure*}

We analyze a case where the source is dimmer than the fiducial, with an AB magnitude of 24 instead of 22 (the amount of light we receive is reduced roughly by a factor of 6).  We find, as expected, that the uncertainties of the model parameters are larger. The posteriors are shown in Fig. \ref{fig:dimmer_posterior}. We see in \ref{fig:dimmer_error} that the RMO has shifted to larger radii compared to the fiducial case, due to the change in source brightness, which agrees with the steeper power-law measurement from modeling. Larger linear errors within the RMO mean that the pixel differences are consistent with a wider range of power-law slopes, which limits our capacity to measure the effective slope of the system. Source brightness is the most important factor that limits our capacity to probe the effective density slope of the perturbers.

\subsection*{Simpler source:}
\begin{figure*}
    \begin{subfigure}[t]{0.49\textwidth}
        \includegraphics[width=\textwidth]{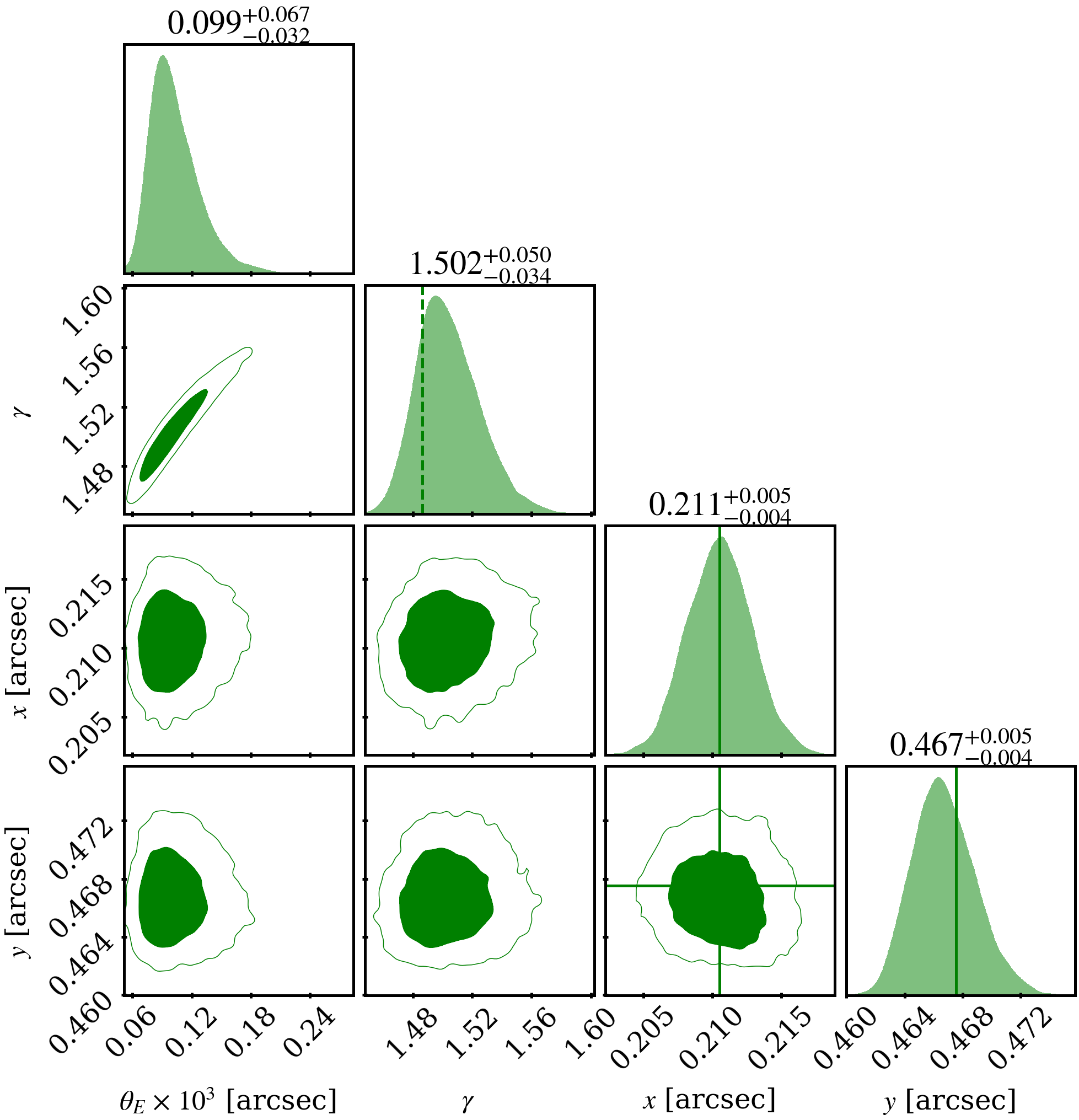}
        \caption{}
        \label{fig:simpler_posterior}
    \end{subfigure}
    \hfill
    \begin{subfigure}[t]{0.49\textwidth}
            \centering
        \includegraphics[width=\textwidth]{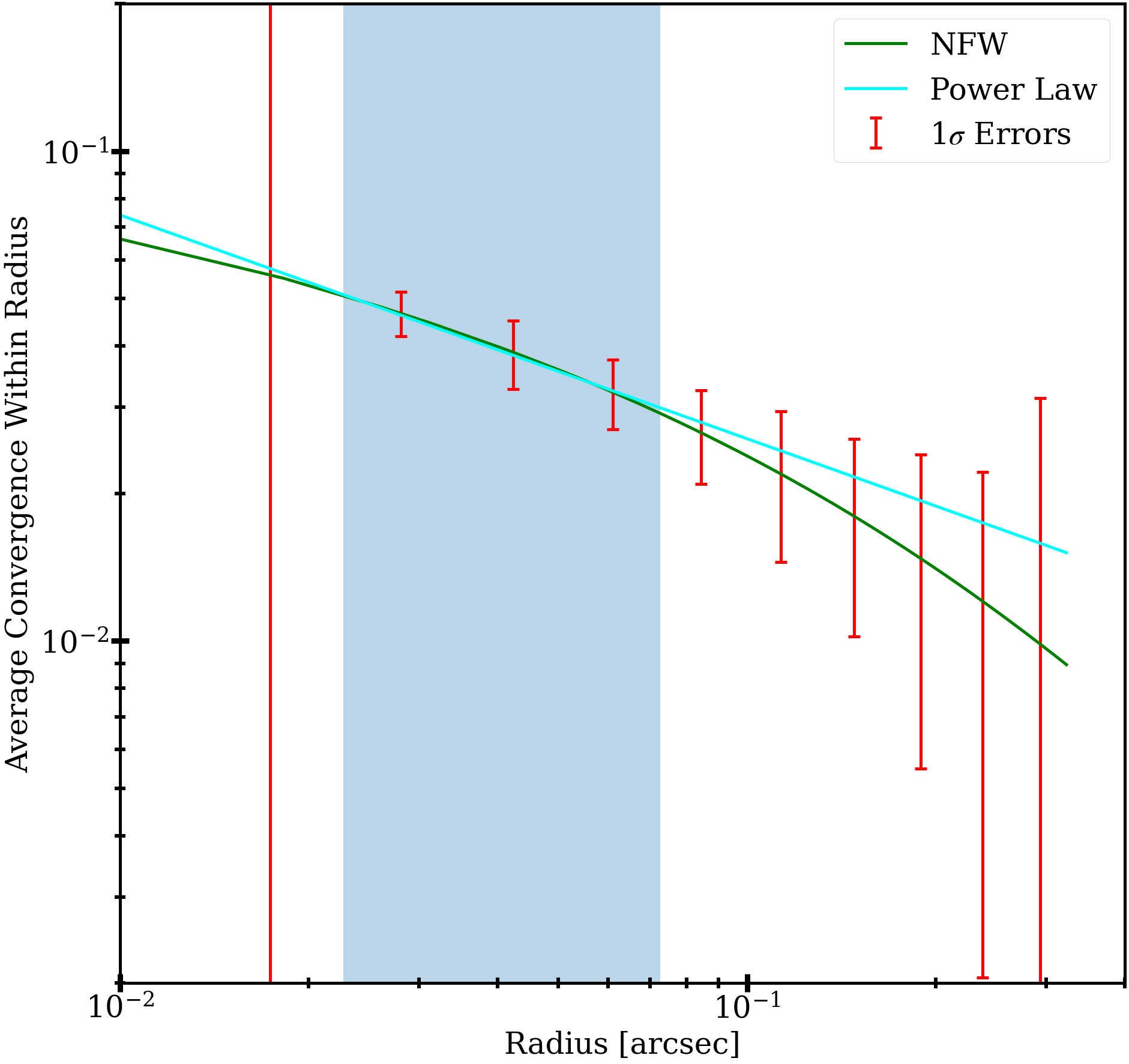}
        \caption{}
        \label{fig:simpler_error}
    \end{subfigure}
    \label{fig:simpler_figures}
    \caption{{\it  Simpler source} model: (a) Posterior probability distribution of the perturber parameters. The true position values are shown in solid green lines. The average power-law slope in the RMO is shown as the dashed green line. (b) Average convergence as a function of radius. The power-law best fit for the {\it Simpler source} model is shown in the cyan line. The true NFW profile is shown with solid green lines. The red error bars show the $1\sigma$ linear errors of the average convergence within each radii. The blue-shaded region corresponds to the RMO.}
\end{figure*}

If instead the source consists of a single Sersic profile instead of the 5 Sersics, we can measure the slope $\gamma = 1.502 \substack{+0.05 \\ -0.034}$.  We see that lowering source complexity does not change our ability to probe the effective power-law slope of the perturber. The posteriors are shown in Fig. \ref{fig:simpler_posterior}. As we can see in Fig. \ref{fig:simpler_error}, the RMO shifts towards smaller radii, which is consistent with our measurement of a shallower slope of the perturber. 

\subsection*{Steeper lens:}
\begin{figure*}
    \begin{subfigure}[t]{0.49\textwidth}
        \includegraphics[width=\textwidth]{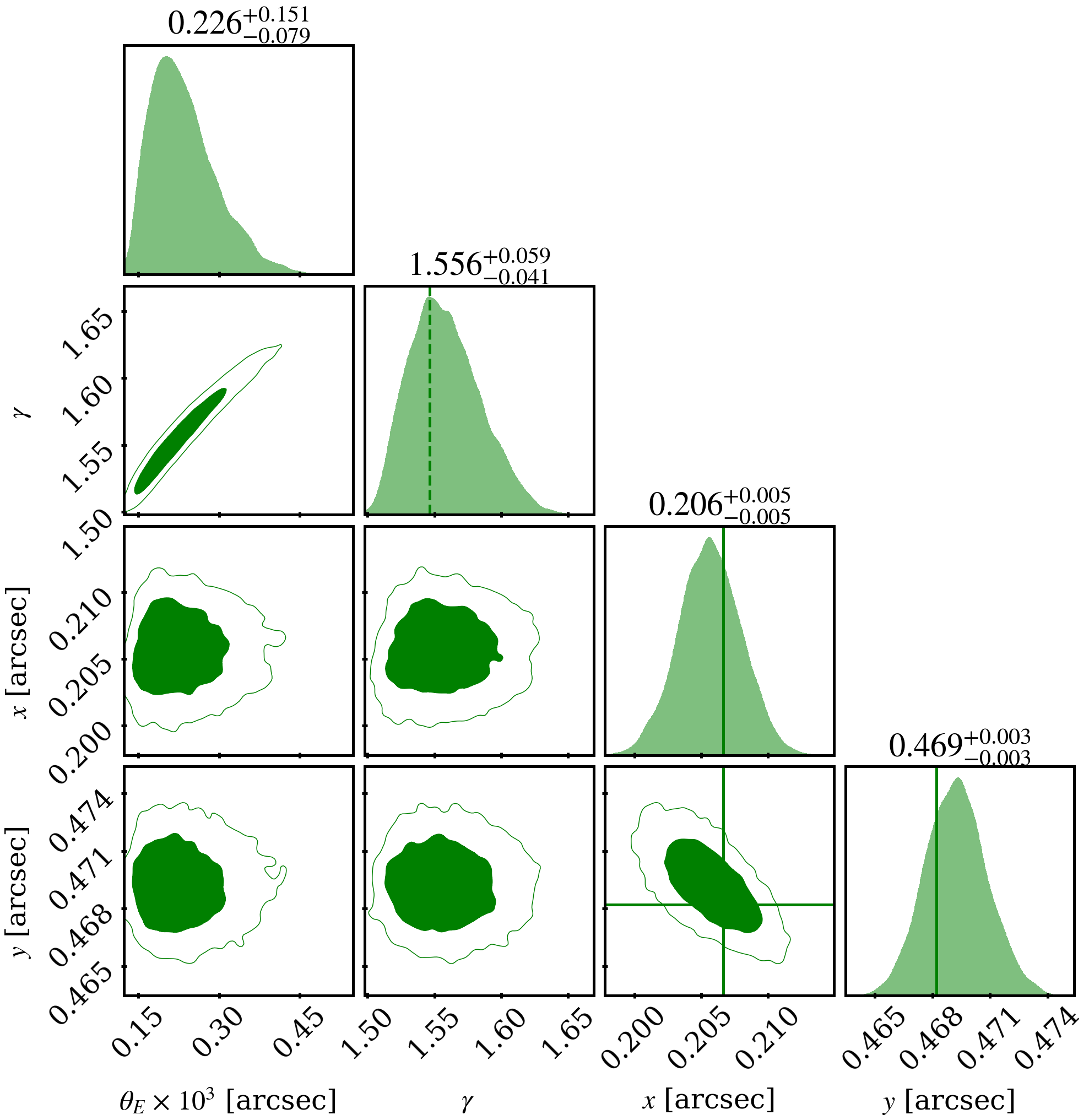}
        \caption{}
        \label{fig:steep_lens_posterior}
    \end{subfigure}
    \hfill
    \begin{subfigure}[t]{0.49\textwidth}
            \centering
        \includegraphics[width=\textwidth]{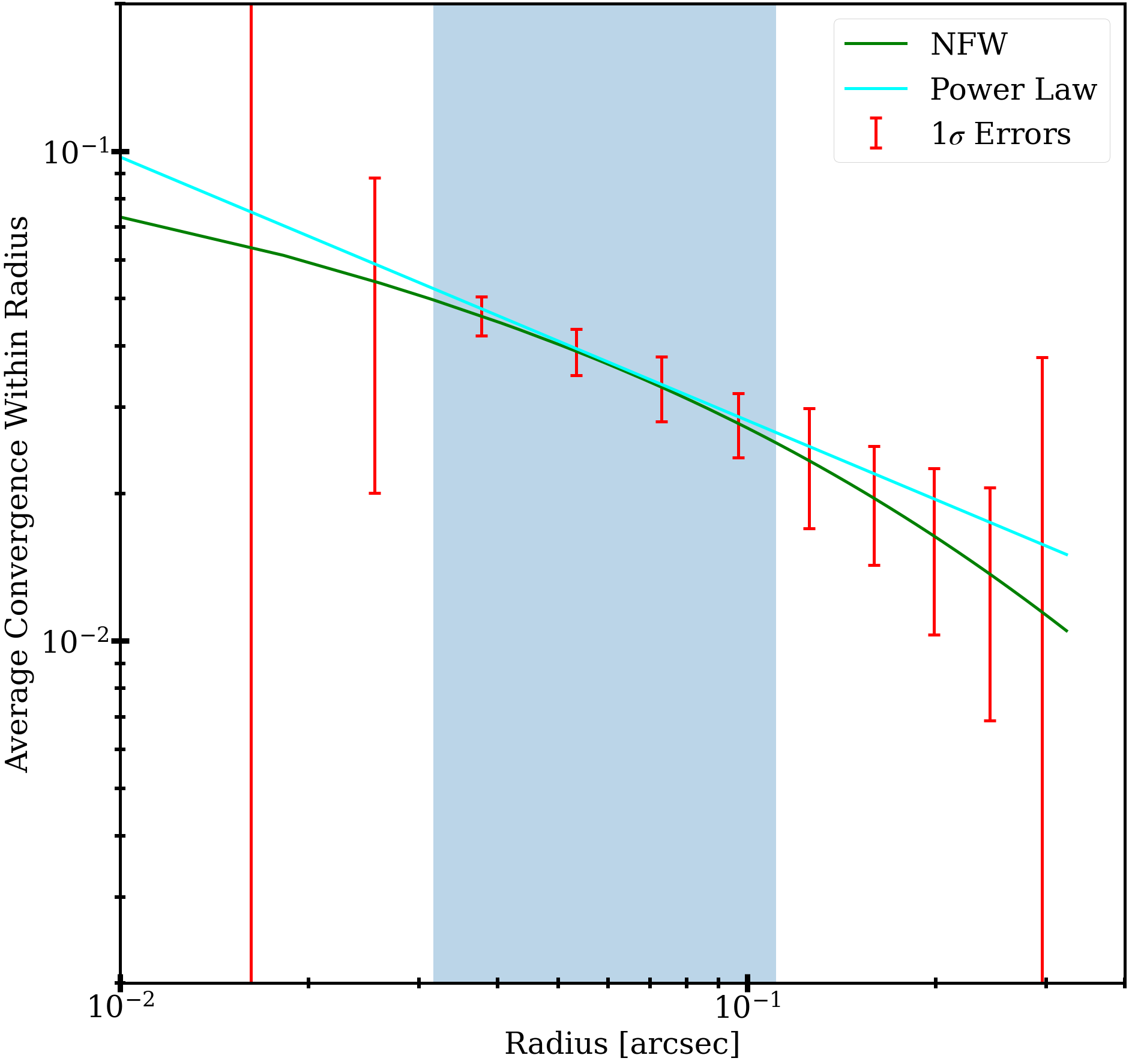}
        \caption{}
        \label{fig:steep_lens_error}
    \end{subfigure}
    \label{fig:steep_lens_figures}
    \caption{{\it  Steeper lens} model: (a) Posterior probability distribution of the perturber parameters. The true position values are shown in solid green lines. The average power-law slope in the RMO is shown as the dashed green line. (b) Average convergence as a function of radius. The power-law best fit for the {\it Steeper lens} model is shown in the cyan line. The true NFW profile is shown with solid green lines. The red error bars show the $1\sigma$ linear errors of the average convergence within each radii. The blue-shaded region corresponds to the RMO.}
\end{figure*}

When the power-law slope of the matter distribution of the main lens is steeper (we increased it to $2.3$), we measure a slope $\gamma = 1.556 \substack{+0.059 \\ -0.041}$ for the perturber. The uncertainty is slightly higher than that of the fiducial model, but the measurement is still precise enough to distinguish between the different DM scenarios analyzed in this work. The posteriors are shown in Fig. \ref{fig:steep_lens_posterior}. We find that the RMO does not get significantly affected by the change in the slope of the main lens, which can be seen by comparing Figs. \ref{fig:fiducial_error} and \ref{fig:steep_lens_error}.
Note that the perturber in this case has a larger $M_{200,\mr{perturber}}$ compared to the fiducial one to keep the effective perturber lensing mass fixed.

\subsection*{High-concentration perturber:}
\begin{figure*}
    \begin{subfigure}[t]{0.49\textwidth}
        \includegraphics[width=\textwidth]{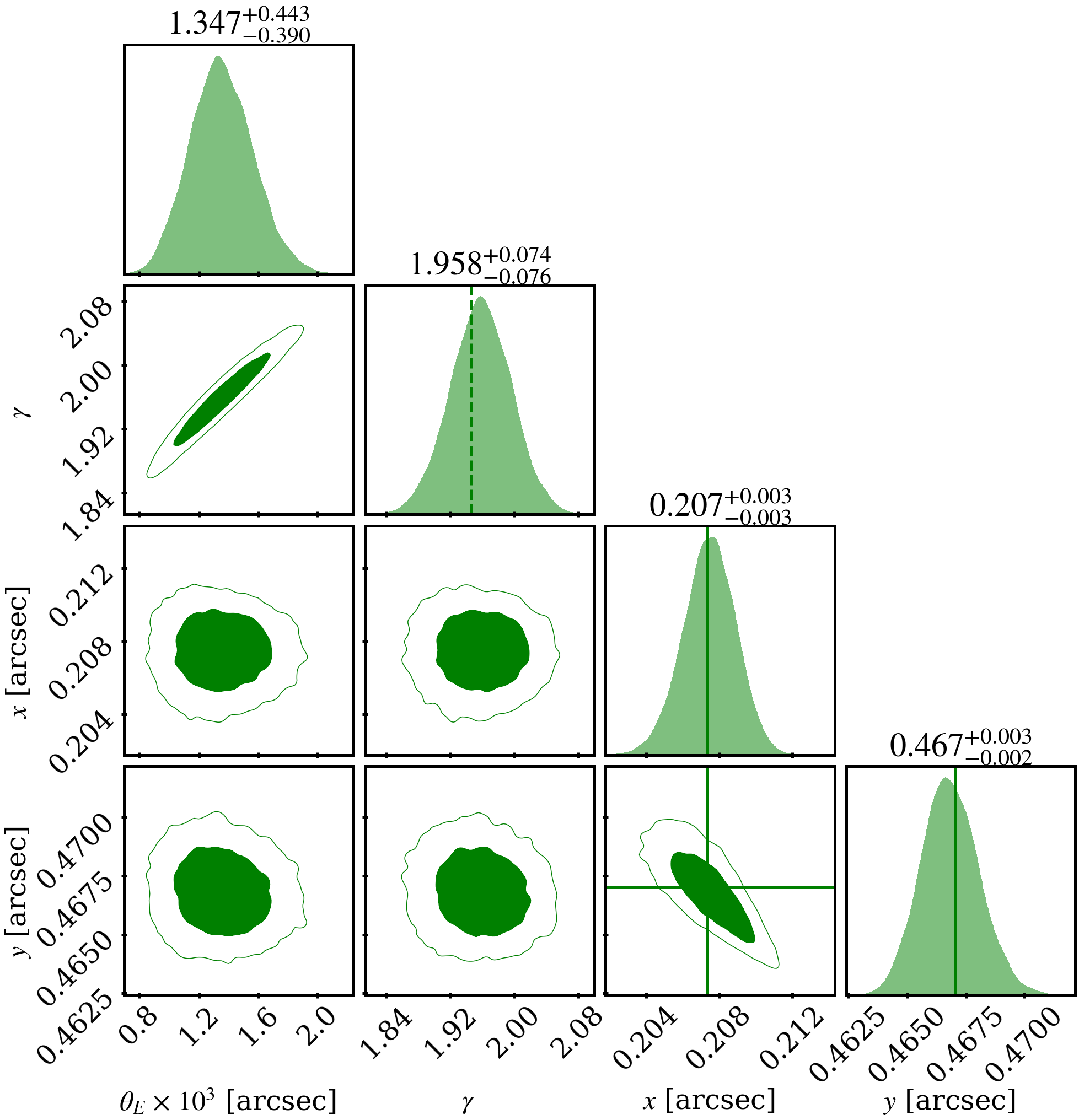}
        \caption{}
        \label{fig:high_c_posterior}
    \end{subfigure}
    \hfill
    \begin{subfigure}[t]{0.49\textwidth}
            \centering
        \includegraphics[width=\textwidth]{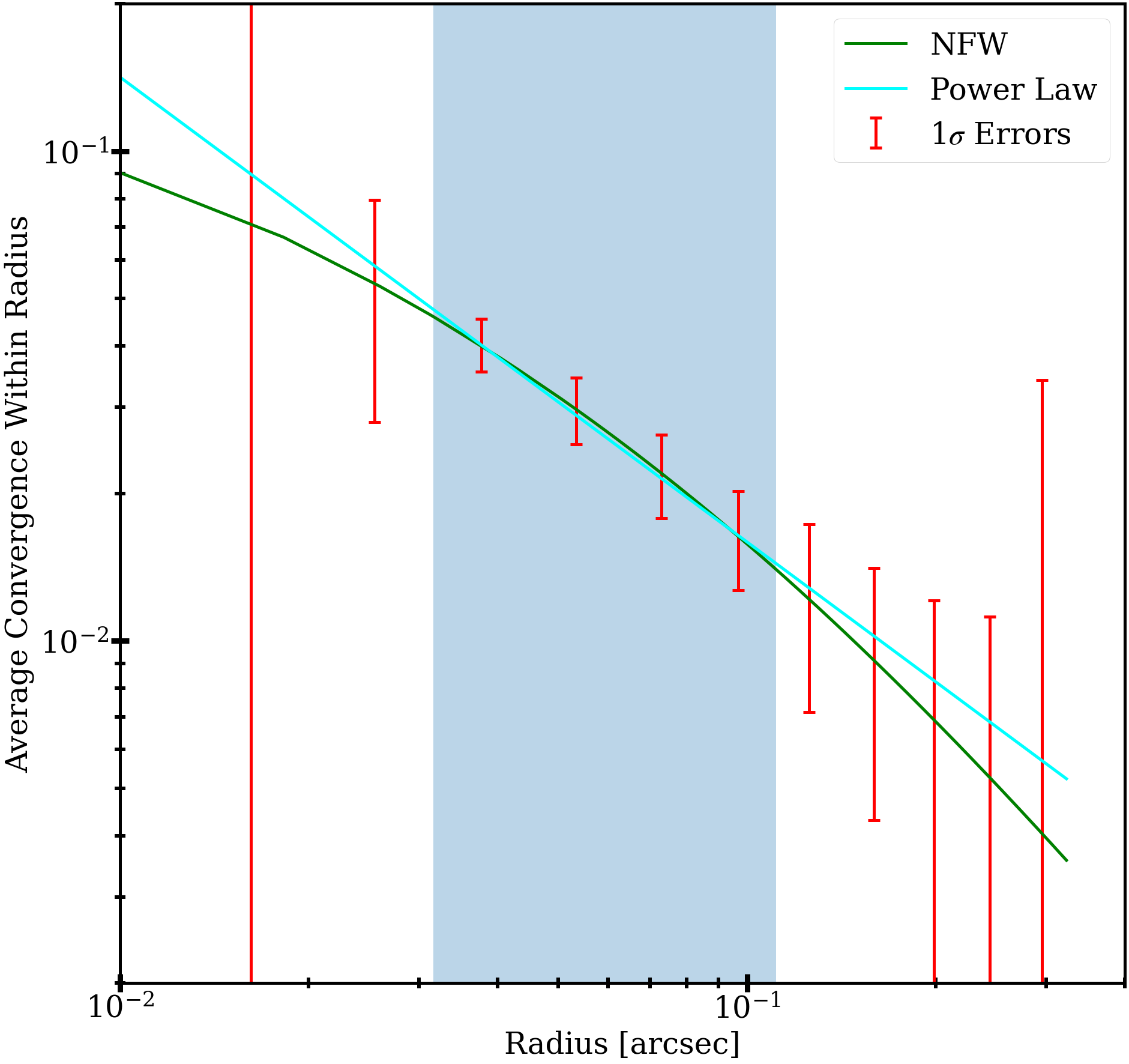}
        \caption{}
        \label{fig:high_c_error}
    \end{subfigure}
    \label{fig:high_c_figures}
    \caption{{\it  High-concentration perturber} model: (a) Posterior probability distribution of the perturber parameters. The true position values are shown in solid green lines. The average power-law slope in the RMO is shown as the dashed green line. (b) Average convergence as a function of radius. The power-law best fit for the {\it High-concentration perturber} model is shown in the cyan line. The true NFW profile is shown with solid green lines. The red error bars show the $1\sigma$ linear errors of the average convergence within each radii. The blue-shaded region corresponds to the RMO.}
\end{figure*}

If the concentration of the perturber is raised to $c_{200}=30$, we find that we can measure a slope of $\gamma = 1.958 \substack{+0.074 \\ -0.076}$. The posteriors are shown in Fig. \ref{fig:high_c_posterior}. Our slope measurement is consistent with the average slope within the RMO, which is not significantly affected by the change of the perturber profile, shown in Fig. \ref{fig:high_c_error}. An NFW perturber with a higher concentration has a lower scale radius. Around the scale radius, the effective slope is roughly $\gamma = 2$. Since the RMO for this case (shown in Fig. \ref{fig:high_c_error}) includes the scale radius, which is $r_s \approx 0.1''$, we measure an effective slope close to 2. The best fit power-law is compared with the true NFW profile in Fig. \ref{fig:high_c_error}. In this case, the perturber also has a larger $M_{200,\mr{perturber}}$ compared to the fiducial one to keep the effective perturber lensing mass fixed.

\subsection*{Smaller lens:}
\begin{figure*}
    \begin{subfigure}[t]{0.49\textwidth}
        \includegraphics[width=\textwidth]{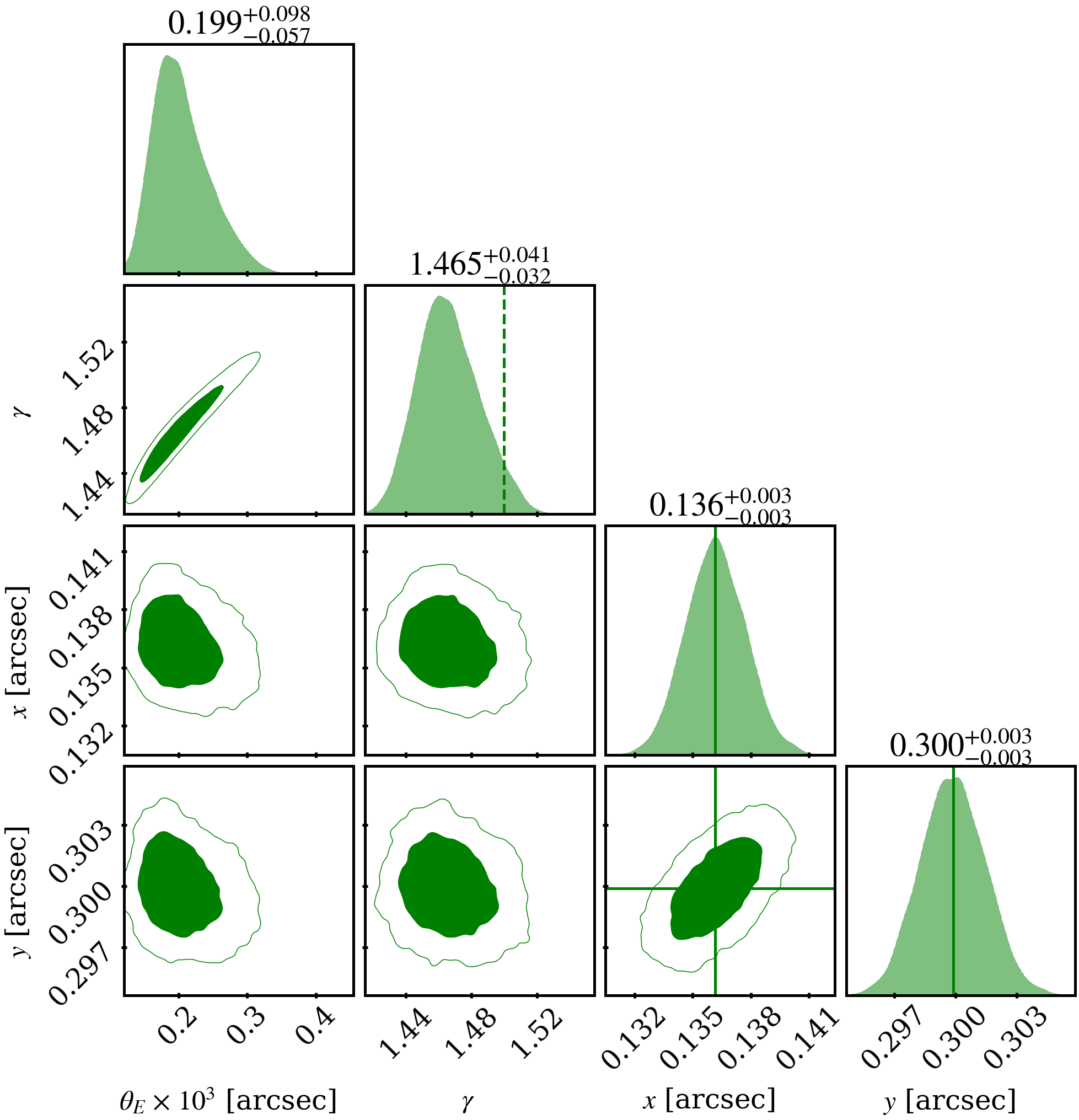}
        \caption{}
        \label{fig:small_lens_posterior}
    \end{subfigure}
    \hfill
    \begin{subfigure}[t]{0.49\textwidth}
            \centering
        \includegraphics[width=\textwidth]{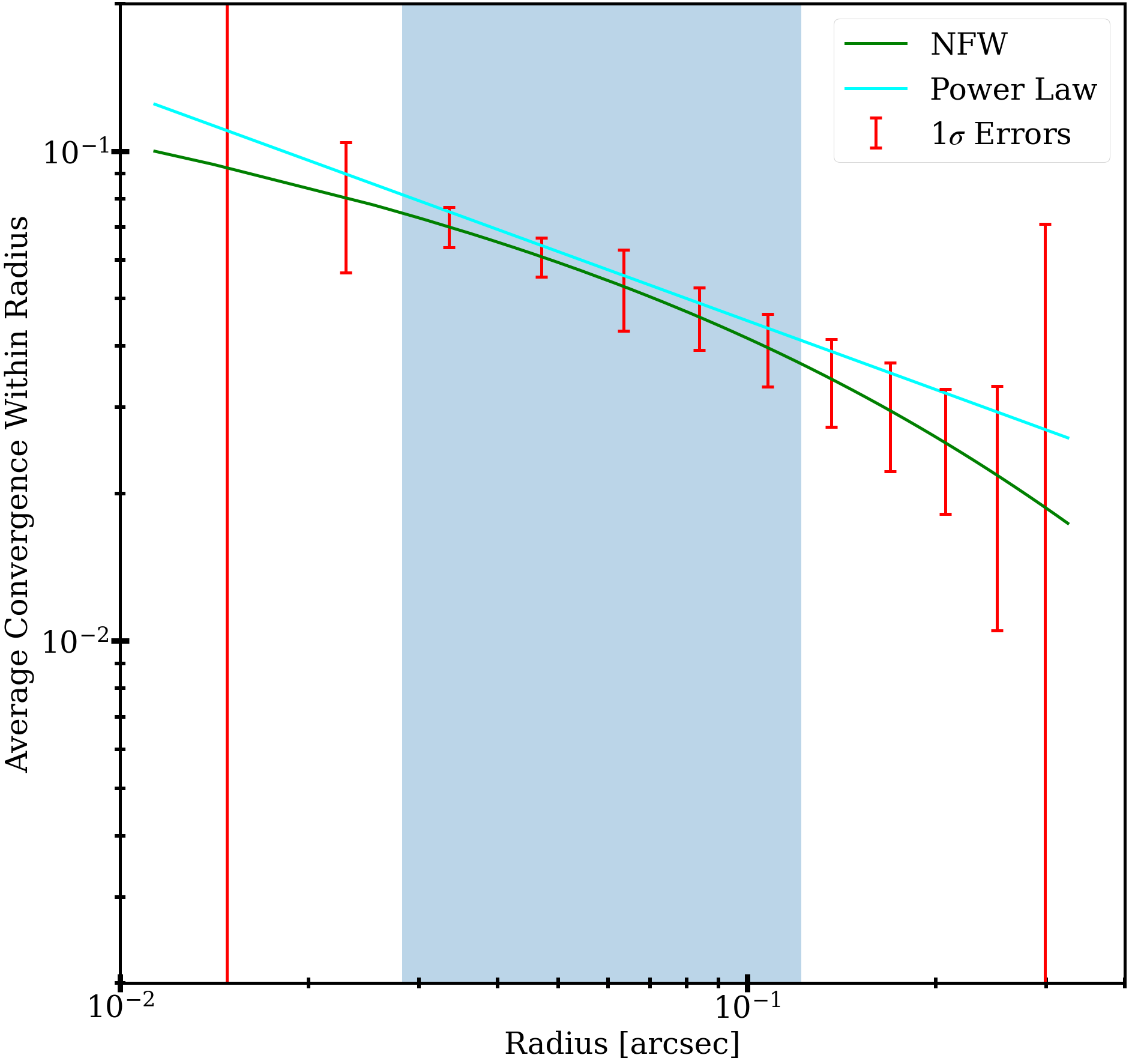}
        \caption{}
        \label{fig:small_lens_error}
    \end{subfigure}
    \label{fig:small_lens_figures}
    \caption{{\it  Smaller lens} model: (a) Posterior probability distribution of the perturber parameters. The true position values are shown in solid green lines. The average power-law slope in the RMO is shown as the dashed green line. (b) Average convergence as a function of radius. The power-law best fit for the {\it Smaller lens} model is shown in the cyan line. The true NFW profile is shown with solid green lines. The red error bars show the $1\sigma$ linear errors of the average convergence within each radii. The blue-shaded region corresponds to the RMO.}
\end{figure*}

Furthermore, we tested making the Einstein radius of the main lens smaller ( $0.3''$), and we found that in this case we can measure the slope $\gamma = 1.465 \substack{+0.041 \\ -0.032}$. The posteriors are shown in Fig. \ref{fig:small_lens_posterior}. The slope is consistent with the expected one from the RMO, shown in Fig. \ref{fig:small_lens_error}. This case outperforms even the {\it Fiducial} model because, with a smaller Einstein radius, the source light (which is kept at the same magnitude) is now concentrated in a smaller number of pixels. Since we enforce that the perturber has to be close to those pixels, the deflections caused by the perturber are now measured more accurately. Another contributing factor is the perturber being more massive relative to the main lens (since now the latter has a smaller mass). The same perturber mass is now more dominant in the overall shape of the image. However, we note that in lenses with smaller Einstein radii, the probability of finding a perturber at bright enough positions is also smaller. Here again, the perturber has a larger $M_{200, {\rm perturber}}$ value compared to the fiducial one to keep the effective lensing mass of the perturber fixed.

\subsection{Robustness tests}
\label{sec:robustness_tests}

\begin{figure}
    \centering
    \includegraphics[width=0.99\linewidth]{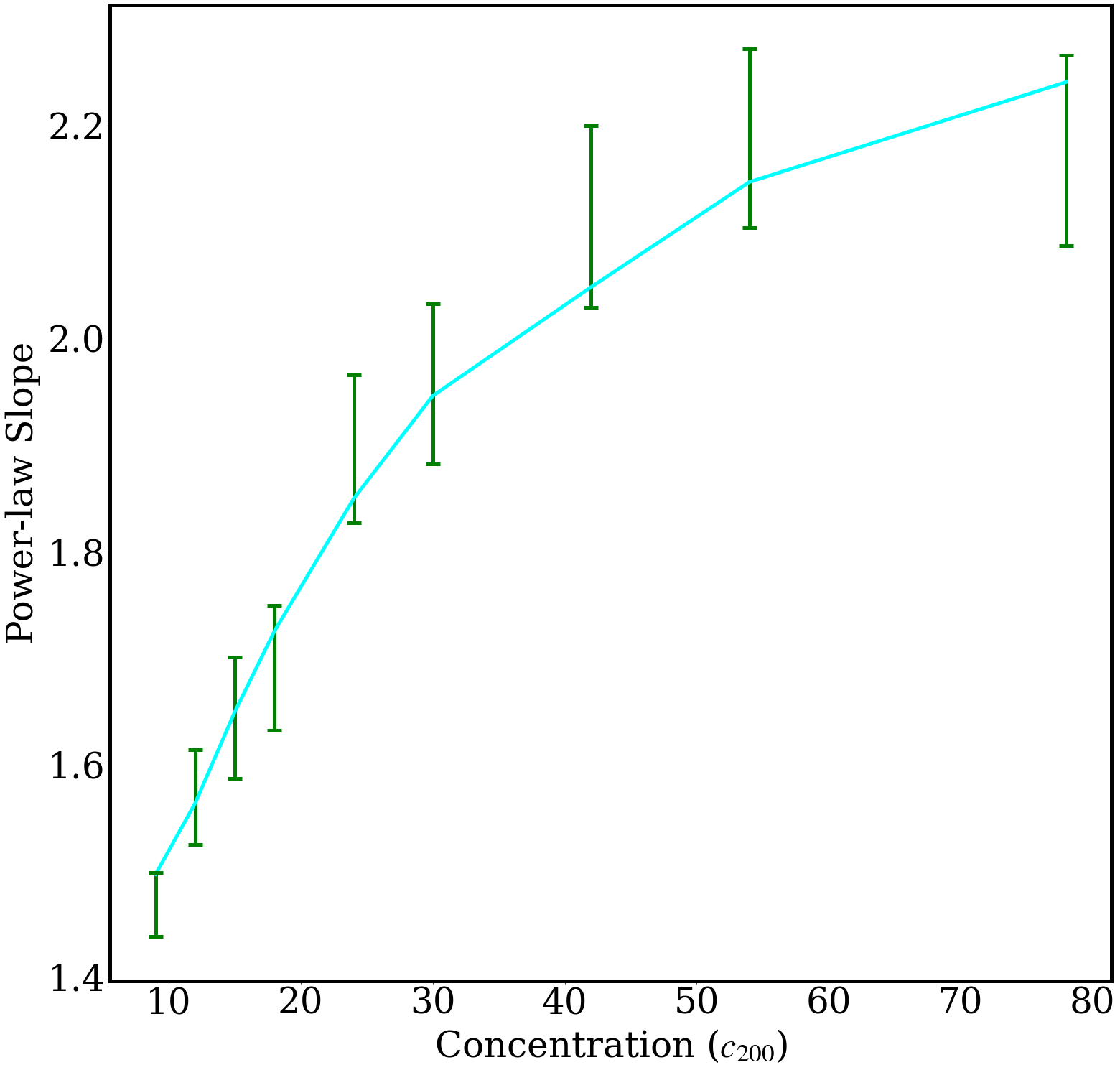}
    \caption{The expected and measured power-law slope for perturbers with varying concentration. The power-law slope of the average convergence within the RMO is shown in cyan line. The green error bars correspond to the 1$\sigma$ confidence interval of the slope measurement obtained modeling the mock lens with a power-law perturber.} 
    \label{fig:concen_test}
\end{figure}

To confirm that we are indeed measuring the power-law slope of the average convergence within the RMO, we performed robustness tests where we create a set of mock lenses with an NFW perturber where the concentration of the perturber varies from $c = 9$ to $c=78$, while the mass within the RMO is kept fixed at $1.5\,\times \,10^{7}\,\msun$. The choices we made for the rest of the lens parameters are the same as in the {\it Fiducial} case in \S \ref{sec:HSTsims}. We expect to measure a steeper slope for the higher concentration case because an increase in concentration results in a decrease in the scale radius for an NFW halo. At the same scale, an NFW halo with a smaller scale radius has a steeper power-law slope than an NFW halo with a larger scale radius. In Fig. \ref{fig:concen_test}, we show the slope measurements as the green error bars obtained following the same modeling procedure described in \S \ref{sec:HSTsims}, which involves sampling the posteriors of a model that has a power-law perturber. The cyan lines show the average power-law slope of the true NFW profile of each perturber within the RMO. We see that the measured slope agrees with the expected slope from the RMO for all the cases.

We have also tested for changes in the truncation radius, $r_t$, of an NFW with a fixed mass and scale radius. We measure (again by modeling a mock lens with a power-law perturber) the effective slope of a truncated NFW perturber \citep{Baltz:2007vq} with $\tau \equiv r_t/r_s = 1,2,3,4,5$. In Fig. \ref{fig:concen_test}, we show with green error bars the $1\sigma$ confidence interval of the slope measurements from mock lenses by fitting a power-law. The cyan line shows the average power-law slope of the true truncated NFW profile of each perturber within the RMO. For each case, we find that the measured slope matches the expected slope from the RMO.

\begin{figure}
    \centering
    \includegraphics[width=0.99\linewidth]{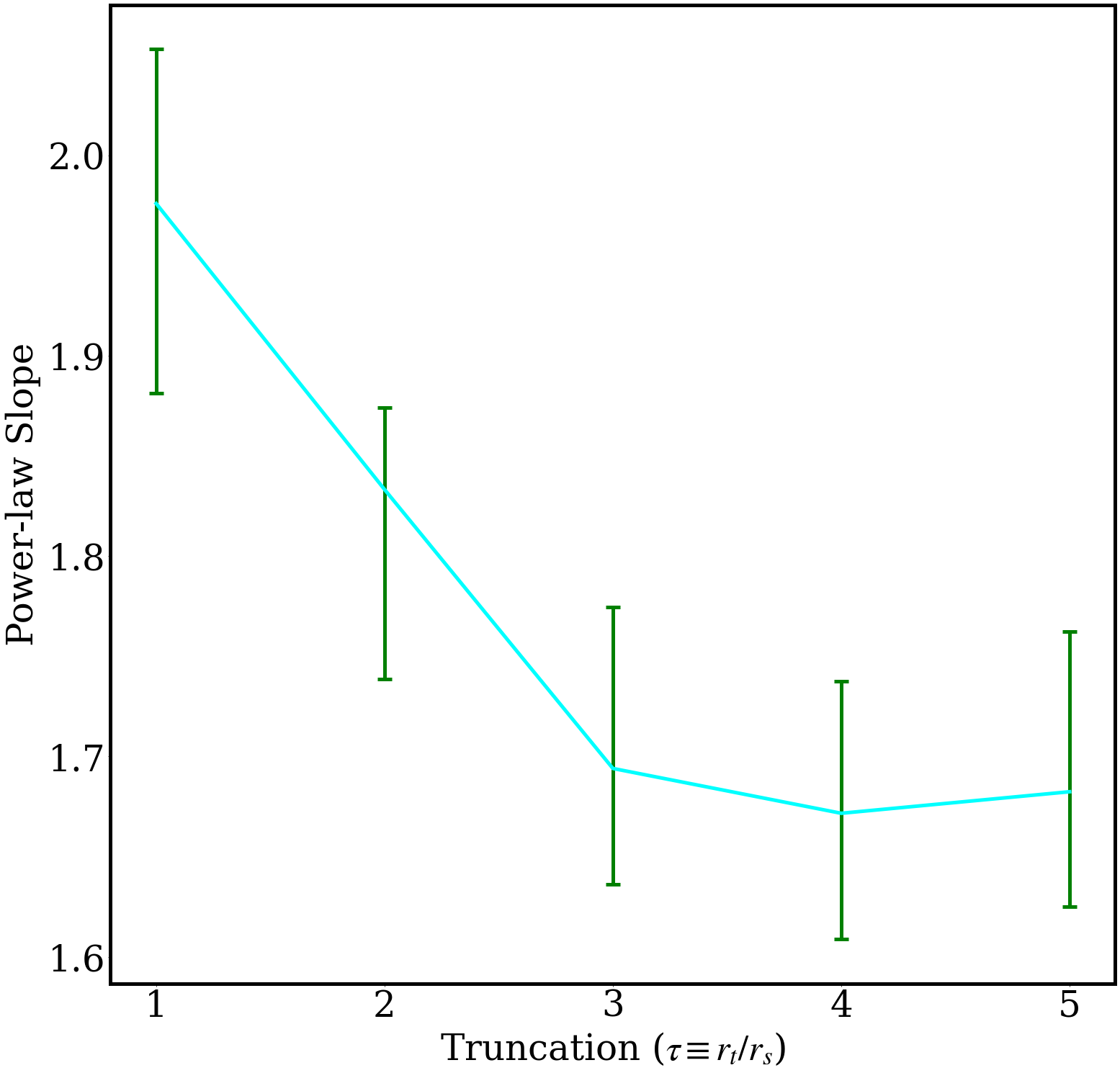}
    \caption{The expected and measured power-law slope for perturbers with varying truncation radius. The power-law slope of the average convergence within the RMO is shown in cyan line. The green error bars correspond to the 1$\sigma$ confidence interval of the slope measurement obtained modeling the mock lens with a power-law perturber.}
    \label{fig:trunc_test}
\end{figure}

\section{Measuring the Effective Density Slope of the Perturber in JVAS B1938+666}
\label{sec:JVAS}

We will focus here on a perturber that has been previously detected in the strong lens system JVAS B1938+666 \citep{lagatutta,2012Natur.481..341V}, using the gravitational imaging technique \citep{gravitational_imaging}. The main lens in this system is at redshift $z = 0.881$ \citep{z_lens} and the source is at redshift $z = 2.059$ \citep{z_source}. The main lens has an Einstein radius of roughly $\theta_E \approx 0.45''$. We have re-analyzed this system and found that the perturber that was previously assumed to be a subhalo is likely a line-of-sight halo. We refer to Ref. \cite{sengul2021} for a detailed discussion of this analysis. In this section, we will present our measurement of the effective density slope of the perturber in JVAS B1938+666. 
In Ref. \cite{2012Natur.481..341V}, this perturber was shown to be consistent with a singular isothermal sphere (SIS) profile, which is a power-law with a slope of $\gamma = 2$. A subsequent analysis \citep{vegetti_vogelsberger} found that for this system, a family of density profiles is consistent with an SIS profile over a large region in parameter space. Considering the sensitivity and the source-light distribution,  they show that the data are not sensitive to the size of the inner density core. Motivated by these findings and our results in the previous section, we model the main lens and the perturber as an Elliptical Power-Law, shown in Eq. \eqref{eq:epl}, and an external shear. The perturber is modeled with a cylindrically symmetric power-law mass profile (by fixing $q=1$ in Eq. \ref{eq:epl}). Although the true density profile of the perturber might deviate from this power-law at smaller and larger radii, the data are not currently sensitive to such deviations. 

The source light is modeled with shapelets with an optimal shapelet index of $n_\mr{max} = 10$. The shapelet index optimization is done by minimizing the BIC over models with different $n_{\rm max}$, as in the previous section. The background and Poisson noise are set to match the noise level of the image used, following our previous analysis \citep{sengul2021}.

We perform a likelihood analysis using \texttt{dynesty} \citep{2020MNRAS.493.3132S}, and show the posteriors of the perturber parameters in Fig. \ref{fig:posteriors}. The data, reconstruction, and residuals are shown in Appendix \ref{sec:residuals_JVAS}. We find an effective density slope of $\gamma=1.96\substack{+0.12 \\ -0.12}$. Assuming that the underlying slope distributions are Gaussian, our measurement is a $>2.1\sigma$ outlier for a CDM slope distribution from simulations at $z=1.47$ shown in Fig. \ref{fig:z_dependence}.  This result is also in $3.9\sigma, 4.5\sigma, 5.1\sigma$, and $3.2\sigma$ tension with ETHOS1-4, respectively. If the perturber were a subhalo, instead of a line-of-sight halo as Ref. \cite{sengul2021} shows, its slope would have to be compared to the $z=0.88$ simulation snapshots. In this case the slope measurement would be a $1.9 \sigma$ outlier for CDM and a $3.4\sigma, 4.1\sigma, 4.6\sigma$, and $2.7\sigma$ outlier for the ETHOS1-4 models, respectively.

\begin{figure}
    \centering
    \includegraphics[width=0.99\linewidth]{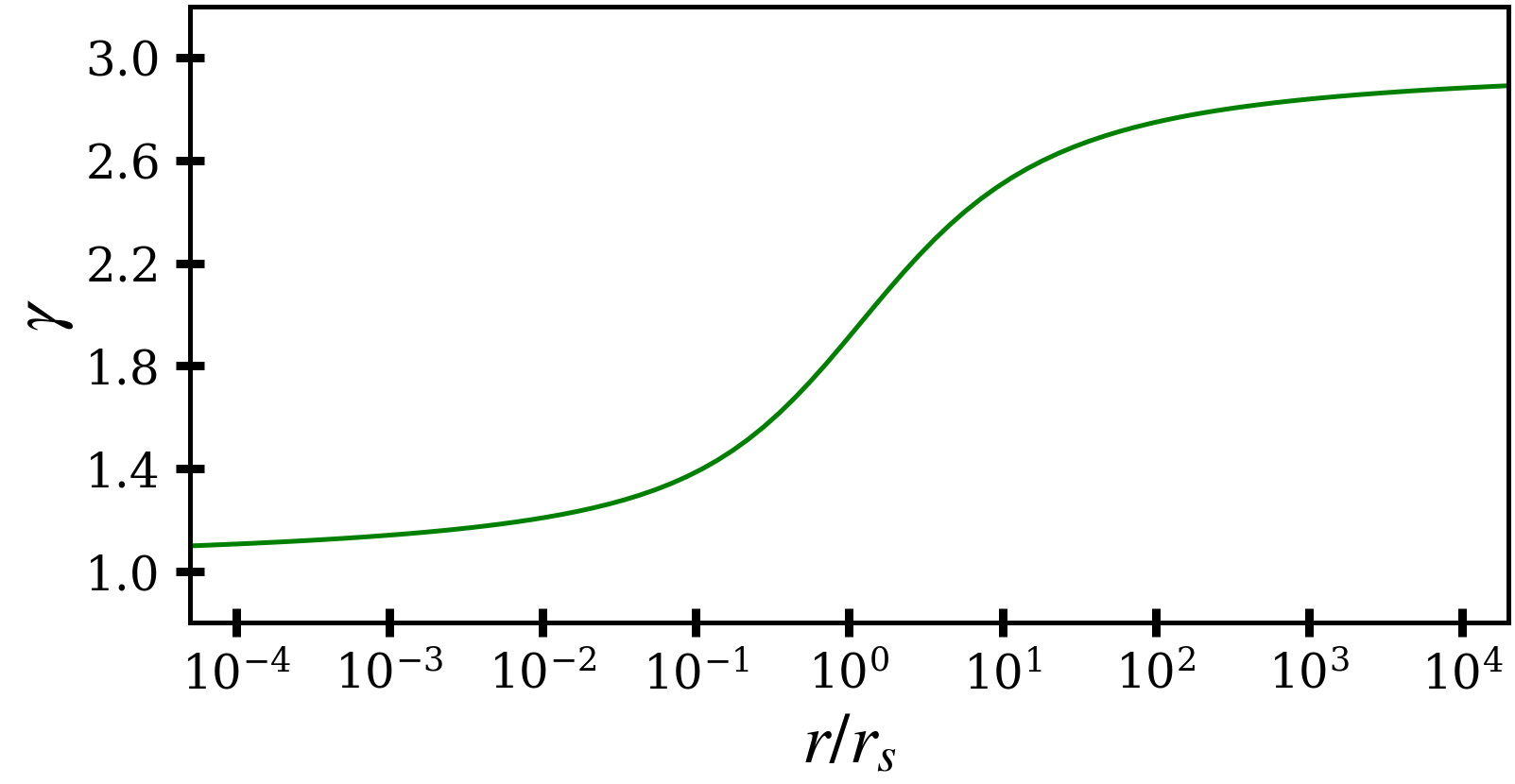}
    \caption{The effective slope of an NFW profile as a function of radius in units of the scale radius $r_s$.}
    \label{fig:nfw_effective_Slope}
\end{figure}

\begin{figure}
    \centering
    \includegraphics[width=0.99\linewidth]{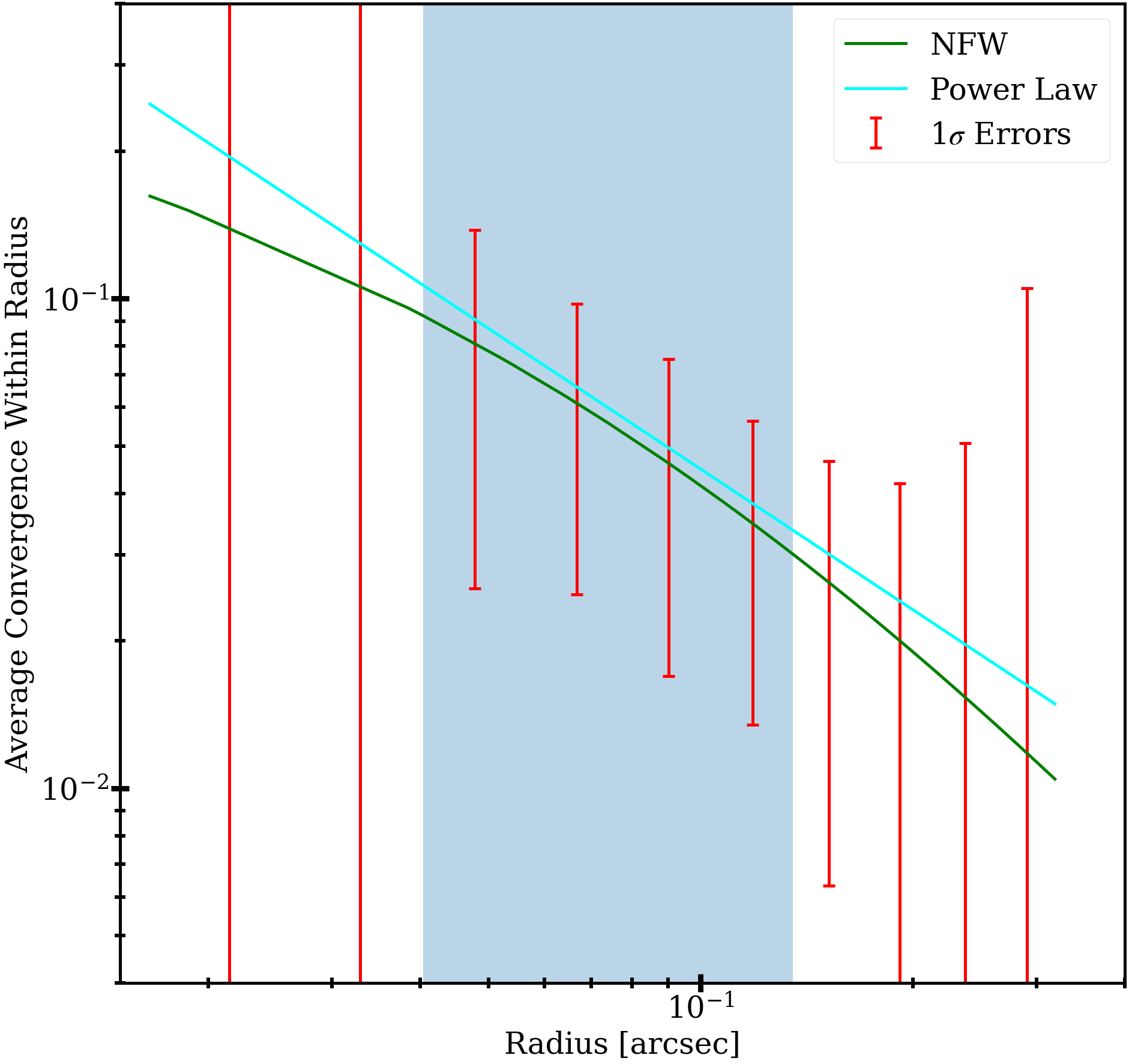}
    \caption{The linear errors for the perturber of the system JVAS B1938+666, calculated with the best fits of the main lens, shear, and source light from Ref. \protect\cite{sengul2021}. The power-law best fit with $\gamma = 1.95$ is shown in cyan line. The NFW best fit is shown with the green line.}    
    \label{fig:errors}
\end{figure}

From our measurement of the power-law slope within the RMO and assuming that the mass distribution of the perturber follows an NFW profile, we can estimate the scale radius and the concentration of the perturber. Around the scale radius, the NFW profile slowly transitions from a slope of $\gamma \approx 3$ to $\gamma \approx 1$, as we move to smaller radii, as can be seen in Fig. \ref{fig:nfw_effective_Slope}. Therefore, the density profile of an NFW perturber that has a scale radius that is much smaller than the RMO will effectively behave like a power-law with a slope of $\gamma \approx 3$. Conversely, when the scale radius is much larger than the RMO, the density profile of the perturber will behave like a power-law with a slope of $\gamma \approx 1$.  In our case, we are measuring a power-law slope of $\gamma \approx 2$, which means that we are probing the density profile at roughly the NFW scale radius. In contrast to our previous study with mocks, for the real JVAS B1938+666 system, we do not know the true source-light distribution, main lens profile, and the perturber position. Instead, we use the best fits from our analysis of the system outlined in Ref. \cite{sengul2021} to calculate the RMO. In Fig. \ref{fig:errors}, we show the linear errors for the average convergence of the perturber. The RMO is shown in the blue-shaded region. We see that we are most sensitive to the average convergence between roughly $0.04''-0.13''$. This corresponds to a physical scale of roughly $0.17 - 0.55\, \kpc$ at the redshift of the perturber. The scale radius of the perturber needs to be within the RMO so that the power-law slope of the perturber at our measurement scale is close to $2$. If $r_s$ were any bigger (smaller), we would expect to measure a slope that is closer to $\gamma = 1 (3)$. Considering the measured mass of the perturber, a scale radius of $r_s = 0.25 \,\kpc$  implies a concentration of roughly $c_{200} \approx 60$, which is much higher than the expected value of $c_{200} \approx 10$ given by the mass-concentration relation found in CDM simulations \citep{mass_concen}. This is consistent with our measurement of $c_{200} = 54 \substack{+21 \\ -15}$ in Ref. \cite{sengul2021} and also matches the expectation in Ref. \cite{Minor:2020bmp} that only high-concentration perturbers with masses $\lesssim 10^9\msun$ are expected to be detected at the HST resolution.

\begin{figure}
    \centering
    \includegraphics[width=0.99\linewidth]{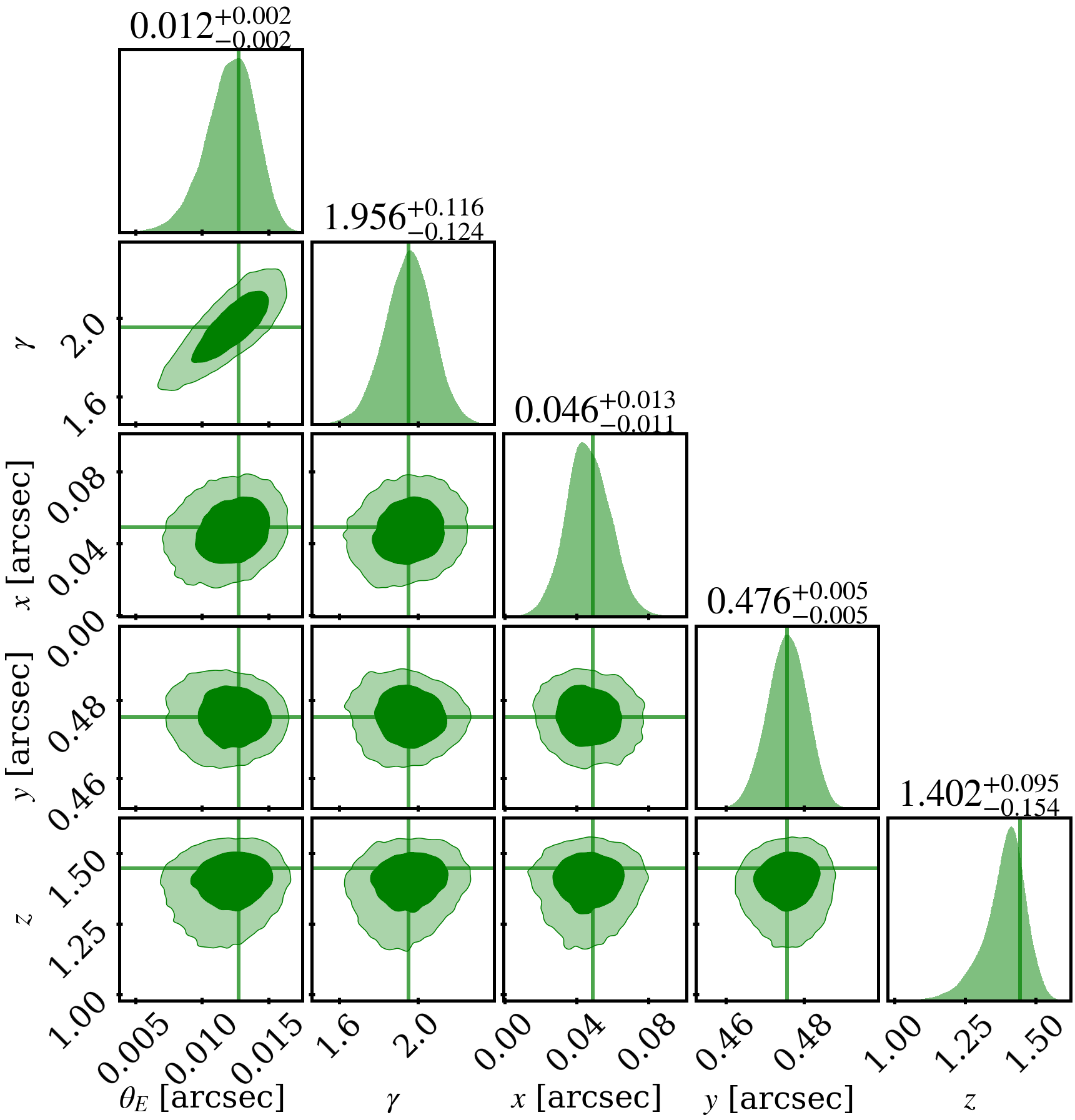}
    \caption{Posterior probability distribution of the perturber parameters in the system JVAS B1938+666. The vertical lines correspond to the best fit values of the parameters.}    
    \label{fig:posteriors}
\end{figure}

\section{Conclusions}

It is standard practice, in strong gravitational lensing studies, to provide a measurement of a system by making assumptions about its extended profile. In some cases, these assumptions can change the inferred mass by an order of magnitude, which can, in turn, prevent us from accurately probing the (sub)halo mass function, a derived quantity that is commonly used to distinguish between different DM scenarios. On the other hand, perturbations in strong lensing images due to low-mass subhalos and line-of-sight halos are quite sensitive to the power-law slope of the average convergence of the perturbers within an observable range (we call this the {\it effective} density slope in this work). We proposed in this work to use this effective density slope as a more direct parameter from strong gravitational lensing measurements, and we recommend future studies to report this parameter, as it will allow for a more accurate comparison with simulations of different dark matter scenarios.

Making predictions of structure formation at the relevant scales considered in this work is only possible by relying on simulations since structure formation is a highly non-linear process. For this purpose, we use the ETHOS zoomed-in simulations \citep{2016MNRAS.460.1399V} of a Milky Way-sized halo under different DM scenarios. We find that these scenarios show different distributions of effective density slopes of the DM perturbers, which can be used as a benchmark to test measurements of the density slopes of real perturbers. Although there is a significant overlap in the slope distribution of different scenarios, in general, CDM perturbers present a steeper slope compared to that of the more exotic DM scenarios (ETHOS1-4) analyzed in this work. We also see that, for subhalos, the slope is slightly correlated with its distance to the center of the host. Subhalos that are closer to their host, on average, have shallower effective slopes, populating the lower tails of the slope distributions in all the DM scenarios we considered. 

We demonstrate the feasibility of measuring the effective density slope by making mock measurements of Hubble Space Telescope-like data. We find that if the source galaxy is sufficiently bright (an AB magnitude of 22 or brighter), measuring the slope of a perturber with a mass of $1\times 10^{9}\, \msun$ is possible with a $3\%$ uncertainty. This measurement also depends on accurate modeling of the main lens profile. Deviations of the main lens mass distribution from an elliptical power-law can result in a bias in the inferred effective power-law slopes for perturbers. The differences in the means of the distributions observed in different DM populations are generally larger than our measurement accuracy, suggesting that we should be able to distinguish between different DM scenarios as we accrue more measurements of the slopes of the perturbers in strong lensing systems.

As a step towards this effort, we measure in this work the effective slope of the density profile of the perturber in the strong lens system JVAS B1938+666. We find that its slope is steeper than what is expected from simulations, being in strong tension with ETHOS1-3. It is also a $\approx 3\sigma$ outlier for ETHOS4 and a $\approx 2\sigma$ outlier for CDM. This tension can also be seen as a tension in concentration values. For an NFW halo, a steep slope indicates a high concentration and vice versa. More measurements of the slopes of perturbers are needed to tell if this indicates new physics, as perturbers with steeper effective slopes (higher concentrations) have a stronger lensing signal and are more likely to be detected.

In the upcoming years, we expect to find tens of thousands of new strong lensing systems with experiments such as the Vera C. Rubin Observatory, {\it Euclid}, the Dark Energy Survey (DES), the Dark Energy Spectroscopic Instrument (DESI), the Square Kilometre Array (SKA), and the Nancy Grace Roman Telescope \citep{2010MNRAS.405.2579O,Collett:2015roa,2015aska.confE..84M,2021ApJ...909...27H}. 
As the capabilities to observe strong lens systems increase, we expect to find a large number of perturbers. Follow-up analyses of these perturbers with high-resolution imaging from the Hubble Space Telescope, the James Webb Space Telescope, and the Extreme Large Telescope will provide us with a distribution of effective density slopes that can be then compared with simulations. 

\label{sec:conclusions}

\vspace{0.5cm}
\subsection*{Acknowledgments} 
CD and ACS are partially supported by the Department of Energy (DOE) Grant No. DE-SC0020223.
We thank Mark Vogelsberger and Jes\'{u}s Zavala for providing us with the CDM and ETHOS simulations analyzed in this work. We would also like to thank Simon Birrer, Adam Coogan, Akaxia Cruz, Arthur Tsang, and Sebastian Wagner-Carena for valuable discussions.

\subsection*{Data Availability} 

The system analyzed in this work is JVAS B1938+666. The data of this system are publicly available on \url{https://mast.stsci.edu/portal/Mashup/Clients/Mast/Portal.html}. The code used for our analysis is available at \url{https://github.com/acagansengul/interlopers_with_lenstronomy}. The data from the particle simulations will be shared on request to the corresponding author with permission of Mark Vogelsberger and Jes\'{u}s Zavala.

\begin{appendix}

\section{Extracting the Amplitude and Slope of the Average Convergence from Simulations}
\label{sec:inner profile}

This appendix shows the details behind the extraction of the effective density slopes from N-body simulations.

We sort the particles of each (sub)halo in the simulations in radial bins of equal width $\delta r$, proportional to the $r_{200}$ of that (sub)halo, around the center. The expected number of particles $\mu_i$ in each bin $i$ will be proportional to the density of the (sub)halo $\rho(r)$ at that radius,
\begin{equation}\label{eq:density}
    \mu_i = \frac{1}{m_p}\int^{r_i+\Delta r}_{r_i}dr\,4\pi r^2 \rho(r) \approx \frac{4\pi r_i^2 \Delta r \rho(r_i)}{m_p},
\end{equation}
where $m_p$ is the mass of a particle in the simulation. Using Eq. \eqref{eq:density}, we obtain the density of each halo at various radii. The uncertainties of these densities are calculated assuming that the number of particles in each bin follows a Poisson distribution with expectation value $\mu_i$. The 2D surface mass density for each (sub)halo can be written in terms of the 3D density as
\begin{equation}
    \Sigma(r) = \int^{\infty}_{0} dz\,\rho\left(\sqrt{r^2 + z^2}\right).
\end{equation}
We are interested in the average convergence, denoted by $\tilde \kappa(r)$, which is proportional to the mass within the 2D radius $r$ divided by $r^2$:
\begin{equation}
    \tilde \kappa(r) \propto  \frac{1}{r^2}\int^{r}_{0} dr'\, 2\pi r'\,\Sigma(r').
\end{equation}
A 3D power-law density profile $\rho \propto r^{-\gamma}$ implies $\tilde \kappa \propto r^{-\gamma + 1}$. So we can extract $\gamma$ by calculating,
\begin{equation}
    \gamma = 1 - \dfrac{r (d\tilde \kappa/dr)}{\tilde \kappa(r)}
\end{equation}
which we do numerically from the bin populations $n_i$ in the simulations.

\section{Analysis of the Mock Images}
\label{sec:model_params}

In this appendix, we show a complete list of the model parameters of the mock images analyzed in \S\ref{sec:HSTsims}. In Table \ref{tab:table_2}, we show the mean and 1$\sigma$ uncertainties of the lens-model parameters of the different cases listed in Table \ref{tab:table_1}. We see that the means and the uncertainties of the main lens and shear parameters are consistent with the true values for all six cases considered here.

\subsection*{Degeneracy Between Main Lens Model and the Perturber}

\begin{figure*}
    \centering
    \includegraphics[width=\linewidth]{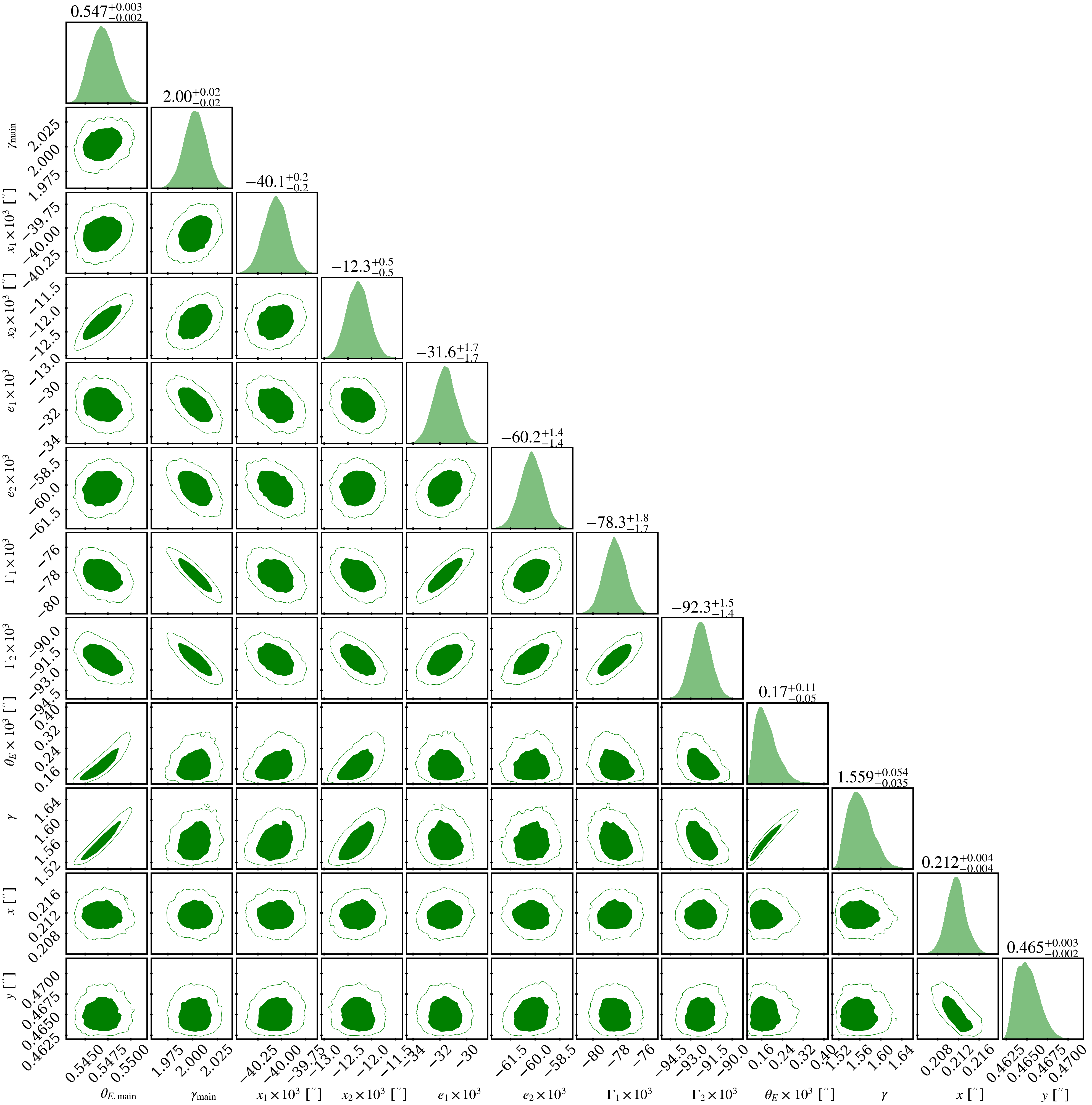}
    \caption{Posterior probability distribution of the model parameters of the \textit{Fiducial} case described in Table \ref{tab:table_1}. Model parameters are described in Table \ref{tab:table_2} We see that the Einstein radius of the main lens ($\theta_{E,\mr{main}}$) is correlated with the effective power-law slope of the perturber ($\gamma$).}
    \label{fig:fid_full_post}
\end{figure*}

We present in Fig. \ref{fig:fid_full_post} the posterior probability distributions of the model parameters for the \textit{Fiducial} case. We see a strong correlation between the Einstein radius of the main lens model ($\theta_{E,\mr{main}}$) and the effective power-law slope of the perturber ($\gamma$). This indicates that failing to correctly model the main lens mass profile can result in a biased measurement of the power-law slope for perturbers.

\setlength{\tabcolsep}{2pt}
\begin{table*}
    \centering
    \begin{tabular}{||m{1.6cm}||m{1.3cm}|m{1.05cm}|m{1.4cm}|m{1.4cm}|m{1.1cm}|m{1.1cm}|m{1.1cm}|m{1.1cm}|m{1.4cm}|m{1.3cm}|m{1.3cm}|m{1.3cm}|}
    \hline
        Model &\multicolumn{6}{|c|}{Main Lens}&\multicolumn{2}{|c|}{Shear}&\multicolumn{4}{|c|}{Perturber} \\
    \hline
        &$\theta_{E,\mr{main}}['']$& $\gamma_{\mr{main}}$ & $c_x\times 10^{3}['']$ & $c_y\times 10^{3}['']$& $e_x \times 10^{3}$& $e_y \times 10^{3}$& $\Gamma_x \times 10^{3}$&$\Gamma_y\times 10^{3}$&$\theta_E  \times 10^{3}['']$ &$\gamma$& $x['']$ & $y['']$\\
    \hline
    \hline
    \multirow{2}{4em}{\it Fiducial} &$0.550$ & $2.0$ & $-40.0$ & $-12.1$& $-31.5$& $-59.8$&$-78.4$&$-92.6$&  & & 0.189 & 0.481\\
    \cline{2-13}
    &${0.547}_{-0.002}^{+0.003}$&${2.00}_{-0.02}^{+0.02}$&${-40.1}_{-0.2}^{+0.2}$&${-12.3}_{-0.5}^{+0.5}$&${-31.6}_{-1.7}^{+1.7}$&${-60.2}_{-1.4}^{+1.4}$&${-78.3}_{-1.7}^{+1.8}$&${-92.3}_{-1.4}^{+1.5}$&${0.17}_{-0.05}^{+0.11}$&${1.559}_{-0.035}^{+0.054}$&${0.212}_{-0.004}^{+0.004}$&${0.465}_{-0.002}^{+0.003}$\\
    \hline
    \hline
    \multirow{2}{4em}{\it Dimmer Source}
    &$0.550$ & $2.0$ & $-40.0$ & $-12.1$& $-31.5$& $-59.8$&$-78.4$&$-92.6$&  & & 0.207 & 0.468\\
    \cline{2-13}
    &${0.548}_{-0.005}^{+0.005}$&${2.05}_{-0.06}^{+0.07}$&${-39.7}_{-1.0}^{+1.0}$&${-12.0}_{-1.5}^{+1.5}$&${-34.0}_{-6.6}^{+6.6}$&${-61.1}_{-5.5}^{+5.3}$&${-82.0}_{-5.3}^{+5.3}$&${-94.8}_{-4.4}^{+4.4}$&${0.51}_{-0.29}^{+0.71}$&${1.683}_{-0.102}^{+0.153}$&${0.194}_{-0.020}^{+0.027}$&${0.478}_{-0.012}^{+0.010}$\\
    \hline
    \hline
    \multirow{2}{4em}{\it Simpler Source}
    &$0.550$ & $2.0$ & $-40.0$ & $-12.1$& $-31.5$& $-59.8$&$-78.4$&$-92.6$&  & & 0.211 & 0.468\\
    \cline{2-13}
    &${0.546}_{-0.004}^{+0.004}$&${1.97}_{-0.02}^{+0.02}$&${-40.3}_{-0.4}^{+0.4}$&${-12.6}_{-0.6}^{+0.6}$&${-30.6}_{-1.7}^{+1.7}$&${-58.8}_{-2.2}^{+2.2}$&${-76.3}_{-1.4}^{+1.4}$&${-90.3}_{-1.5}^{+1.4}$&${0.10}_{-0.03}^{+0.07}$&${1.502}_{-0.034}^{+0.050}$&${0.211}_{-0.004}^{+0.005}$&${0.467}_{-0.004}^{+0.005}$\\
    \hline
    \hline
    \multirow{2}{4em}{\it Steeper Lens}  &$0.550$ & $2.3$ & $-40.0$ & $-12.1$& $-31.5$& $-59.8$&$-78.4$&$-92.6$&  & & 0.207 & 0.468\\
    \cline{2-13}
    &${0.547}_{-0.002}^{+0.003}$&${2.30}_{-0.03}^{+0.03}$&${-40.1}_{-0.2}^{+0.2}$&${-12.3}_{-0.4}^{+0.5}$&${-31.9}_{-2.2}^{+2.2}$&${-59.7}_{-2.3}^{+2.1}$&${-78.2}_{-1.8}^{+1.7}$&${-91.9}_{-1.9}^{+1.7}$&${0.23}_{-0.08}^{+0.15}$&${1.556}_{-0.041}^{+0.059}$&${0.206}_{-0.005}^{+0.005}$&${0.469}_{-0.003}^{+0.003}$\\
    \hline
    \hline
    \multirow{2}{4em}{\it High-concentration Perturber}  &$0.550$ & $2.00$ & $-40.0$ & $-12.1$& $-31.5$& $-59.8$&$-78.4$&$-92.6$&  & & 0.207 & 0.467\\
    \cline{2-13}
    &${0.549}_{-0.001}^{+0.001}$&${2.00}_{-0.02}^{+0.02}$&${-40.0}_{-0.2}^{+0.2}$&${-12.3}_{-0.4}^{+0.4}$&${-31.5}_{-1.7}^{+1.6}$&${-60.0}_{-1.4}^{+1.4}$&${-78.3}_{-1.6}^{+1.6}$&${-92.5}_{-1.3}^{+1.3}$&${1.35}_{-0.39}^{+0.44}$&${1.958}_{-0.076}^{+0.074}$&${0.207}_{-0.003}^{+0.003}$&${0.467}_{-0.002}^{+0.003}$
    \\
    \hline
    \hline
    \multirow{2}{4em}{\it Smaller Lens}   &$0.350$ & $2.00$ & $-40.0$ & $-12.1$& $-31.5$& $-59.8$&$-78.4$&$-92.6$&  & & 0.136 & 0.300\\
    \cline{2-13}
    &${0.345}_{-0.004}^{+0.004}$&${2.00}_{-0.02}^{+0.02}$&${-40.0}_{-0.3}^{+0.3}$&${-12.2}_{-0.5}^{+0.5}$&${-32.3}_{-2.7}^{+2.7}$&${-59.0}_{-2.7}^{+2.7}$&${-77.8}_{-2.3}^{+2.4}$&${-90.7}_{-2.2}^{+2.1}$&${0.20}_{-0.06}^{+0.10}$&${1.465}_{-0.032}^{+0.041}$&${0.136}_{-0.003}^{+0.003}$&${0.300}_{-0.003}^{+0.003}$\\
    \hline
    \end{tabular}
    \caption{Lens-model properties of each of the models shown in Fig. \ref{fig:mock_images}. The first row of each model shows the true values, while the second row shows the means and standard deviations of the measurements coming from a likelihood analysis, as described in \S\ref{sec:HSTsims}. The parameters $c_x$ and $c_y$ are the x and y coordinates of the centroid of the main lens, and $e_x$ $e_y$ are the x and y component of the  ellipticity of the main lens profile. The true $\theta_E$ and $\gamma$ for the perturbers are not applicable because they follow NFW profiles instead of EPL profiles.}
    \label{tab:table_2}
\end{table*}

\subsection*{Source Reconstruction}

\begin{figure*}
    \centering
    \includegraphics[width=0.70\linewidth]{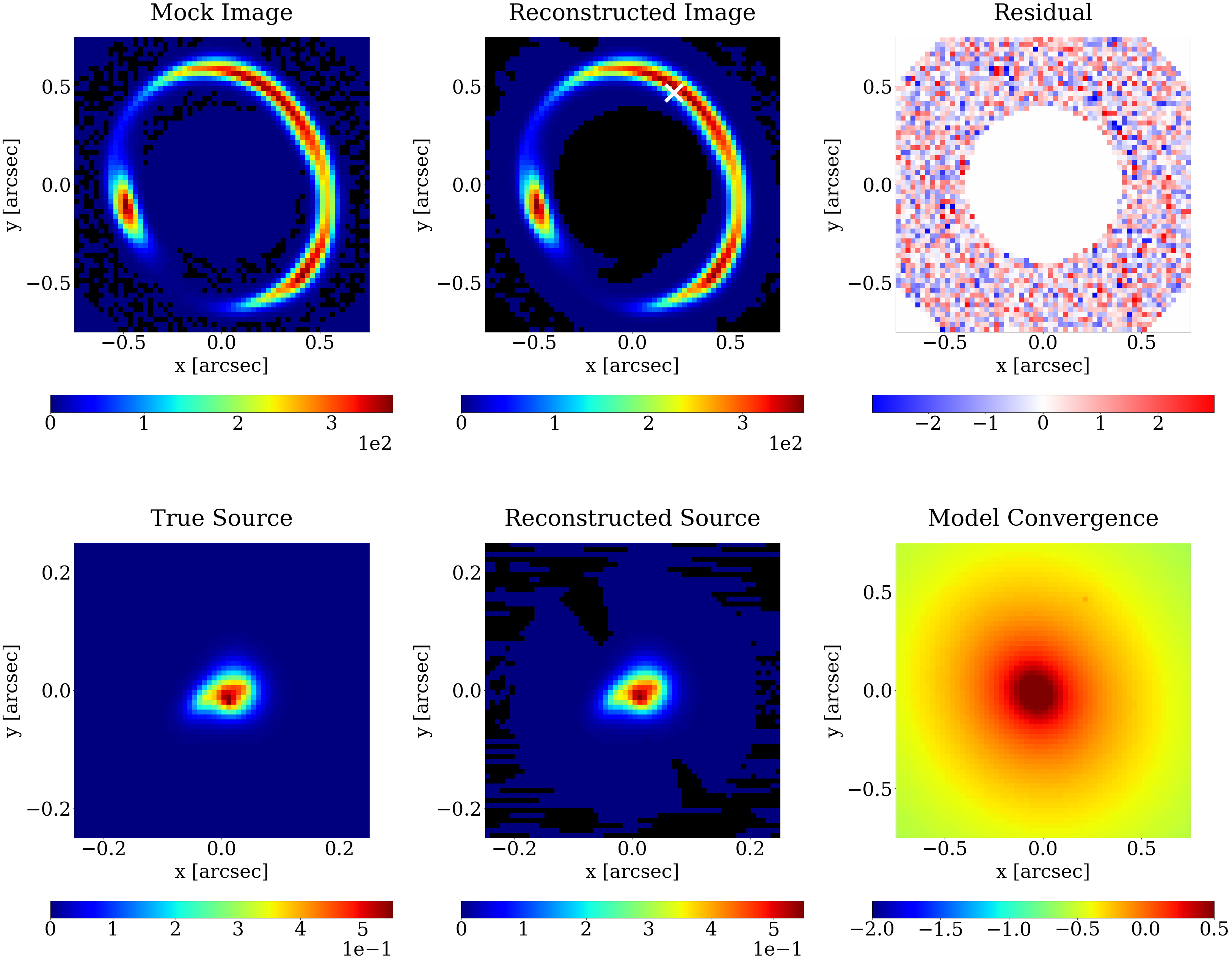}
    \caption{Best fit of the {\it Fiducial} mock in Table \ref{tab:table_1}. The top left panel shows the mock image with noise. The top middle panel shows the model reconstruction from the best fit, where the perturber position is marked with the white cross. The top right panel shows the residuals divided by the noise in each pixel. The bottom left panel shows the true source distribution that is made up of 5 Sersic components. The bottom middle panel shows the shapelet reconstruction of the source. The bottom right panel shows the convergence of the best fit model in log scale, where the perturber can be seen as a small peak.}
    \label{fig:fid_vis}
\end{figure*}

\begin{figure*}
    \centering
    \includegraphics[width=0.70\linewidth]{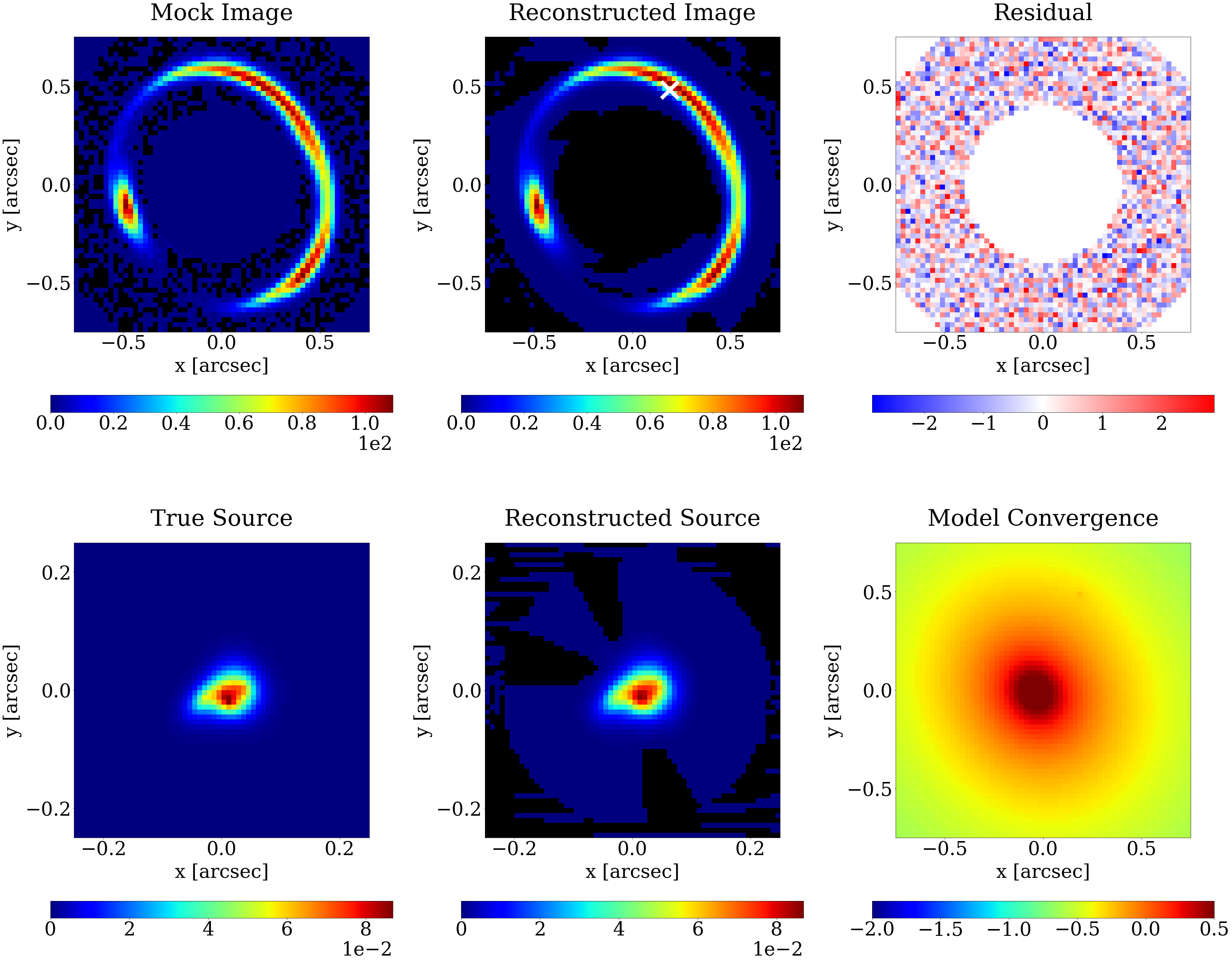}
    \caption{Same as in Fig. \ref{fig:fid_vis} but for the {\it Dimmer source} mock (see Table \ref{tab:table_1}).}
    \label{fig:dim_vis}
\end{figure*}

\begin{figure*}
    \centering
    \includegraphics[width=0.70\linewidth]{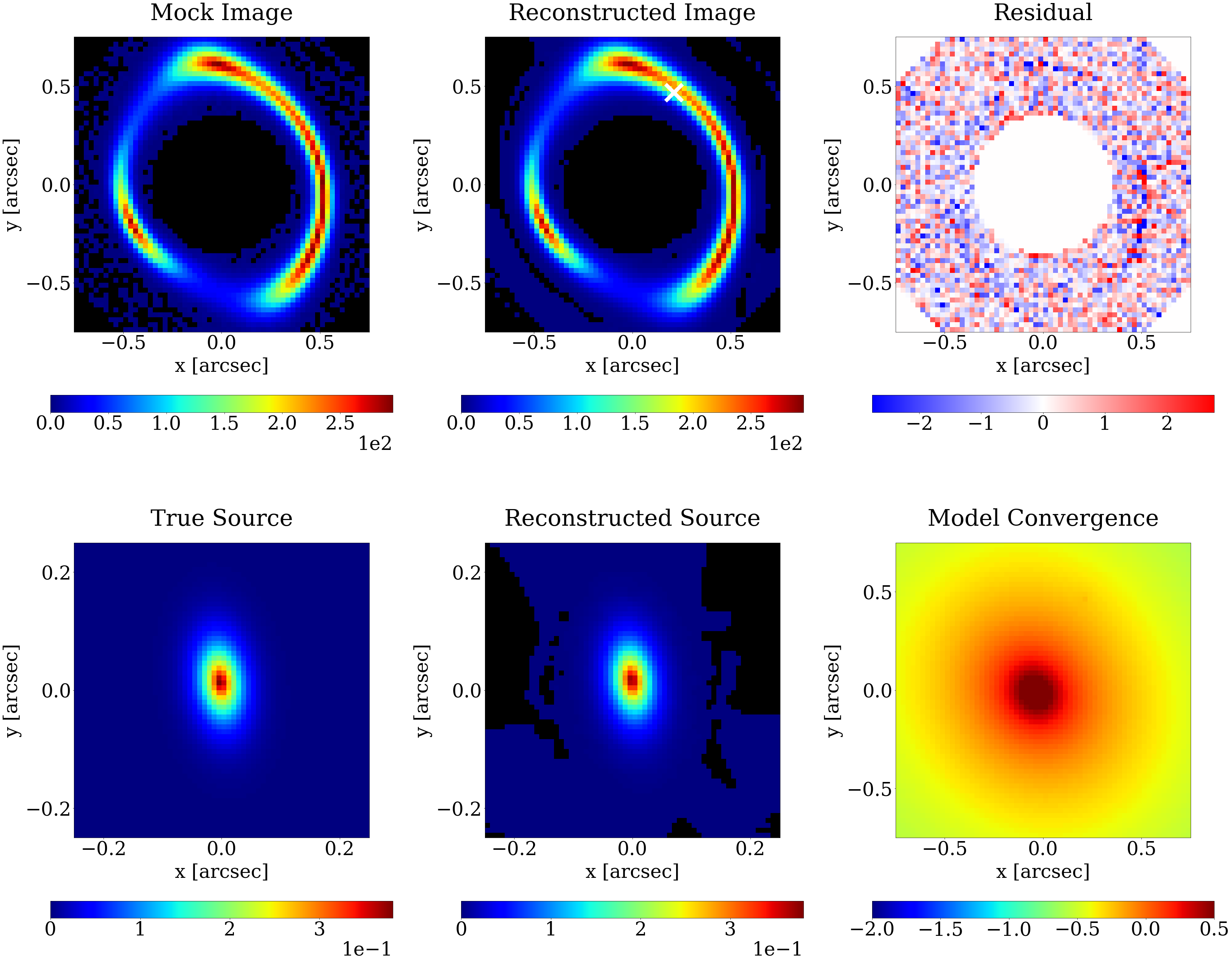}
    \caption{Same as in Fig. \ref{fig:fid_vis} but for the {\it Simpler source} mock (see Table \ref{tab:table_1}).}
    \label{fig:sim_vis}
\end{figure*}

\begin{figure*}
    \centering
    \includegraphics[width=0.70\linewidth]{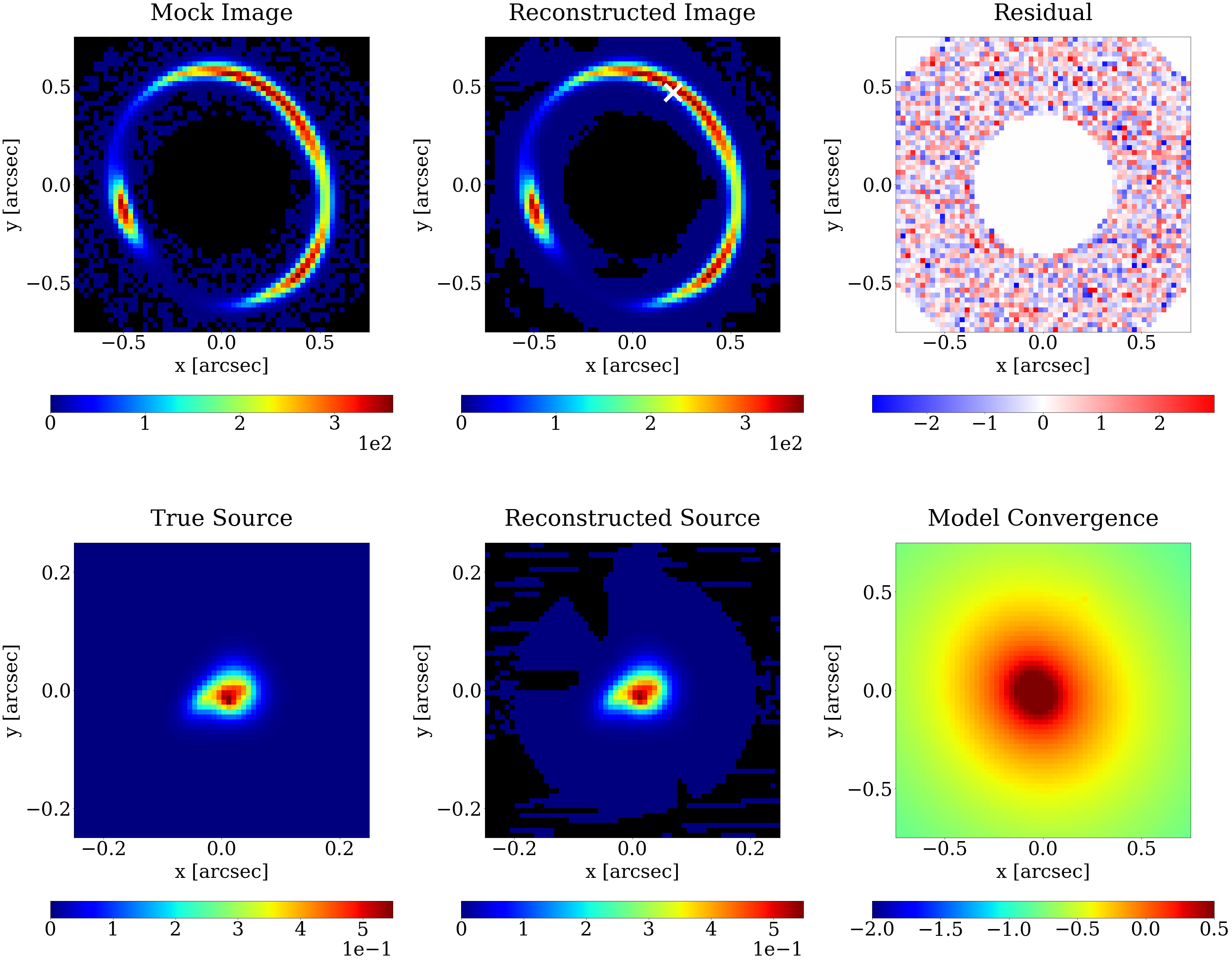}
    \caption{Same as in Fig. \ref{fig:fid_vis} but for the {\it Steeper lens} mock (see Table \ref{tab:table_1}).}
    \label{fig:stl_vis}
\end{figure*}

\begin{figure*}
    \centering
    \includegraphics[width=0.70\linewidth]{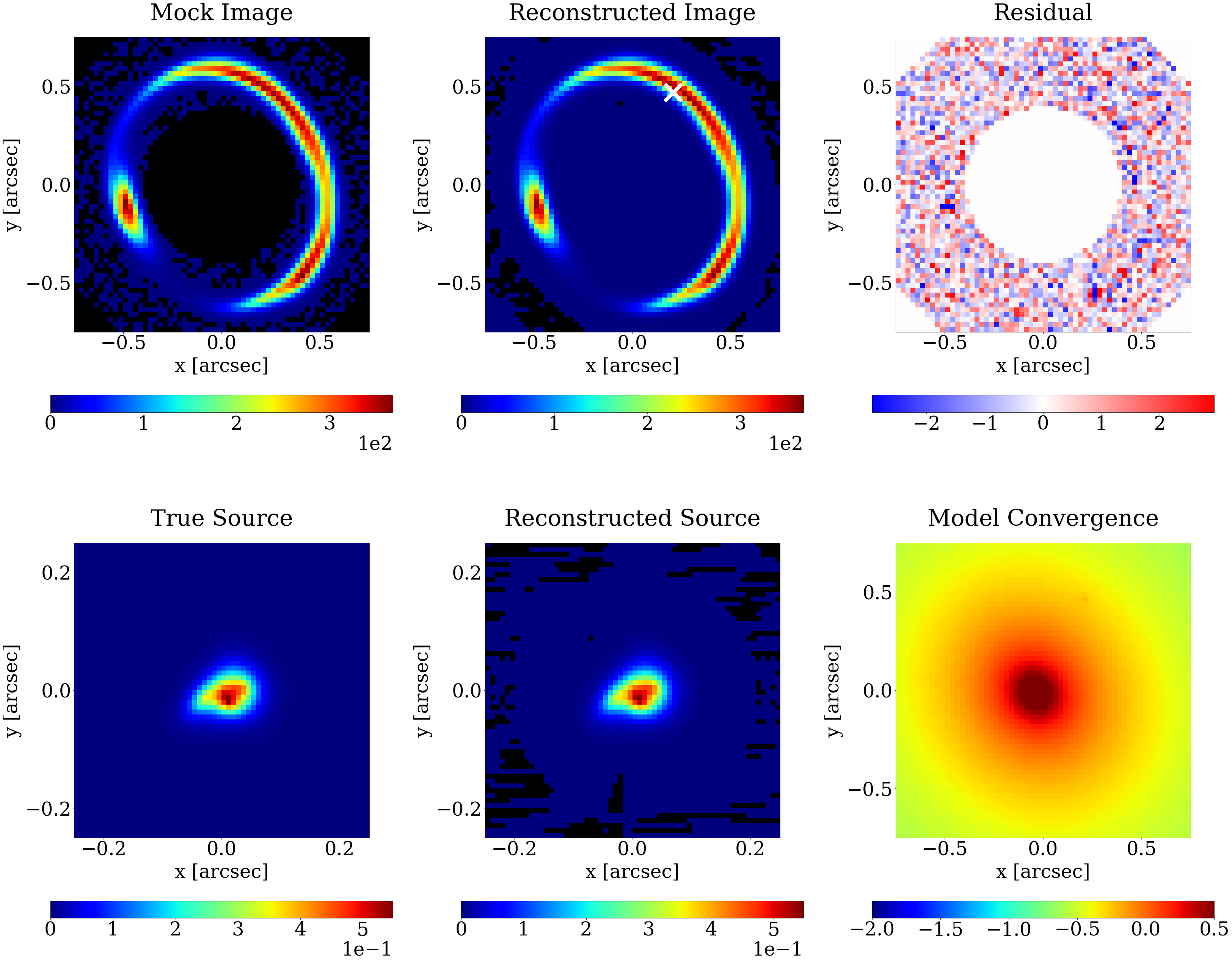}
    \caption{Same as in Fig. \ref{fig:fid_vis} but for the  {\it High-concentration perturber} (see Table \ref{tab:table_1}).}
    \label{fig:stp_vis}
\end{figure*}

\begin{figure*}
    \centering
    \includegraphics[width=0.70\linewidth]{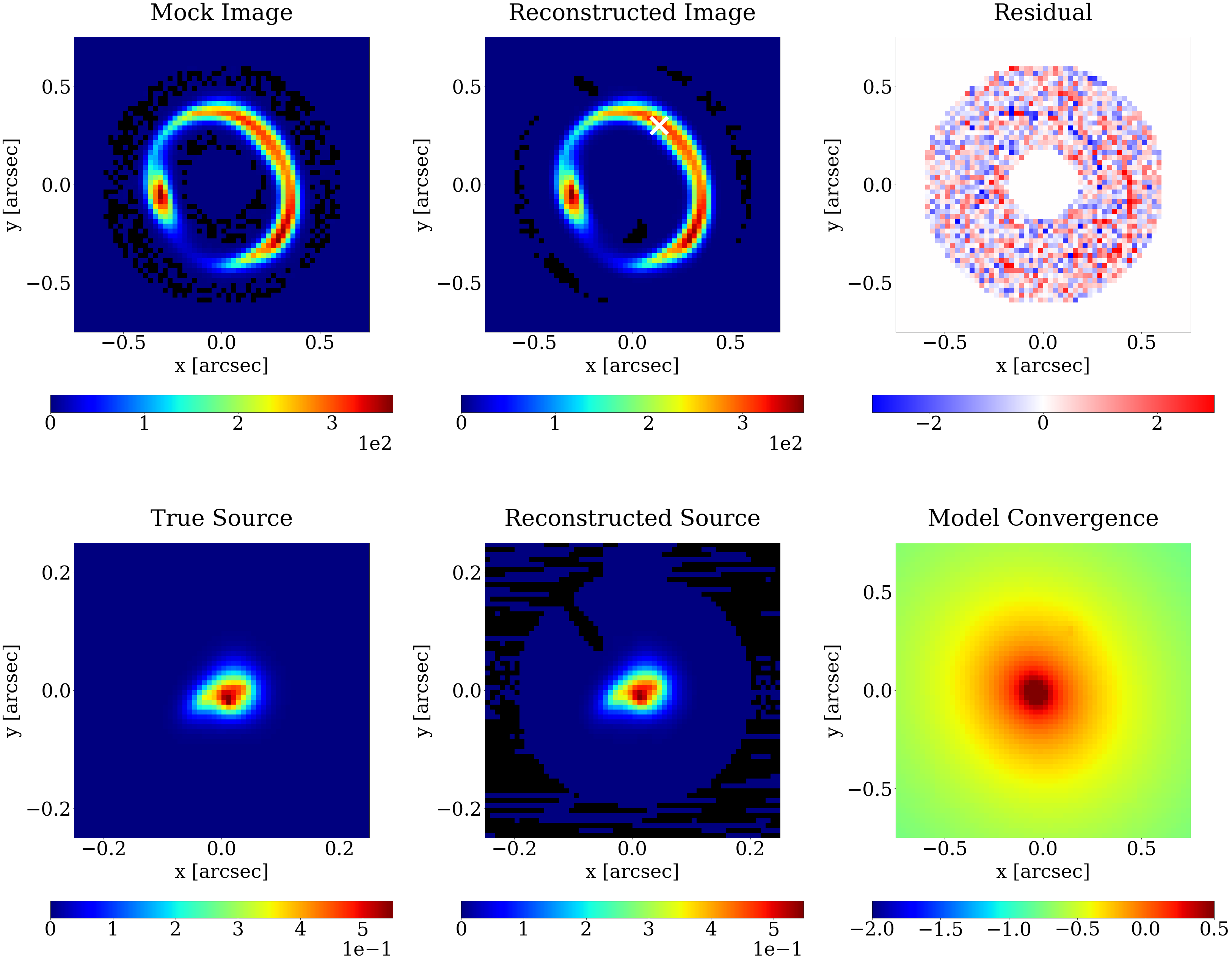}
    \caption{Same as in Fig. \ref{fig:fid_vis} but for the  {\it Smaller lens} (see Table \ref{tab:table_1}).}
    \label{fig:sml_vis}
\end{figure*}

We present in Figs. \ref{fig:fid_vis}, \ref{fig:dim_vis}, \ref{fig:sim_vis}, \ref{fig:stl_vis}, \ref{fig:stp_vis}, and \ref{fig:sml_vis} the source used for the models analyzed in \S\ref{sec:HSTsims}, along with its reconstructions.

\section{Data, Reconstruction, and Residuals in JVAS B1938+666}
\label{sec:residuals_JVAS}

\begin{figure*}
    \centering
    \includegraphics[width=0.99\linewidth]{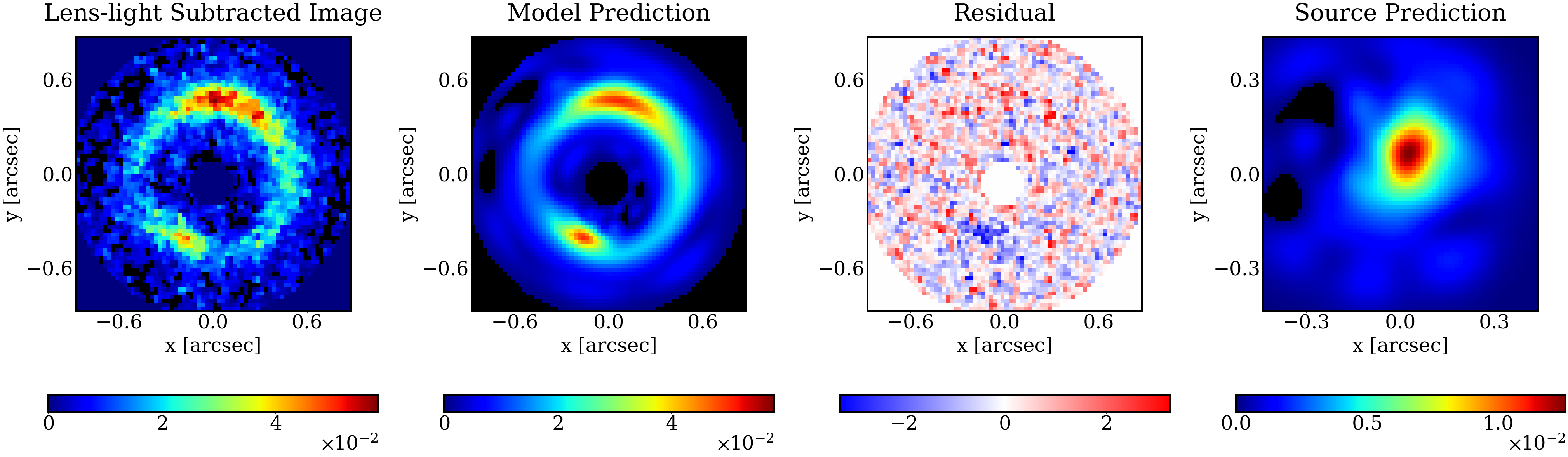}
    \caption{Lens-light subtracted Hubble Space Telescope image of the lens JVAS B1938+666, the model prediction (where the mass profile of the perturber is modeled with a free power-law slope, as in \S\ref{sec:JVAS}), the residuals of the model with a perturber, and the reconstructed source (left to right panels).}
    \label{fig:rec_JVAS}
\end{figure*}

\begin{figure*}
    \centering
    \includegraphics[width=\linewidth]{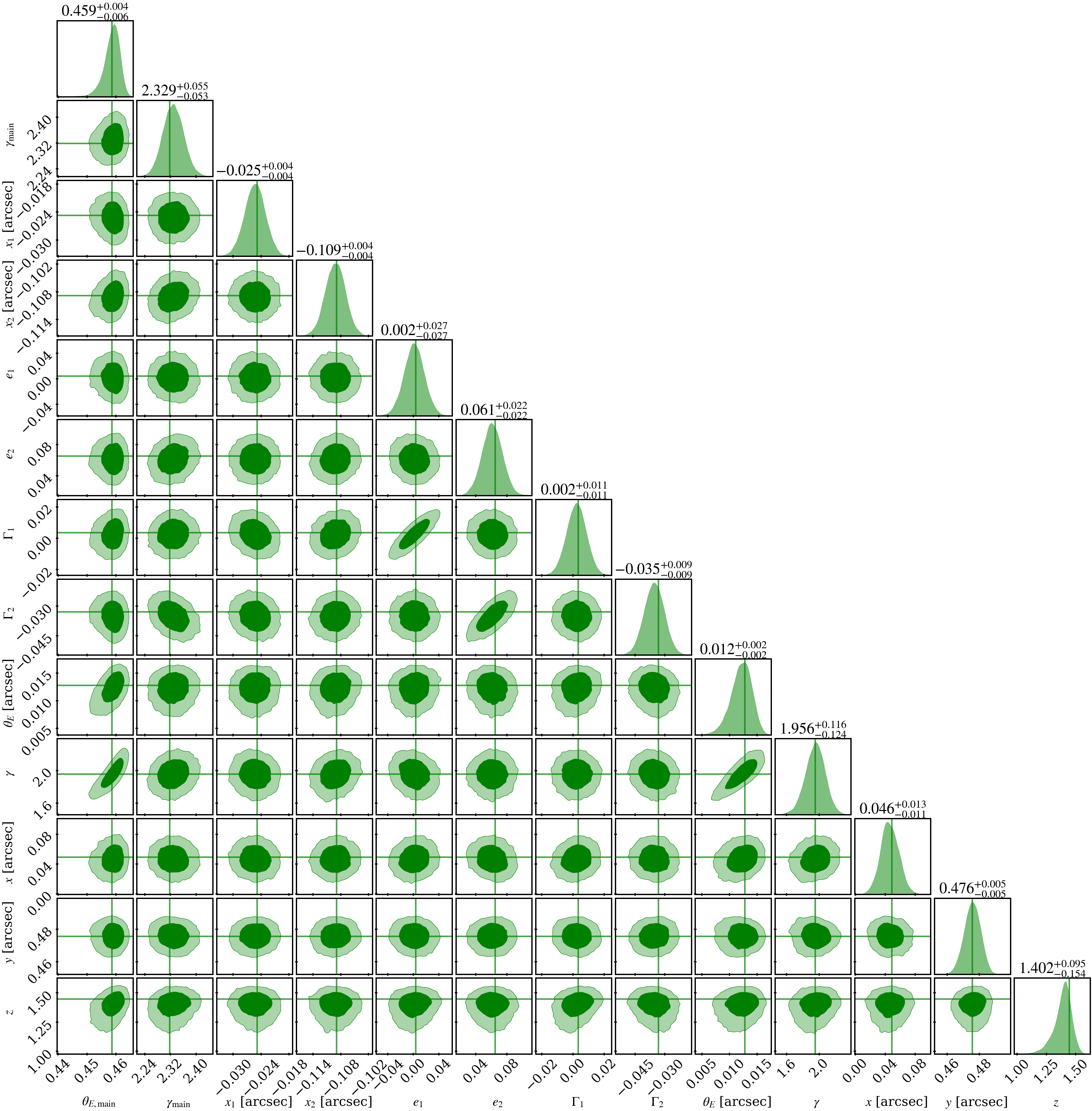}
    \caption{Posterior probability distribution of the model parameters in the system JVAS B1938+666. The parameters $\theta_{E,\mr{main}}$, $\gamma_{\mr{main}}$, $x_1$, $x_2$, $e_1$, and $e_2$ are the Einstein radius, power-law slope, the x and y coordinate of the center, and the Cartesian ellipticity components of the main lens profile, respectively. The parameters $\Gamma_1$ and $\Gamma_2$ are the Cartesian components of the external shear. The parameters $\theta_E$, $\gamma$, $x$, $y$, and $z$ are the Einstein radius, the power-law slope, the x and y coordinate of the center and the redshift of the perturber, respectively. The vertical lines correspond to the best fit values of the parameters.}
    \label{fig:post_JVAS}
\end{figure*}

The Hubble Space Telescope image of JVAS B1938+666 used in this study, its reconstruction and residuals are shown in Fig. \ref{fig:rec_JVAS}. The posterior probability distributions of the model parameters obtained by nested sampling are shown in Fig. \ref{fig:post_JVAS}

\end{appendix}

\clearpage
\bibliographystyle{mnras}
\bibliography{main} 

\begin{thebibliography}{}
\makeatletter
\relax
\def\mn@urlcharsother{\let\do\@makeother \do\$\do\&\do\#\do\^\do\_\do\%\do\~}
\def\mn@doi{\begingroup\mn@urlcharsother \@ifnextchar [ {\mn@doi@}
  {\mn@doi@[]}}
\def\mn@doi@[#1]#2{\def\@tempa{#1}\ifx\@tempa\@empty \href
  {http://dx.doi.org/#2} {doi:#2}\else \href {http://dx.doi.org/#2} {#1}\fi
  \endgroup}
\def\mn@eprint#1#2{\mn@eprint@#1:#2::\@nil}
\def\mn@eprint@arXiv#1{\href {http://arxiv.org/abs/#1} {{\tt arXiv:#1}}}
\def\mn@eprint@dblp#1{\href {http://dblp.uni-trier.de/rec/bibtex/#1.xml}
  {dblp:#1}}
\def\mn@eprint@#1:#2:#3:#4\@nil{\def\@tempa {#1}\def\@tempb {#2}\def\@tempc
  {#3}\ifx \@tempc \@empty \let \@tempc \@tempb \let \@tempb \@tempa \fi \ifx
  \@tempb \@empty \def\@tempb {arXiv}\fi \@ifundefined
  {mn@eprint@\@tempb}{\@tempb:\@tempc}{\expandafter \expandafter \csname
  mn@eprint@\@tempb\endcsname \expandafter{\@tempc}}}

\bibitem[\protect\citeauthoryear{Abbott et~al.,}{Abbott
  et~al.}{2018}]{des_matter_power}
Abbott T. M.~C.,  et~al., 2018, \mn@doi [Phys. Rev. D]
  {10.1103/PhysRevD.98.043526}, 98, 043526

\bibitem[\protect\citeauthoryear{{Abolfathi} et~al.,}{{Abolfathi}
  et~al.}{2018}]{lyman_alpha}
{Abolfathi} B.,  et~al., 2018, \mn@doi [\apjs] {10.3847/1538-4365/aa9e8a},
  \href {https://ui.adsabs.harvard.edu/abs/2018ApJS..235...42A} {235, 42}

\bibitem[\protect\citeauthoryear{{Amara}, {Metcalf}, {Cox}  \&
  {Ostriker}}{{Amara} et~al.}{2006}]{2006MNRAS.367.1367A}
{Amara} A.,  {Metcalf} R.~B.,  {Cox} T.~J.,   {Ostriker} J.~P.,  2006, \mn@doi
  [\mnras] {10.1111/j.1365-2966.2006.10053.x}, \href
  {https://ui.adsabs.harvard.edu/abs/2006MNRAS.367.1367A} {367, 1367}

\bibitem[\protect\citeauthoryear{{Anderhalden}, {Schneider}, {Macci{\`o}},
  {Diemand}  \& {Bertone}}{{Anderhalden} et~al.}{2013}]{milky_way_satellites}
{Anderhalden} D.,  {Schneider} A.,  {Macci{\`o}} A.~V.,  {Diemand} J.,
  {Bertone} G.,  2013, \mn@doi [\jcap] {10.1088/1475-7516/2013/03/014}, \href
  {https://ui.adsabs.harvard.edu/abs/2013JCAP...03..014A} {2013, 014}

\bibitem[\protect\citeauthoryear{{Angulo}, {Hahn}  \& {Abel}}{{Angulo}
  et~al.}{2013}]{wdm_mass_function}
{Angulo} R.~E.,  {Hahn} O.,   {Abel} T.,  2013, \mn@doi [\mnras]
  {10.1093/mnras/stt1246}, \href
  {https://ui.adsabs.harvard.edu/abs/2013MNRAS.434.3337A} {434, 3337}

\bibitem[\protect\citeauthoryear{Baltz, Marshall  \& Oguri}{Baltz
  et~al.}{2009}]{Baltz:2007vq}
Baltz E.~A.,  Marshall P.,   Oguri M.,  2009, \mn@doi [JCAP]
  {10.1088/1475-7516/2009/01/015}, 01, 015

\bibitem[\protect\citeauthoryear{{Barkana}}{{Barkana}}{1999}]{barkana_spemd}
{Barkana} R.,  1999, {FASTELL: Fast calculation of a family of elliptical mass
  gravitational lens models} (\mn@eprint {ascl} {9910.003})

\bibitem[\protect\citeauthoryear{{Bechtol} et~al.,}{{Bechtol}
  et~al.}{2019}]{Bechtol:2019acd}
{Bechtol} K.,  et~al., 2019, \baas, \href
  {https://ui.adsabs.harvard.edu/abs/2019BAAS...51c.207B} {51, 207}

\bibitem[\protect\citeauthoryear{{Birrer}, {Amara}  \& {Refregier}}{{Birrer}
  et~al.}{2015}]{lenstronomy_shape}
{Birrer} S.,  {Amara} A.,   {Refregier} A.,  2015, \mn@doi [\apj]
  {10.1088/0004-637X/813/2/102}, \href
  {https://ui.adsabs.harvard.edu/abs/2015ApJ...813..102B} {813, 102}

\bibitem[\protect\citeauthoryear{Birrer, Amara  \& Refregier}{Birrer
  et~al.}{2017}]{Birrer:2017rpp}
Birrer S.,  Amara A.,   Refregier A.,  2017, \mn@doi [JCAP]
  {10.1088/1475-7516/2017/05/037}, 05, 037

\bibitem[\protect\citeauthoryear{Birrer et~al.}{Birrer
  et~al.}{2019}]{Birrer:2018vtm}
Birrer S.,  et~al., 2019, \mn@doi [\mnras] {10.1093/mnras/stz200}, 484, 4726

\bibitem[\protect\citeauthoryear{Birrer et~al.}{Birrer
  et~al.}{2021}]{Birrer:2021wjl}
Birrer S.,  et~al., 2021, \mn@doi [J. Open Source Softw.]
  {10.21105/joss.03283}, 6, 3283

\bibitem[\protect\citeauthoryear{{Bode}, {Ostriker}  \& {Turok}}{{Bode}
  et~al.}{2001}]{2001ApJ...556...93B}
{Bode} P.,  {Ostriker} J.~P.,   {Turok} N.,  2001, \mn@doi [\apj]
  {10.1086/321541}, \href
  {https://ui.adsabs.harvard.edu/abs/2001ApJ...556...93B} {556, 93}

\bibitem[\protect\citeauthoryear{Brehmer, Mishra-Sharma, Hermans, Louppe  \&
  Cranmer}{Brehmer et~al.}{2019}]{Brehmer:2019jyt}
Brehmer J.,  Mishra-Sharma S.,  Hermans J.,  Louppe G.,   Cranmer K.,  2019,
  \mn@doi [Astrophys. J.] {10.3847/1538-4357/ab4c41}, 886, 49

\bibitem[\protect\citeauthoryear{Brennan, Benson, Cyr-Racine, Keeton, Moustakas
   \& Pullen}{Brennan et~al.}{2019}]{Brennan:2018jhq}
Brennan S.,  Benson A.~J.,  Cyr-Racine F.-Y.,  Keeton C.~R.,  Moustakas L.~A.,
   Pullen A.~R.,  2019, \mn@doi [Mon. Not. Roy. Astron. Soc.]
  {10.1093/mnras/stz1607}, 488, 5085

\bibitem[\protect\citeauthoryear{Brewer, Huijser  \& Lewis}{Brewer
  et~al.}{2016}]{Brewer:2015yya}
Brewer B.~J.,  Huijser D.,   Lewis G.~F.,  2016, \mn@doi [Mon. Not. Roy.
  Astron. Soc.] {10.1093/mnras/stv2370}, 455, 1819

\bibitem[\protect\citeauthoryear{{Brook}, {Stinson}, {Gibson}, {Wadsley}  \&
  {Quinn}}{{Brook} et~al.}{2012}]{2012MNRAS.424.1275B}
{Brook} C.~B.,  {Stinson} G.,  {Gibson} B.~K.,  {Wadsley} J.,   {Quinn} T.,
  2012, \mn@doi [\mnras] {10.1111/j.1365-2966.2012.21306.x}, \href
  {https://ui.adsabs.harvard.edu/abs/2012MNRAS.424.1275B} {424, 1275}

\bibitem[\protect\citeauthoryear{{Buckley} \& {Peter}}{{Buckley} \&
  {Peter}}{2018}]{2018PhR...761....1B}
{Buckley} M.~R.,  {Peter} A. H.~G.,  2018, \mn@doi [\physrep]
  {10.1016/j.physrep.2018.07.003}, \href
  {https://ui.adsabs.harvard.edu/abs/2018PhR...761....1B} {761, 1}

\bibitem[\protect\citeauthoryear{Bullock \& Boylan-Kolchin}{Bullock \&
  Boylan-Kolchin}{2017}]{Bullock:2017xww}
Bullock J.~S.,  Boylan-Kolchin M.,  2017, \mn@doi [Ann. Rev. Astron.
  Astrophys.] {10.1146/annurev-astro-091916-055313}, 55, 343

\bibitem[\protect\citeauthoryear{{Chan}, {Kere{\v{s}}}, {O{\~n}orbe},
  {Hopkins}, {Muratov}, {Faucher-Gigu{\`e}re}  \& {Quataert}}{{Chan}
  et~al.}{2015}]{2015MNRAS.454.2981C}
{Chan} T.~K.,  {Kere{\v{s}}} D.,  {O{\~n}orbe} J.,  {Hopkins} P.~F.,  {Muratov}
  A.~L.,  {Faucher-Gigu{\`e}re} C.~A.,   {Quataert} E.,  2015, \mn@doi [\mnras]
  {10.1093/mnras/stv2165}, \href
  {https://ui.adsabs.harvard.edu/abs/2015MNRAS.454.2981C} {454, 2981}

\bibitem[\protect\citeauthoryear{Collett}{Collett}{2015}]{Collett:2015roa}
Collett T.~E.,  2015, \mn@doi [Astrophys. J.] {10.1088/0004-637X/811/1/20},
  811, 20

\bibitem[\protect\citeauthoryear{Cyr-Racine, Sigurdson, Zavala, Bringmann,
  Vogelsberger  \& Pfrommer}{Cyr-Racine et~al.}{2016a}]{Cyr-Racine:2015ihg}
Cyr-Racine F.-Y.,  Sigurdson K.,  Zavala J.,  Bringmann T.,  Vogelsberger M.,
  Pfrommer C.,  2016a, \mn@doi [Phys. Rev. D] {10.1103/PhysRevD.93.123527}, 93,
  123527

\bibitem[\protect\citeauthoryear{Cyr-Racine, Moustakas, Keeton, Sigurdson  \&
  Gilman}{Cyr-Racine et~al.}{2016b}]{Cyr-Racine:2015jwa}
Cyr-Racine F.-Y.,  Moustakas L.~A.,  Keeton C.~R.,  Sigurdson K.,   Gilman
  D.~A.,  2016b, \mn@doi [Phys. Rev. D] {10.1103/PhysRevD.94.043505}, 94,
  043505

\bibitem[\protect\citeauthoryear{Cyr-Racine, Keeton  \& Moustakas}{Cyr-Racine
  et~al.}{2019}]{Cyr-Racine:2018htu}
Cyr-Racine F.-Y.,  Keeton C.~R.,   Moustakas L.~A.,  2019, \mn@doi [Phys. Rev.
  D] {10.1103/PhysRevD.100.023013}, 100, 023013

\bibitem[\protect\citeauthoryear{{Dalal} \& {Kochanek}}{{Dalal} \&
  {Kochanek}}{2002}]{2002ApJ...572...25D}
{Dalal} N.,  {Kochanek} C.~S.,  2002, \mn@doi [\apj] {10.1086/340303}, \href
  {https://ui.adsabs.harvard.edu/abs/2002ApJ...572...25D} {572, 25}

\bibitem[\protect\citeauthoryear{{Davis}, {Efstathiou}, {Frenk}  \&
  {White}}{{Davis} et~al.}{1985}]{1985ApJ...292..371D}
{Davis} M.,  {Efstathiou} G.,  {Frenk} C.~S.,   {White} S.~D.~M.,  1985,
  \mn@doi [\apj] {10.1086/163168}, \href
  {https://ui.adsabs.harvard.edu/abs/1985ApJ...292..371D} {292, 371}

\bibitem[\protect\citeauthoryear{Daylan, Cyr-Racine, Diaz~Rivero, Dvorkin  \&
  Finkbeiner}{Daylan et~al.}{2018}]{Daylan:2017kfh}
Daylan T.,  Cyr-Racine F.-Y.,  Diaz~Rivero A.,  Dvorkin C.,   Finkbeiner D.~P.,
   2018, \mn@doi [Astrophys. J.] {10.3847/1538-4357/aaaa1e}, 854, 141

\bibitem[\protect\citeauthoryear{Diaz~Rivero, Cyr-Racine  \&
  Dvorkin}{Diaz~Rivero et~al.}{2018a}]{DiazRivero:2017xkd}
Diaz~Rivero A.,  Cyr-Racine F.-Y.,   Dvorkin C.,  2018a, \mn@doi [Phys. Rev. D]
  {10.1103/PhysRevD.97.023001}, 97, 023001

\bibitem[\protect\citeauthoryear{D\'\i{}az~Rivero, Dvorkin, Cyr-Racine, Zavala
  \& Vogelsberger}{D\'\i{}az~Rivero et~al.}{2018b}]{DiazRivero:2018oxk}
D\'\i{}az~Rivero A.,  Dvorkin C.,  Cyr-Racine F.-Y.,  Zavala J.,   Vogelsberger
  M.,  2018b, \mn@doi [Phys. Rev. D] {10.1103/PhysRevD.98.103517}, 98, 103517

\bibitem[\protect\citeauthoryear{{Dutton} \& {Macci{\`o}}}{{Dutton} \&
  {Macci{\`o}}}{2014}]{mass_concen}
{Dutton} A.~A.,  {Macci{\`o}} A.~V.,  2014, \mn@doi [\mnras]
  {10.1093/mnras/stu742}, \href
  {https://ui.adsabs.harvard.edu/abs/2014MNRAS.441.3359D} {441, 3359}

\bibitem[\protect\citeauthoryear{{Elbert}, {Bullock}, {Garrison-Kimmel},
  {Rocha}, {O{\~n}orbe}  \& {Peter}}{{Elbert} et~al.}{2015}]{sidm_core2}
{Elbert} O.~D.,  {Bullock} J.~S.,  {Garrison-Kimmel} S.,  {Rocha} M.,
  {O{\~n}orbe} J.,   {Peter} A. H.~G.,  2015, \mn@doi [\mnras]
  {10.1093/mnras/stv1470}, \href
  {https://ui.adsabs.harvard.edu/abs/2015MNRAS.453...29E} {453, 29}

\bibitem[\protect\citeauthoryear{Fadely \& Keeton}{Fadely \&
  Keeton}{2011}]{10.1111/j.1365-2966.2011.19729.x}
Fadely R.,  Keeton C.~R.,  2011, \mn@doi [Monthly Notices of the Royal
  Astronomical Society] {10.1111/j.1365-2966.2011.19729.x}, 419, 936

\bibitem[\protect\citeauthoryear{Fitts et~al.}{Fitts
  et~al.}{2017}]{Fitts:2016usl}
Fitts A.,  et~al., 2017, \mn@doi [Mon. Not. Roy. Astron. Soc.]
  {10.1093/mnras/stx1757}, 471, 3547

\bibitem[\protect\citeauthoryear{{Flores} \& {Primack}}{{Flores} \&
  {Primack}}{1994}]{cusp_vs_core2}
{Flores} R.~A.,  {Primack} J.~R.,  1994, \mn@doi [\apjl] {10.1086/187350},
  \href {https://ui.adsabs.harvard.edu/abs/1994ApJ...427L...1F} {427, L1}

\bibitem[\protect\citeauthoryear{Fry et~al.,}{Fry et~al.}{2015}]{Fry:2015rta}
Fry A.~B.,  et~al., 2015, \mn@doi [Mon. Not. Roy. Astron. Soc.]
  {10.1093/mnras/stv1330}, 452, 1468

\bibitem[\protect\citeauthoryear{Gentile, Salucci, Klein, Vergani  \&
  Kalberla}{Gentile et~al.}{2004}]{Gentile:2004tb}
Gentile G.,  Salucci P.,  Klein U.,  Vergani D.,   Kalberla P.,  2004, \mn@doi
  [Mon. Not. Roy. Astron. Soc.] {10.1111/j.1365-2966.2004.07836.x}, 351, 903

\bibitem[\protect\citeauthoryear{Gil-Marín et~al.,}{Gil-Marín
  et~al.}{2016}]{boss_matter_power}
Gil-Marín H.,  et~al., 2016, \mn@doi [Monthly Notices of the Royal
  Astronomical Society] {10.1093/mnras/stw1096}, 460, 4188

\bibitem[\protect\citeauthoryear{Gilman, Birrer, Treu, Nierenberg  \&
  Benson}{Gilman et~al.}{2019}]{Gilman:2019vca}
Gilman D.,  Birrer S.,  Treu T.,  Nierenberg A.,   Benson A.,  2019, \mn@doi
  [Mon. Not. Roy. Astron. Soc.] {10.1093/mnras/stz1593}, 487, 5721

\bibitem[\protect\citeauthoryear{Gnedin \& Zhao}{Gnedin \&
  Zhao}{2002}]{Gnedin:2001ec}
Gnedin O.~Y.,  Zhao H.,  2002, \mn@doi [Mon. Not. Roy. Astron. Soc.]
  {10.1046/j.1365-8711.2002.05361.x}, 333, 299

\bibitem[\protect\citeauthoryear{{Governato} et~al.,}{{Governato}
  et~al.}{2012}]{cuspy_no_more}
{Governato} F.,  et~al., 2012, \mn@doi [\mnras]
  {10.1111/j.1365-2966.2012.20696.x}, \href
  {https://ui.adsabs.harvard.edu/abs/2012MNRAS.422.1231G} {422, 1231}

\bibitem[\protect\citeauthoryear{Green, Hofmann  \& Schwarz}{Green
  et~al.}{2004}]{Green:2003un}
Green A.~M.,  Hofmann S.,   Schwarz D.~J.,  2004, \mn@doi [Mon. Not. Roy.
  Astron. Soc.] {10.1111/j.1365-2966.2004.08232.x}, 353, L23

\bibitem[\protect\citeauthoryear{Hezaveh, Dalal, Holder, Kisner, Kuhlen  \&
  Perreault~Levasseur}{Hezaveh et~al.}{2016a}]{Hezaveh:2014aoa}
Hezaveh Y.,  Dalal N.,  Holder G.,  Kisner T.,  Kuhlen M.,
  Perreault~Levasseur L.,  2016a, \mn@doi [JCAP]
  {10.1088/1475-7516/2016/11/048}, 11, 048

\bibitem[\protect\citeauthoryear{{Hezaveh} et~al.,}{{Hezaveh}
  et~al.}{2016b}]{hezaveh_alma_substructure}
{Hezaveh} Y.~D.,  et~al., 2016b, \mn@doi [\apj] {10.3847/0004-637X/823/1/37},
  \href {https://ui.adsabs.harvard.edu/abs/2016ApJ...823...37H} {823, 37}

\bibitem[\protect\citeauthoryear{Hu, Barkana  \& Gruzinov}{Hu
  et~al.}{2000}]{PhysRevLett.85.1158}
Hu W.,  Barkana R.,   Gruzinov A.,  2000, \mn@doi [Phys. Rev. Lett.]
  {10.1103/PhysRevLett.85.1158}, 85, 1158

\bibitem[\protect\citeauthoryear{{Huang} et~al.,}{{Huang}
  et~al.}{2021}]{2021ApJ...909...27H}
{Huang} X.,  et~al., 2021, \mn@doi [\apj] {10.3847/1538-4357/abd62b}, \href
  {https://ui.adsabs.harvard.edu/abs/2021ApJ...909...27H} {909, 27}

\bibitem[\protect\citeauthoryear{{Jacobs} et~al.,}{{Jacobs}
  et~al.}{2019}]{2019ApJS..243...17J}
{Jacobs} C.,  et~al., 2019, \mn@doi [\apjs] {10.3847/1538-4365/ab26b6}, \href
  {https://ui.adsabs.harvard.edu/abs/2019ApJS..243...17J} {243, 17}

\bibitem[\protect\citeauthoryear{Kamada, Kaplinghat, Pace  \& Yu}{Kamada
  et~al.}{2017}]{sidm_solves_div}
Kamada A.,  Kaplinghat M.,  Pace A.~B.,   Yu H.-B.,  2017, \mn@doi [Phys. Rev.
  Lett.] {10.1103/PhysRevLett.119.111102}, 119, 111102

\bibitem[\protect\citeauthoryear{Kaplinghat}{Kaplinghat}{2005}]{PhysRevD.72.063510}
Kaplinghat M.,  2005, \mn@doi [Phys. Rev. D] {10.1103/PhysRevD.72.063510}, 72,
  063510

\bibitem[\protect\citeauthoryear{Kaplinghat, Tulin  \& Yu}{Kaplinghat
  et~al.}{2016}]{sidm_cross_section}
Kaplinghat M.,  Tulin S.,   Yu H.-B.,  2016, \mn@doi [Phys. Rev. Lett.]
  {10.1103/PhysRevLett.116.041302}, 116, 041302

\bibitem[\protect\citeauthoryear{Kaplinghat, Ren  \& Yu}{Kaplinghat
  et~al.}{2020}]{Kaplinghat:2019dhn}
Kaplinghat M.,  Ren T.,   Yu H.-B.,  2020, \mn@doi [JCAP]
  {10.1088/1475-7516/2020/06/027}, 06, 027

\bibitem[\protect\citeauthoryear{{Koopmans}}{{Koopmans}}{2005}]{gravitational_imaging}
{Koopmans} L.~V.~E.,  2005, \mn@doi [\mnras]
  {10.1111/j.1365-2966.2005.09523.x}, \href
  {https://ui.adsabs.harvard.edu/abs/2005MNRAS.363.1136K} {363, 1136}

\bibitem[\protect\citeauthoryear{Lagattuta, Vegetti, Fassnacht, Auger, Koopmans
   \& McKean}{Lagattuta et~al.}{2012}]{lagatutta}
Lagattuta D.~J.,  Vegetti S.,  Fassnacht C.~D.,  Auger M.~W.,  Koopmans L.
  V.~E.,   McKean J.~P.,  2012, \mn@doi [Monthly Notices of the Royal
  Astronomical Society] {10.1111/j.1365-2966.2012.21406.x}, 424, 2800

\bibitem[\protect\citeauthoryear{{Lovell} et~al.,}{{Lovell}
  et~al.}{2012}]{wdm_halos2}
{Lovell} M.~R.,  et~al., 2012, \mn@doi [\mnras]
  {10.1111/j.1365-2966.2011.20200.x}, \href
  {https://ui.adsabs.harvard.edu/abs/2012MNRAS.420.2318L} {420, 2318}

\bibitem[\protect\citeauthoryear{{Lovell}, {Frenk}, {Eke}, {Jenkins}, {Gao}  \&
  {Theuns}}{{Lovell} et~al.}{2014}]{wdm_at_small}
{Lovell} M.~R.,  {Frenk} C.~S.,  {Eke} V.~R.,  {Jenkins} A.,  {Gao} L.,
  {Theuns} T.,  2014, \mn@doi [\mnras] {10.1093/mnras/stt2431}, \href
  {https://ui.adsabs.harvard.edu/abs/2014MNRAS.439..300L} {439, 300}

\bibitem[\protect\citeauthoryear{MacLeod, Jones, Agol  \& Kochanek}{MacLeod
  et~al.}{2013}]{MacLeod_2013}
MacLeod C.~L.,  Jones R.,  Agol E.,   Kochanek C.~S.,  2013, \mn@doi [The
  Astrophysical Journal] {10.1088/0004-637x/773/1/35}, 773, 35

\bibitem[\protect\citeauthoryear{Mao \& Schneider}{Mao \&
  Schneider}{1998}]{10.1046/j.1365-8711.1998.01319.x}
Mao S.,  Schneider P.,  1998, \mn@doi [Monthly Notices of the Royal
  Astronomical Society] {10.1046/j.1365-8711.1998.01319.x}, 295, 587

\bibitem[\protect\citeauthoryear{Marasco, Oman, Navarro, Frenk  \&
  Oosterloo}{Marasco et~al.}{2018}]{10.1093/mnras/sty354}
Marasco A.,  Oman K.~A.,  Navarro J.~F.,  Frenk C.~S.,   Oosterloo T.,  2018,
  \mn@doi [Monthly Notices of the Royal Astronomical Society]
  {10.1093/mnras/sty354}, 476, 2168

\bibitem[\protect\citeauthoryear{{Mashchenko}, {Couchman}  \&
  {Wadsley}}{{Mashchenko} et~al.}{2006}]{2006Natur.442..539M}
{Mashchenko} S.,  {Couchman} H.~M.~P.,   {Wadsley} J.,  2006, \mn@doi [\nat]
  {10.1038/nature04944}, \href
  {https://ui.adsabs.harvard.edu/abs/2006Natur.442..539M} {442, 539}

\bibitem[\protect\citeauthoryear{{McKean} et~al.,}{{McKean}
  et~al.}{2015}]{2015aska.confE..84M}
{McKean} J.,  et~al., 2015, in Advancing Astrophysics with the Square Kilometre
  Array (AASKA14). p.~84 (\mn@eprint {arXiv} {1502.03362})

\bibitem[\protect\citeauthoryear{{Metcalf} \& {Madau}}{{Metcalf} \&
  {Madau}}{2001}]{2001ApJ...563....9M}
{Metcalf} R.~B.,  {Madau} P.,  2001, \mn@doi [\apj] {10.1086/323695}, \href
  {https://ui.adsabs.harvard.edu/abs/2001ApJ...563....9M} {563, 9}

\bibitem[\protect\citeauthoryear{{Minor}, {Kaplinghat}  \& {Li}}{{Minor}
  et~al.}{2017}]{better_mass}
{Minor} Q.~E.,  {Kaplinghat} M.,   {Li} N.,  2017, \mn@doi [\apj]
  {10.3847/1538-4357/aa7fee}, \href
  {https://ui.adsabs.harvard.edu/abs/2017ApJ...845..118M} {845, 118}

\bibitem[\protect\citeauthoryear{Minor, Kaplinghat, Chan  \& Simon}{Minor
  et~al.}{2021a}]{Minor:2020bmp}
Minor Q.~E.,  Kaplinghat M.,  Chan T.~H.,   Simon E.,  2021a, \mn@doi [Mon.
  Not. Roy. Astron. Soc.] {10.1093/mnras/stab2209}, 507, 1202

\bibitem[\protect\citeauthoryear{{Minor}, {Gad-Nasr}, {Kaplinghat}  \&
  {Vegetti}}{{Minor} et~al.}{2021b}]{sdss_slope}
{Minor} Q.,  {Gad-Nasr} S.,  {Kaplinghat} M.,   {Vegetti} S.,  2021b, \mn@doi
  [\mnras] {10.1093/mnras/stab2247}, \href
  {https://ui.adsabs.harvard.edu/abs/2021MNRAS.507.1662M} {507, 1662}

\bibitem[\protect\citeauthoryear{{Moore}}{{Moore}}{1994}]{cusp_vs_core1}
{Moore} B.,  1994, \mn@doi [\nat] {10.1038/370629a0}, \href
  {https://ui.adsabs.harvard.edu/abs/1994Natur.370..629M} {370, 629}

\bibitem[\protect\citeauthoryear{{Moore}, {Ghigna}, {Governato}, {Lake},
  {Quinn}, {Stadel}  \& {Tozzi}}{{Moore} et~al.}{1999}]{1999ApJ...524L..19M}
{Moore} B.,  {Ghigna} S.,  {Governato} F.,  {Lake} G.,  {Quinn} T.,  {Stadel}
  J.,   {Tozzi} P.,  1999, \mn@doi [\apjl] {10.1086/312287}, \href
  {https://ui.adsabs.harvard.edu/abs/1999ApJ...524L..19M} {524, L19}

\bibitem[\protect\citeauthoryear{Moustakas \& Metcalf}{Moustakas \&
  Metcalf}{2003}]{10.1046/j.1365-8711.2003.06055.x}
Moustakas L.~A.,  Metcalf R.~B.,  2003, \mn@doi [Monthly Notices of the Royal
  Astronomical Society] {10.1046/j.1365-8711.2003.06055.x}, 339, 607

\bibitem[\protect\citeauthoryear{Navarro, Eke  \& Frenk}{Navarro
  et~al.}{1996a}]{Navarro:1996bv}
Navarro J.~F.,  Eke V.~R.,   Frenk C.~S.,  1996a, \mn@doi [Mon. Not. Roy.
  Astron. Soc.] {10.1093/mnras/283.3.L72}, 283, L72

\bibitem[\protect\citeauthoryear{{Navarro}, {Frenk}  \& {White}}{{Navarro}
  et~al.}{1996b}]{NFW}
{Navarro} J.~F.,  {Frenk} C.~S.,   {White} S. D.~M.,  1996b, \mn@doi [\apj]
  {10.1086/177173}, \href
  {https://ui.adsabs.harvard.edu/abs/1996ApJ...462..563N} {462, 563}

\bibitem[\protect\citeauthoryear{{Navarro}, {Frenk}  \& {White}}{{Navarro}
  et~al.}{1997}]{NFW2}
{Navarro} J.~F.,  {Frenk} C.~S.,   {White} S. D.~M.,  1997, \mn@doi [\apj]
  {10.1086/304888}, \href
  {https://ui.adsabs.harvard.edu/abs/1997ApJ...490..493N} {490, 493}

\bibitem[\protect\citeauthoryear{{Nierenberg}, {Treu}, {Wright}, {Fassnacht}
  \& {Auger}}{{Nierenberg} et~al.}{2014}]{nierenberg_subhalo}
{Nierenberg} A.~M.,  {Treu} T.,  {Wright} S.~A.,  {Fassnacht} C.~D.,   {Auger}
  M.~W.,  2014, \mn@doi [\mnras] {10.1093/mnras/stu862}, \href
  {https://ui.adsabs.harvard.edu/abs/2014MNRAS.442.2434N} {442, 2434}

\bibitem[\protect\citeauthoryear{Nierenberg et~al.,}{Nierenberg
  et~al.}{2017}]{10.1093/mnras/stx1400}
Nierenberg A.~M.,  et~al., 2017, \mn@doi [Monthly Notices of the Royal
  Astronomical Society] {10.1093/mnras/stx1400}, 471, 2224

\bibitem[\protect\citeauthoryear{{Ogiya}, {van den Bosch}, {Hahn}, {Green},
  {Miller}  \& {Burkert}}{{Ogiya} et~al.}{2019}]{2019MNRAS.485..189O}
{Ogiya} G.,  {van den Bosch} F.~C.,  {Hahn} O.,  {Green} S.~B.,  {Miller}
  T.~B.,   {Burkert} A.,  2019, \mn@doi [\mnras] {10.1093/mnras/stz375}, \href
  {https://ui.adsabs.harvard.edu/abs/2019MNRAS.485..189O} {485, 189}

\bibitem[\protect\citeauthoryear{{Oguri} \& {Marshall}}{{Oguri} \&
  {Marshall}}{2010}]{2010MNRAS.405.2579O}
{Oguri} M.,  {Marshall} P.~J.,  2010, \mn@doi [\mnras]
  {10.1111/j.1365-2966.2010.16639.x}, \href
  {https://ui.adsabs.harvard.edu/abs/2010MNRAS.405.2579O} {405, 2579}

\bibitem[\protect\citeauthoryear{{Oman} et~al.,}{{Oman}
  et~al.}{2015}]{diversity}
{Oman} K.~A.,  et~al., 2015, \mn@doi [\mnras] {10.1093/mnras/stv1504}, \href
  {https://ui.adsabs.harvard.edu/abs/2015MNRAS.452.3650O} {452, 3650}

\bibitem[\protect\citeauthoryear{Oman, Marasco, Navarro, Frenk, Schaye  \&
  Bentez-Llambay}{Oman et~al.}{2018}]{10.1093/mnras/sty2687}
Oman K.~A.,  Marasco A.,  Navarro J.~F.,  Frenk C.~S.,  Schaye J.,
  Bentez-Llambay A.,  2018, \mn@doi [Monthly Notices of the Royal Astronomical
  Society] {10.1093/mnras/sty2687}, 482, 821

\bibitem[\protect\citeauthoryear{Ostdiek, Diaz~Rivero  \& Dvorkin}{Ostdiek
  et~al.}{2022a}]{Ostdiek:2020cqz}
Ostdiek B.,  Diaz~Rivero A.,   Dvorkin C.,  2022a, \mn@doi [Astron. Astrophys.]
  {10.1051/0004-6361/202142030}, 657, L14

\bibitem[\protect\citeauthoryear{Ostdiek, Diaz~Rivero  \& Dvorkin}{Ostdiek
  et~al.}{2022b}]{Ostdiek:2020mvo}
Ostdiek B.,  Diaz~Rivero A.,   Dvorkin C.,  2022b, \mn@doi [Astrophys. J.]
  {10.3847/1538-4357/ac2d8d}, 927, 83

\bibitem[\protect\citeauthoryear{{Pe{\~n}arrubia}, {Benson}, {Walker},
  {Gilmore}, {McConnachie}  \& {Mayer}}{{Pe{\~n}arrubia}
  et~al.}{2010}]{2010MNRAS.406.1290P}
{Pe{\~n}arrubia} J.,  {Benson} A.~J.,  {Walker} M.~G.,  {Gilmore} G.,
  {McConnachie} A.~W.,   {Mayer} L.,  2010, \mn@doi [\mnras]
  {10.1111/j.1365-2966.2010.16762.x}, \href
  {https://ui.adsabs.harvard.edu/abs/2010MNRAS.406.1290P} {406, 1290}

\bibitem[\protect\citeauthoryear{{Peebles}}{{Peebles}}{2000}]{2000ApJ...534L.127P}
{Peebles} P.~J.~E.,  2000, \mn@doi [\apjl] {10.1086/312677}, \href
  {https://ui.adsabs.harvard.edu/abs/2000ApJ...534L.127P} {534, L127}

\bibitem[\protect\citeauthoryear{{Pereira Wilson}, {Navarro}, {Santos Santos}
  \& {Benitez Llambay}}{{Pereira Wilson} et~al.}{2022}]{2022arXiv220605338P}
{Pereira Wilson} M.,  {Navarro} J.,  {Santos Santos} I.,   {Benitez Llambay}
  A.,  2022, arXiv e-prints, \href
  {https://ui.adsabs.harvard.edu/abs/2022arXiv220605338P} {p. arXiv:2206.05338}

\bibitem[\protect\citeauthoryear{{Planck Collaboration} et~al.,}{{Planck
  Collaboration} et~al.}{2020}]{planck2018}
{Planck Collaboration} et~al., 2020, \mn@doi [\aap]
  {10.1051/0004-6361/201833910}, \href
  {https://ui.adsabs.harvard.edu/abs/2020A&A...641A...6P} {641, A6}

\bibitem[\protect\citeauthoryear{{Pontzen} \& {Governato}}{{Pontzen} \&
  {Governato}}{2012}]{2012MNRAS.421.3464P}
{Pontzen} A.,  {Governato} F.,  2012, \mn@doi [\mnras]
  {10.1111/j.1365-2966.2012.20571.x}, \href
  {https://ui.adsabs.harvard.edu/abs/2012MNRAS.421.3464P} {421, 3464}

\bibitem[\protect\citeauthoryear{Read \& Gilmore}{Read \&
  Gilmore}{2005}]{10.1111/j.1365-2966.2004.08424.x}
Read J.~I.,  Gilmore G.,  2005, \mn@doi [Monthly Notices of the Royal
  Astronomical Society] {10.1111/j.1365-2966.2004.08424.x}, 356, 107

\bibitem[\protect\citeauthoryear{{Read}, {Iorio}, {Agertz}  \&
  {Fraternali}}{{Read} et~al.}{2017}]{2017MNRAS.467.2019R}
{Read} J.~I.,  {Iorio} G.,  {Agertz} O.,   {Fraternali} F.,  2017, \mn@doi
  [\mnras] {10.1093/mnras/stx147}, \href
  {https://ui.adsabs.harvard.edu/abs/2017MNRAS.467.2019R} {467, 2019}

\bibitem[\protect\citeauthoryear{{Refregier}}{{Refregier}}{2003}]{shapelets}
{Refregier} A.,  2003, \mn@doi [\mnras] {10.1046/j.1365-8711.2003.05901.x},
  \href {https://ui.adsabs.harvard.edu/abs/2003MNRAS.338...35R} {338, 35}

\bibitem[\protect\citeauthoryear{{Riechers}}{{Riechers}}{2011}]{z_source}
{Riechers} D.~A.,  2011, \mn@doi [\apj] {10.1088/0004-637X/730/2/108}, \href
  {https://ui.adsabs.harvard.edu/abs/2011ApJ...730..108R} {730, 108}

\bibitem[\protect\citeauthoryear{{Ritondale}, {Vegetti}, {Despali}, {Auger},
  {Koopmans}  \& {McKean}}{{Ritondale}
  et~al.}{2019}]{ritondale_lack_of_detection}
{Ritondale} E.,  {Vegetti} S.,  {Despali} G.,  {Auger} M.~W.,  {Koopmans}
  L.~V.~E.,   {McKean} J.~P.,  2019, \mn@doi [\mnras] {10.1093/mnras/stz464},
  \href {https://ui.adsabs.harvard.edu/abs/2019MNRAS.485.2179R} {485, 2179}

\bibitem[\protect\citeauthoryear{{Rocha}, {Peter}, {Bullock}, {Kaplinghat},
  {Garrison-Kimmel}, {O{\~n}orbe}  \& {Moustakas}}{{Rocha}
  et~al.}{2013}]{sidm_core}
{Rocha} M.,  {Peter} A. H.~G.,  {Bullock} J.~S.,  {Kaplinghat} M.,
  {Garrison-Kimmel} S.,  {O{\~n}orbe} J.,   {Moustakas} L.~A.,  2013, \mn@doi
  [\mnras] {10.1093/mnras/sts514}, \href
  {https://ui.adsabs.harvard.edu/abs/2013MNRAS.430...81R} {430, 81}

\bibitem[\protect\citeauthoryear{{Santos-Santos} et~al.,}{{Santos-Santos}
  et~al.}{2020}]{baryons_on_diversity}
{Santos-Santos} I. M.~E.,  et~al., 2020, \mn@doi [\mnras]
  {10.1093/mnras/staa1072}, \href
  {https://ui.adsabs.harvard.edu/abs/2020MNRAS.495...58S} {495, 58}

\bibitem[\protect\citeauthoryear{{Schneider}}{{Schneider}}{2015}]{small_scale_hmf}
{Schneider} A.,  2015, \mn@doi [\mnras] {10.1093/mnras/stv1169}, \href
  {https://ui.adsabs.harvard.edu/abs/2015MNRAS.451.3117S} {451, 3117}

\bibitem[\protect\citeauthoryear{Shajib et~al.}{Shajib
  et~al.}{2020}]{DES:2019fny}
Shajib A.~J.,  et~al., 2020, \mn@doi [\mnras] {10.1093/mnras/staa828}, 494,
  6072

\bibitem[\protect\citeauthoryear{{Speagle}}{{Speagle}}{2020}]{2020MNRAS.493.3132S}
{Speagle} J.~S.,  2020, \mn@doi [\mnras] {10.1093/mnras/staa278}, \href
  {https://ui.adsabs.harvard.edu/abs/2020MNRAS.493.3132S} {493, 3132}

\bibitem[\protect\citeauthoryear{Spergel \& Steinhardt}{Spergel \&
  Steinhardt}{2000}]{sidm_base}
Spergel D.~N.,  Steinhardt P.~J.,  2000, \mn@doi [Phys. Rev. Lett.]
  {10.1103/PhysRevLett.84.3760}, 84, 3760

\bibitem[\protect\citeauthoryear{{Springel}, {Frenk}  \& {White}}{{Springel}
  et~al.}{2006}]{large_scales}
{Springel} V.,  {Frenk} C.~S.,   {White} S. D.~M.,  2006, \mn@doi [\nat]
  {10.1038/nature04805}, \href
  {https://ui.adsabs.harvard.edu/abs/2006Natur.440.1137S} {440, 1137}

\bibitem[\protect\citeauthoryear{Syer \& White}{Syer \&
  White}{1998}]{10.1046/j.1365-8711.1998.01285.x}
Syer D.,  White S. D.~M.,  1998, \mn@doi [Monthly Notices of the Royal
  Astronomical Society] {10.1046/j.1365-8711.1998.01285.x}, 293, 337

\bibitem[\protect\citeauthoryear{{Tonry} \& {Kochanek}}{{Tonry} \&
  {Kochanek}}{2000}]{z_lens}
{Tonry} J.~L.,  {Kochanek} C.~S.,  2000, \mn@doi [\aj] {10.1086/301273}, \href
  {https://ui.adsabs.harvard.edu/abs/2000AJ....119.1078T} {119, 1078}

\bibitem[\protect\citeauthoryear{Troxel et~al.,}{Troxel
  et~al.}{2018}]{cosmic_shear}
Troxel M.~A.,  et~al., 2018, \mn@doi [Phys. Rev. D]
  {10.1103/PhysRevD.98.043528}, 98, 043528

\bibitem[\protect\citeauthoryear{Tulin \& Yu}{Tulin \&
  Yu}{2018}]{Tulin:2017ara}
Tulin S.,  Yu H.-B.,  2018, \mn@doi [Phys. Rept.]
  {10.1016/j.physrep.2017.11.004}, 730, 1

\bibitem[\protect\citeauthoryear{{Vegetti} \& {Vogelsberger}}{{Vegetti} \&
  {Vogelsberger}}{2014}]{vegetti_vogelsberger}
{Vegetti} S.,  {Vogelsberger} M.,  2014, \mn@doi [\mnras]
  {10.1093/mnras/stu1284}, \href
  {https://ui.adsabs.harvard.edu/abs/2014MNRAS.442.3598V} {442, 3598}

\bibitem[\protect\citeauthoryear{{Vegetti}, {Koopmans}, {Bolton}, {Treu}  \&
  {Gavazzi}}{{Vegetti} et~al.}{2010}]{sdss_substructure}
{Vegetti} S.,  {Koopmans} L.~V.~E.,  {Bolton} A.,  {Treu} T.,   {Gavazzi} R.,
  2010, \mn@doi [\mnras] {10.1111/j.1365-2966.2010.16865.x}, \href
  {https://ui.adsabs.harvard.edu/abs/2010MNRAS.408.1969V} {408, 1969}

\bibitem[\protect\citeauthoryear{{Vegetti}, {Lagattuta}, {McKean}, {Auger},
  {Fassnacht}  \& {Koopmans}}{{Vegetti} et~al.}{2012}]{2012Natur.481..341V}
{Vegetti} S.,  {Lagattuta} D.~J.,  {McKean} J.~P.,  {Auger} M.~W.,  {Fassnacht}
  C.~D.,   {Koopmans} L.~V.~E.,  2012, \mn@doi [\nat] {10.1038/nature10669},
  \href {https://ui.adsabs.harvard.edu/abs/2012Natur.481..341V} {481, 341}

\bibitem[\protect\citeauthoryear{{Vegetti}, {Koopmans}, {Auger}, {Treu}  \&
  {Bolton}}{{Vegetti} et~al.}{2014}]{unrealistic_tidal_assumption}
{Vegetti} S.,  {Koopmans} L.~V.~E.,  {Auger} M.~W.,  {Treu} T.,   {Bolton}
  A.~S.,  2014, \mn@doi [\mnras] {10.1093/mnras/stu943}, \href
  {https://ui.adsabs.harvard.edu/abs/2014MNRAS.442.2017V} {442, 2017}

\bibitem[\protect\citeauthoryear{{Vogelsberger}, {Zavala}  \&
  {Loeb}}{{Vogelsberger} et~al.}{2012}]{velocity_dependent_sidm}
{Vogelsberger} M.,  {Zavala} J.,   {Loeb} A.,  2012, \mn@doi [\mnras]
  {10.1111/j.1365-2966.2012.21182.x}, \href
  {https://ui.adsabs.harvard.edu/abs/2012MNRAS.423.3740V} {423, 3740}

\bibitem[\protect\citeauthoryear{{Vogelsberger}, {Zavala}, {Cyr-Racine},
  {Pfrommer}, {Bringmann}  \& {Sigurdson}}{{Vogelsberger}
  et~al.}{2016}]{2016MNRAS.460.1399V}
{Vogelsberger} M.,  {Zavala} J.,  {Cyr-Racine} F.-Y.,  {Pfrommer} C.,
  {Bringmann} T.,   {Sigurdson} K.,  2016, \mn@doi [\mnras]
  {10.1093/mnras/stw1076}, \href
  {https://ui.adsabs.harvard.edu/abs/2016MNRAS.460.1399V} {460, 1399}

\bibitem[\protect\citeauthoryear{{Wagner-Carena}, {Aalbers}, {Birrer},
  {Nadler}, {Darragh-Ford}, {Marshall}  \& {Wechsler}}{{Wagner-Carena}
  et~al.}{2022}]{Wagner-Carena:2022mrn}
{Wagner-Carena} S.,  {Aalbers} J.,  {Birrer} S.,  {Nadler} E.~O.,
  {Darragh-Ford} E.,  {Marshall} P.~J.,   {Wechsler} R.~H.,  2022, arXiv
  e-prints, \href {https://ui.adsabs.harvard.edu/abs/2022arXiv220300690W} {p.
  arXiv:2203.00690}

\bibitem[\protect\citeauthoryear{Wang, Bose, Frenk, Gao, Jenkins, Springel  \&
  White}{Wang et~al.}{2020}]{Wang:2019ftp}
Wang J.,  Bose S.,  Frenk C.~S.,  Gao L.,  Jenkins A.,  Springel V.,   White S.
  D.~M.,  2020, \mn@doi [Nature] {10.1038/s41586-020-2642-9}, 585, 39

\bibitem[\protect\citeauthoryear{Xu, Mao, Cooper, Gao, Frenk, Angulo  \&
  Helly}{Xu et~al.}{2012}]{10.1111/j.1365-2966.2012.20484.x}
Xu D.~D.,  Mao S.,  Cooper A.~P.,  Gao L.,  Frenk C.~S.,  Angulo R.~E.,   Helly
  J.,  2012, \mn@doi [Monthly Notices of the Royal Astronomical Society]
  {10.1111/j.1365-2966.2012.20484.x}, 421, 2553

\bibitem[\protect\citeauthoryear{\c{S}eng\"ul, Tsang, Diaz~Rivero, Dvorkin, Zhu
   \& Seljak}{\c{S}eng\"ul et~al.}{2020}]{CaganSengul:2020nat}
\c{S}eng\"ul A.~c.,  Tsang A.,  Diaz~Rivero A.,  Dvorkin C.,  Zhu H.-M.,
  Seljak U.,  2020, \mn@doi [Phys. Rev. D] {10.1103/PhysRevD.102.063502}, 102,
  063502

\bibitem[\protect\citeauthoryear{\c{S}eng\"ul, Dvorkin, Ostdiek  \&
  Tsang}{\c{S}eng\"ul et~al.}{2021}]{sengul2021}
\c{S}eng\"ul A.~c.,  Dvorkin C.,  Ostdiek B.,   Tsang A.,  2021, arXiv
  e-prints, \href {https://ui.adsabs.harvard.edu/abs/2021arXiv211200749C} {p.
  arXiv:2112.00749}

\bibitem[\protect\citeauthoryear{{de Blok}}{{de
  Blok}}{2010}]{2010AdAst2010E...5D}
{de Blok} W.~J.~G.,  2010, \mn@doi [Advances in Astronomy]
  {10.1155/2010/789293}, \href
  {https://ui.adsabs.harvard.edu/abs/2010AdAst2010E...5D} {2010, 789293}

\bibitem[\protect\citeauthoryear{de Blok, McGaugh, Bosma  \& Rubin}{de~Blok
  et~al.}{2001}]{deBlok:2001hbg}
de Blok W. J.~G.,  McGaugh S.~S.,  Bosma A.,   Rubin V.~C.,  2001, \mn@doi
  [Astrophys. J. Lett.] {10.1086/320262}, 552, L23

\makeatother
\end{thebibliography}
\label{lastpage}
\end{document}